\documentclass[usenatbib, twocolumn]{mnras}

\usepackage{newtxtext,newtxmath}
\usepackage[T1]{fontenc}

\DeclareRobustCommand{\VAN}[3]{#2}
\let\VANthebibliography\thebibliography
\def\thebibliography{\DeclareRobustCommand{\VAN}[3]{##3}\VANthebibliography}

\usepackage{amsmath, caption, subcaption, gensymb, graphicx, multirow, braket}  % improve math presentation
\usepackage[dvipsnames]{xcolor}
\usepackage{makecell}
\usepackage[normalem]{ulem}

\title[Multipoles and time-delay cosmography]{Multipoles in the IllustrisTNG simulation galaxies and implications for Strong Gravitational Lens modelling in Time-delay Cosmography}

\author[A. Al Zaif et al.]{
Abdullah Al Zaif$^{1}$\thanks{E-mail: aaz26@cantab.ac.uk},
Suhail Dhawan$^{3}$,
Vasily Belokurov$^{1}$
and
Erin E. Hayes$^{1,2}$
\\
$^{1}$Institute of Astronomy, Madingley Road, Cambridge CB3 0HA, UK\\
$^{2}$Kavli Institute for Cosmology, University of Cambridge, Madingley Road, CB3 0HA, UK\\
$^{3}$School of Physics and Astronomy, University of Birmingham, Birmingham, B15 2TT, UK
}

\date{Accepted XXX. Received YYY; in original form ZZZ}

\pubyear{\the\year{}}

\begin{document}
\label{firstpage}
\pagerange{\pageref{firstpage}--\pageref{lastpage}}
\maketitle

\begin{abstract}

Time-delay cosmography is a powerful independent probe of the Hubble constant ($H_0$), yet reaching the sub-2\% precision required to resolve the Hubble tension demands a rigorous treatment of mass model systematics. We use realistic massive elliptical galaxies from the IllustrisTNG hydrodynamic simulation suite to characterise higher-order angular mass complexity in the form of multipoles ($m=1,3,4$) and quantify their impact on $H_0$ inference. We find that these multipole perturbations are common and emergent features of galaxy assembly; in 51\% of our total samples ($54\%$ TNG100-1 galaxies and $85$\% for TNG50-1 galaxies), the addition of multipole terms to a standard elliptical power-law model is preferred by the Bayesian Information Criterion. Ignoring these structures in lens modelling introduces systematic biases in $H_0$: for perturbations of $\sim2\%$, the biases are of the order of 1--2\% for $m=1$, 4--5\% for $m=3$, and 3--6\% for $m=4$ multipole. While the $m=4$ (boxy/discyness) mode induces a predictable directional bias, odd-order modes ($m=1,3$) introduce significant stochastic scatter. We demonstrate a degeneracy where unmodelled internal multipoles can be effectively absorbed by the external shear component. This combined with low signal-to-noise observations can create ``hidden biases'' where formally perfect fits to imaging data ($\chi_\nu^2 \approx 1$) could harbour significant cosmological errors. Our results provide an astrophysical basis for multipole priors and highlight the necessity of moving beyond rigid parametric models to achieve high-accuracy cosmography in the era of \textit{Euclid} and LSST.

\end{abstract}


\begin{keywords}
{{Strong Lensing}{} --- {Cosmology}{} --- {Galaxies: structure}} %--- {Hubble Tension}
\end{keywords}

\section{Introduction}
Gravitational lensing is the phenomenon wherein light from a background source is distorted by an intervening massive object. In favourable configurations, this can lead to multiple images of the background source.
Strong gravitationally lensed supernovae (gLSNe) are excellent laboratories for cosmography and astrophysics of the high-$z$ universe. For multiply imaged lensed supernovae, the delay in the arrival time between individual images is sensitive to the present-day expansion rate of the universe, i.e. the Hubble constant ($H_0$).
The value of the $H_0$ is currently a topic of contention: the inference from the early universe using the cosmic microwave background (CMB) and assuming the standard $\Lambda$CDM cosmology \citep{Planck2018} and the local Cepheid distance ladder \citep{Riess2022} is discrepant at the $\sim 5 \sigma$ level \citep{verde_tensions_2019, verde_tale_2023}. However, some analyses using older stellar populations as distance indicators find a smaller discrepancy \citep[e.g.,][]{Freedman2021}. This difference, colloquially dubbed as the ``Hubble tension'' is one of the strongest challenges to standard cosmology. If it is not a manifestation of unknown systematics, it is the strongest sign of new physics beyond standard cosmology. As gLSNe probe time-delay distances, which is a ratio of angular diameter distances, they are independent of both the local distance ladder and early universe physics, making them an excellent probe of $H_0$. For time-delay cosmography to meaningfully weigh in on the $H_0$ tension at the 5$\sigma$ level (order of $\sim$10\%) as an independent probe, the total systematic and statistical error budget should be $< 2\%$ \citep{ding_time_2021, suyu_strong_2024}.

The physical quantity of interest in time-delay cosmography is the difference between the Fermat potential $\phi$ at the individual image positions \citep[e.g. see][for a review of time-delay cosmography]{ suyu_strong_2024}. The Fermat potential is inferred from the convergence $\kappa$ which is the projected density of the lensing mass distribution normalised by the critical density ($\Sigma_{\rm crit}$). Recovering $H_0$ from the arrival time delays requires accurate modelling of the gravitational potential of the deflector system. In the simplest case, this deflector is a single lens galaxy. In earlier studies, the mass distribution of lensing galaxies have been modelled parametrically as singular isothermal spheres (SIS), i.e. systems where the surface mass density $\Sigma \propto r^{-2}$. More flexible parametric forms of $\kappa$ include ellipticity and a free radial slope which are referred to as elliptical power law (EPL) models \citep{tessore_elliptical_2015}. Modelling the light of the background source, i.e. the host galaxy of the transient \citep[e.g.][]{2010ApJ...711..201Suyu, birrer_lenstronomy_2018}, provides additional constraints to break degeneracies between the model parameters and accurately infer $H_0$. This has been possible due to the development of modelling algorithms over the last couple of decades, e.g. semi-linear inversion \citep{2003ApJ...590..673Warren} and Bayesian regularisation \citep{2006MNRAS.371..983Suyu}, progressing from ideas in \citet{1992ApJ...401..461Kochanek} and \citet{1996ApJ...465...64Wallington}.

Simple parameterisations of the lens galaxy mass distribution, however, may be insufficient to explain the diversity of the lens population. For example, angular structure is known to play a role in the mass distributions of some lens galaxies \citep{evans_fitting_2003,2018MNRAS.475.1987Gomer, kochanek_overconstrained_2020, 2021MNRAS.504.1340Gomer}. Unaccounted-for deviations of the projected density profile from simple radial power law models can bias $H_0$ at up to the $\sim 10\%$ level \citep{xu_lens_2016} due to the mass-sheet transformation \citep[MST;][]{falco_model-dependent_1985,schneider_mass-sheet_2013}. In cosmographic studies, perturbations from an idealised EPL model are frequently captured with an `external shear' component \citep[e.g.,][]{birrer_tdcosmo_2020,shajib_tdcosmo_2023}. While this component accounts for higher order complexity than a simple EPL, it is physically disconnected from and may not appropriately capture the effect of angular structure in the lens galaxy \citep{etherington_strong_2024}.

Angular structure in the lensing galaxy can affect the inferred Fermat potential difference, $\Delta \phi$, and hence, $H_0$. There are a few different prescriptions of the ways of describing angular structure in elliptical galaxies. One way is to infer the radial variation of the ellipticity or look for ellipticity twists in the mass distribution \citep{2019MNRAS.489.2049Nightingale, 2024MNRAS.532.2441He}. Another method is to look for multipole perturbations to the standard mass profile, namely lopsidedness ($a_1$), triangulation ($a_3$), and boxy/discyness ($a_4$), which are the focus of this work. Observational evidence of higher order multipole structures, particularly $a_3$ and $a_4$, were obtained from fitting stellar isophotes of early-type galaxies (ETGs) by \citet{bender_isophote_1988} using CCD-photometry of 
69 nearby bright ellipticals, followed by a larger sample of 847 SDSS galaxies by \citet{hao_isophotal_2006}. Necessity for angular complexity using multipoles has also been motivated by microwave wavelength data of strong lenses from ALMA \citep[e.g.,][]{stacey_complex_2024}. \citet{amvrosiadis_lopsidedness_2025} find that 5 out of 30 ETGs show strong evidence of lopsidedness, i.e. $a_1$ multipole. These galaxies have a physical companion, which while being a more complex system than only a single galaxy, could likely be the deflector system for some lensed SNe, e.g. potentially for SN2025wny \citep{2025ApJ...995L..17Johansson,Taubenberger2025,2026arXiv260402418Storfer,2026arXiv260216620Ecker}. Models with complexities including angular structure are also important for inference of dark matter substructure in the deflector, in which context they have been extensively explored \citep[e.g, ][]{cohen_general_2024, oriordan_angular_2024, 2025MNRAS.539..704Lange}. \citet{2025MNRAS.539..704Lange} find that for the one of the two lensed quasars in their work, there is a non-zero detection of multipoles corresponding to the lopsidedness ($a_1$), as well as triangulation ($a_3$) and boxy/discyness ($a_4$).

In this paper, we investigate the diversity of angular structure in potential lens galaxies using self-consistent hydrodynamical simulations from the IllustrisTNG project \citep{nelson_illustristng_2019, pillepich_simulating_2018}. We specifically characterise departures from elliptical symmetry through the higher-order multipoles of the lensing potential, namely, the lopsidedness ($a_1$), triangulation ($a_3$) and boxy/discyness ($a_4$). For each of the multipoles we account for ellipticity explicitly in the formalism, as ignoring ellipticity has been shown to lead to unphysical mass distributions \citep{paugnat_elliptical_2025}. By measuring the distributions of these multipoles across a simulated population of massive ETGs, we establish a physically motivated baseline for the angular complexity of potential deflectors of strongly lensed supernovae (gLSNe). Subsequently, based on the input multipoles from these distributions, we test what the impact of not explicitly modelling the angular structure is on the inferred value of $H_0$ for different multipole orders in mock lens systems. This extends previous work by considering multiple orders of angular structure within a physically motivated framework, rather than focusing on a single prescribed perturbation. In particular, while the cosmographic impact of fourth-order multipoles has previously been investigated \citep{vyvere_tdcosmo_2022}, the effects of lopsided and triangular structure remain comparatively unexplored.
%An investigation of how simple radial power law models can bias $H_0$, when there are deviations in the projected density profile, e.g. due to the mass-sheet transformation \citep[MST;][]{falco_model-dependent_1985,schneider_mass-sheet_2013}, demonstrated biases upto few to $\sim 10\%$ \citealp{xu_how_2015}.
While the MST can be broken with independent observables, e.g. lensing magnification if the gLSN is of Type Ia \citep{pascale_sn_2024} or stellar kinematics \citep[demonstrated with quasars by \citealp{shajib_tdcosmo_2023}, and tested with EAGLE simulations in][]{2018MNRAS.474.3403Tagore}, there is an imperative need to investigation of how sources of modelling systematics like higher order multipoles affect cosmography. Exploring multipoles of various different origins is crucial to test their impact on cosmography as future samples of strongly lensed transients grow, particularly when the aspiration is to reduce the total error budget, including systematics, to less than 2$\%$.

%\sdnotetwo{A line here foreshadows the results}

The structure of the paper is as follows. We present the theory of time-delay cosmography in section~\ref{section 2: Theory}, followed by the methodology applied to model the simulated lens galaxies in section~\ref{sec:methodology}. We summarise our results in section~\ref{sec:results}, and their impact on time-delay cosmography in section~\ref{sec:impact_tdcosmo}. We discuss them in section~\ref{sec:discussion} and finally conclude in section~\ref{sec:conclusion}.

\section{Theory}
\label{section 2: Theory}
\subsection{\textbf{Lensing Formalism}}

The basic theory and vernaculars of strong gravitational lensing are summarised first.

The arrival time difference, or time delay, between any two lensed images (e.g. image A and image B) is given by: 
\begin{equation}
    \Delta t_{AB} = \frac{D_{\Delta t}}{c}[\phi(\boldsymbol{\theta}_{A} ) - {\phi}(\boldsymbol{\theta}_{B})].
    \label{equ: time-delay equation}
\end{equation}
Here, $D_{\Delta t}$ is the time-delay distance and $\phi(\boldsymbol{\theta})$ is the Fermat potential \citep{schneider_new_1985, blandford_fermats_1986}. The time-delay distance is a combination of the angular diameter distances between the source, lens and observer \citep{refsdal_possibility_1964, schneider_gravitational_1992}:
\begin{equation}
     D_{\Delta t} = (1+z_{L}) \frac{D_{L}D_{S}}{D_{LS}} 
\end{equation}
where $D_{LS}$, $D_S$, and $D_L$ correspond to the lens-to-source, observer-to-source and the observer-to-lens distances respectively, and $z_L$ is the redshift of the lens.

The Fermat potential is defined as:
\begin{equation}
    \phi(\boldsymbol{\theta}) = \frac{1}{2} (\boldsymbol{\theta}-\boldsymbol{\beta})^2-\psi(\boldsymbol{\theta})
\end{equation}
where $\boldsymbol{\beta}$ is the angular position of the background light source, $\boldsymbol{\theta}$ is the angular position of observed image, and $\psi(\boldsymbol{\theta})$ is the lensing potential. The first term in the Fermat potential describes the \textit{geometric delay} due to the path difference of the light travelling through the lens and the second term accounts for the \textit{Shapiro delay} arising from local spacetime dilation. The lensing potential is the projected Newtonian potential, $\Phi$, of the lens along the line of sight from the source to observer \citep{narayan_lectures_1997} given by: 
\begin{equation}
    \psi(\boldsymbol{\theta}) = \frac{2}{c^2} \frac{D_{LS}}{D_{S}D_{L}} \int \Phi(D_{L}\boldsymbol{\theta}, l) \, dl 
\end{equation}

The lensing convergence is the surface mass density (along line of sight) normalised by the critical lensing surface density:
\begin{equation}
    \kappa(\boldsymbol{\theta}) = \frac{\Sigma(\boldsymbol{\theta})}{\Sigma_{\text{crit}}}
    \label{eqn : lensing kappa}
\end{equation}
where the critical density is defined as:
\begin{equation}
    \Sigma_{\text{crit}} = \frac{c^2 D_{S}}{4\pi G D_{L}D_{LS}} 
    \label{eqn: Sigma_crit}
\end{equation}

The lensing convergence is obtained from the lensing potential using
\begin{equation}
     \kappa(\boldsymbol{\theta}) = \frac{1}{2} \nabla^2 \psi(\boldsymbol{\theta})
     \label{eqn: lensing potential ODE}
\end{equation}

The Hubble constant is inversely proportional to absolute scales of the Universe and scales with the time-delay distance as:
\begin{equation}
     H_{0} \propto \frac{1}{D_{\Delta t}}
     \label{equ: Hubble-time delay disctance scaling}
\end{equation}

Therefore, by measuring the time delays $\Delta t_{AB}$ and modelling the deflection and line-of-sight mass distributions to obtain the Fermat potential difference $\Delta \phi_{AB}$, we can infer the time-delay distance. Given an assumed cosmology type, we can constrain $H_0$ with weak dependence on other cosmological parameters.

\subsection{\textbf{Galaxy Mass  Models}}

\subsubsection{The Elliptic Power Law Profile}

\label{Section 2.3: The Lens Model}

% \begin{figure}
% \begin{subfigure}{0.45\textwidth}
%   \centering
%   \includegraphics[width=1.\linewidth]{Figures/Introduction+Theory/Multipole-Isodensity-contour.png}
%     \caption{}
%     \label{fig: Multipole-Isodensity-contour}
% \end{subfigure}
% \begin{subfigure}{0.45\textwidth}
%   \centering
%   \includegraphics[width=1.\linewidth]{Figures/Introduction+Theory/Circular vs Elliptical multipole.png}
%   \caption{}
%     \label{fig: Circular vs Elliptical multipole}
% \end{subfigure}
% \caption{\rev{\textbf{(a)} Isodensity contours for an EPL mass profile (grey dashed line) and an EPL + fourth-order multipole mass profile (black plain line) of slope $\gamma=2.0$, and amplitudes $a_4=0.03$ (top) and $a_4=0.05$ (bottom) along the EPL main axis to create a discy profile (left), or at $45\degree$ to the main axis to create a boxy profile (right). \textbf{(b)} Fourth-order multipole patterns in the circular (left) and elliptical (right) formulation, applied to reference ellipses with the same radius but varying axis ratio $q$. The circular multipoles yield perturbation shapes that depend on the axis ratio, and that do not behave as expected for high ellipticities (adapted from \citealp{paugnat_elliptical_2025}) \textbf{[Replace Figure]}}} 
% \label{fig: Multipoles Diagram}

% \end{figure}

% More than $50\%$ of the stellar mass in the low-redshift universe is in early-type galaxies (ETGs) \citep{fukugita_cosmic_1998}. 
%Massive elliptical galaxies are thought to be the end product of hierarchical mergers of dark matter haloes along with stars and gas contained within them. 
The total mass density profile of the dark matter and baryonic components of elliptical galaxies is found to be well approximated by a power-law distribution (i.e. $\rho(r)_{\text{tot}} \propto r^{-\gamma}$), with a mean slope of $\langle \gamma \rangle = 2.08 \pm 0.03$ \citep{auger_sloan_2010}.

For an elliptic power-law model (EPL), the lensing convergence is defined as \begin{equation}
    \kappa_{\text{EPL}}(x,y) = \frac{3-\gamma}{2} \left( \frac{\theta_{E}}{\sqrt{ qx^2 + y^2/q }} \right)^{\gamma-1}
    \label{eqn: EPL_convergence}
\end{equation}
where $\theta_E$ is the Einstein radius, $\gamma$ is the logarithmic mass slope of the profile with $1<\gamma<3$, $q$ is the axis ratio of the elliptical profile, and $(x,y)$ are the Cartesian coordinates on the lensing plane \citep{tessore_elliptical_2015}.

\subsubsection{Multipole Mass Profile}

\begin{figure}
    \centering
    \begin{subfigure}[t]{0.25\textwidth}
        \centering
        \includegraphics[height=1.6in]{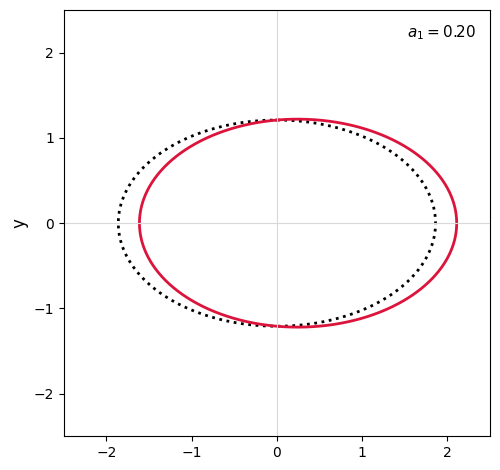}

    \end{subfigure}%
    ~ 
    \begin{subfigure}[t]{0.25\textwidth}
        \centering
        \includegraphics[height=1.6in]{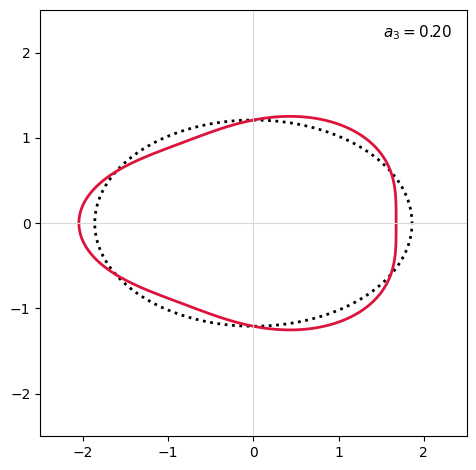}

    \end{subfigure}
    ~ 
    \begin{subfigure}[t]{0.25\textwidth}
        \centering
        \includegraphics[height=1.7in]{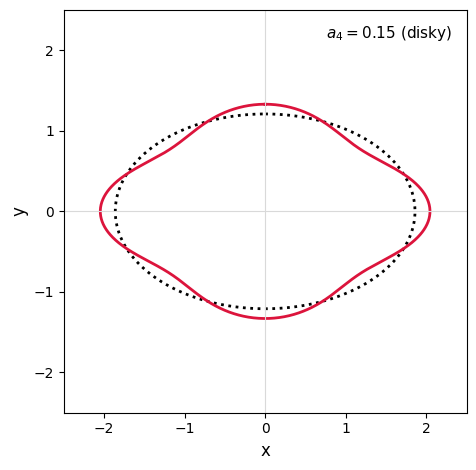}

    \end{subfigure}%
    ~ 
    \begin{subfigure}[t]{0.25\textwidth}
        \centering
        \includegraphics[height=1.7in]{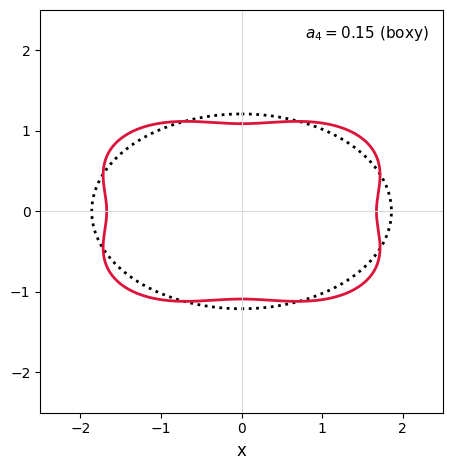}

    \end{subfigure}
    \caption{Perturbations corresponding to the m = 1 (top left) , m = 3 (top right), and m = 4 (bottom panel) multipoles applied to reference ellipses with the same elliptical radius $R = 1.5$ and axis ratio $q=0.65$. The $m=1$ and $m=3$ multipoles are aligned with the ellipse ($\phi_m=0$). The $m=4$ multipole aligned with the reference ellipse ($\phi_m=0$) creates a discy shape (bottom left). A boxy shape is produced if the $m=4$ multipole is aligned at $45^{\degree}$ to the axis of the ellipse. The values for $a_m$ are larger than physical expectations for better visualization of the deviation from perfect ellipticity.}
    \label{fig:Multipole perturbations}
\end{figure}

Deviations from pure ellipticity are considered by multipole expansions of the elliptic profiles of galaxies. Multipoles are Fourier-type perturbations added to the angular part of the density profile while keeping the radial part unchanged. They are defined by a deformation to the 2D projected isodensity ellipse as:
\begin{equation}
    R(\phi) \rightarrow R(\phi)+\delta R_{m}(\phi) \  \text{where} \   \delta R_{m}(\phi) = a_m \cos(m(\phi-\phi_m))
    \label{equ: Multipole isophote deformation}
\end{equation} where $R = \sqrt{qx^2+y^2/q}$ and $\phi = \arctan(qx,y)$ are elliptical polar coordinates, $m$ is the order of the multipole, $a_m$ is the physical multipole strength, and $\phi_m$ is the multipole orientation angle. 
%\citep{keeton_identifying_2003, xu_how_2014}. 
We adopt the elliptical multipole formulation that describes deviations from ellipticity for profiles of any axis ratio according to the formalism in \cite{paugnat_elliptical_2025}.
The corresponding lensing convergence describing multipole perturbations to the power-law mass profile is given by: \begin{equation}
    \kappa_m({x,y}) = \frac{1}{2\sqrt{ qx^2 + y^2/q }} a_{m} \cos(m(\phi-\phi_{m}))
    \label{eqn: EPL-multipoles}
\end{equation}

%The external shear represents a quadrupolar perturbation ($m=2$) to the gravitational potential of the deflector. 
The $m=1$ multipole is an asymmetric perturbation that represents lopsidedness in the mass distribution, while the $m=3$ and $m=4$ multipoles are hexapolar and octupolar moments of the potential respectively and the next most significant in galaxies. 
%The use of multipoles is motivated by photometric observations of elliptical galaxy isophotes.
The third-order multipole gives rise to a non-centrosymmetric, triangular distortion to the base profile of elliptical galaxies, and the fourth-order multipole introduces characteristic ``discy" ($\phi_4 = 0$) or ``boxy" ($\phi_4 = 45\degree$) shapes which have been observed in photometric studies of elliptical \citep{bender_isophote_1988, hao_isophotal_2006}. The $m=2$ quadrupole moment describes an elliptical or elongated distortion of an otherwise perfectly symmetric (spherical or circular) system. This complexity is already accounted for by the axis ratio parameter of the EPL profile or by the external shear in lens modelling, and so is not considered in this work.

\begin{figure*}
    \centering
    \begin{subfigure}[t]{0.5\textwidth}
        \centering
        \includegraphics[height=2.8in]{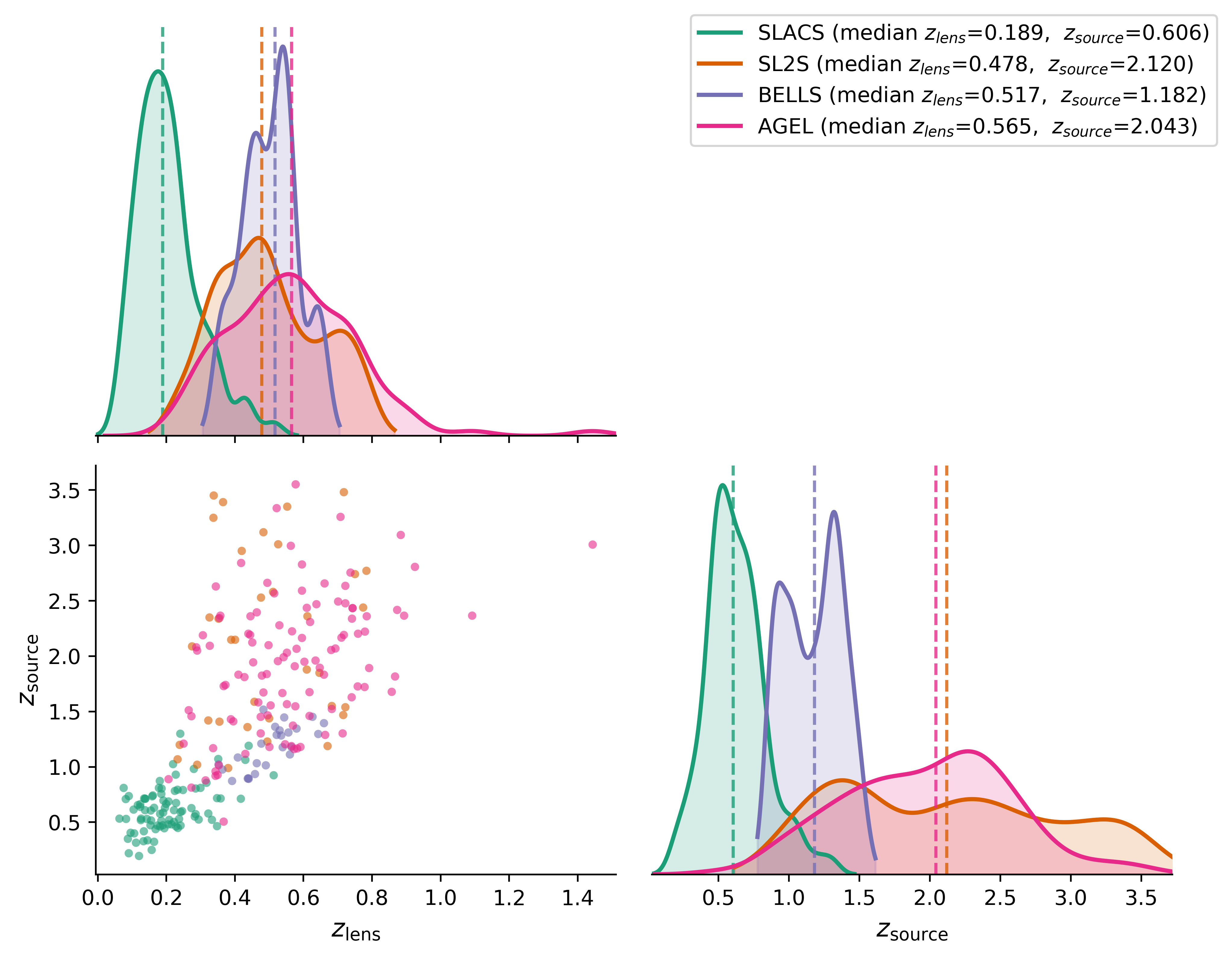}

    \end{subfigure}%
    ~ 
    \begin{subfigure}[t]{0.5\textwidth}
        \centering
        \includegraphics[height=2.8in]{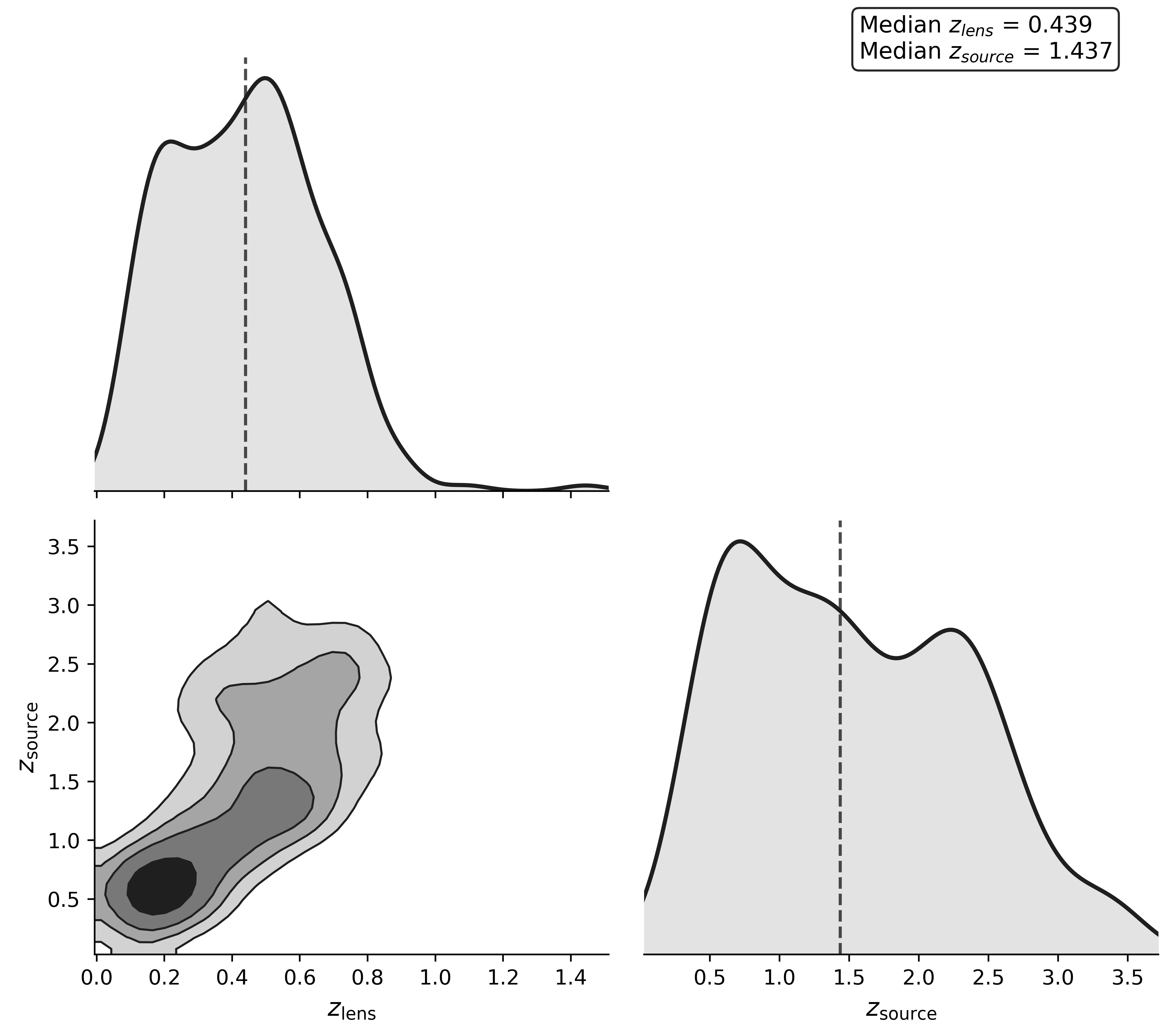}

    \end{subfigure}
    \caption{Lens galaxy and source redshift distributions of galaxy-scale strong lensing systems from major galaxy surveys in the past two decades. \textbf{Left:} The density distributions of redshifts from the Sloan Lens ACS (SLACS) Survey \citep{auger_sloan_2009}, BOSS Emission-Line Lens Survey (BELLS) \citep{brownstein_boss_2011}, Strong Lensing Legacy Survey (SL2S) \citep{sonnenfeld_sl2s_2013} and ASTRO 3D Galaxy Evolution with Lenses (AGEL) Survey \citep{barone_agel_2025}. \textbf{Right:} The combined redshift distribution with the $1\sigma$, $2\sigma$ and $3\sigma$ contour intervals, where the median lens and source redshift of all samples is 0.439 and 1.437 respectively.}
    %\rev{[Note: Euclid lenses have not been considered here, but have observed lens photometric redshifts of between $0.2-0.8$, so median is consistent]}
    \label{fig: Lens galaxy and source redshifts strong lenses in galaxy surveys}
\end{figure*}

% \begin{figure*}
%     \centering
%     \includegraphics[width=0.5\linewidth]{Figures/Einstein_Radius_vs_Redshift.png}
%     \caption{Einstein radius versus lens galaxy redshift in galaxy surveys from the SLACS, SL2S, and BELLS lens samples}
% \end{figure*}

%%%%%%%%%%%%%%%%%%%%%%%%%%%%%%%%%%%%%%%%%%%%%%%%%%%
\section{{Methodology}}
\label{sec:methodology}
\subsection{The simulation suite}

The IllustrisTNG project (hereafter referred to as TNG) is a set of large-volume magnetohydrodynamic cosmological  simulations of galaxy formation run with the moving-mesh code AREPO \citep{springel_e_2010}. The project consists of three simulation volumes: TNG50, TNG100, and TNG300 of sizes (51.7 Mpc)$^3$, (110.7 Mpc)$^3$, and (302.6 Mpc)$^3$ respectively. The simulations use the Planck 2015  cosmology parameters \citep{planck_collaboration_planck_2016}, with a total matter density parameter $\Omega_m=0.3089$, dark energy density parameter $\Omega_{\Lambda}=0.6911$ and reduced Hubble constant $h=0.6774$. Each simulation follows the coupled evolution of dark matter (DM) and gas from the early universe ($z=127$) until today ($z=0$) under the $\Lambda$CDM paradigm, self-consistently modelling gravitational interaction and magnetohydrodynamical interactions of the gaseous component. By taking into account various baryonic processes, such as gas cooling, stellar evolution and feedback, chemical enrichment, supermassive black hole growth and feedback from active galactic nuclei, the TNG simulation resolves several hundred thousand galaxies of diverse morphologies. The full physics TNG simulation suite reproduces many fundamental relations in observed galaxies, such as the galaxy-colour bimodality, the evolution of the mass-metallicity relation, the galaxy size-mass relation, and the Tully-Fisher relations. These verifications of the TNG physics model are strong confirmation of the plausibility of the simulated IllustrisTNG galaxy and galaxy cluster populations \citep{wang_early-type_2020}.

For this paper, we specifically use the TNG100-1 \citep{nelson_first_2018, pillepich_first_2018, springel_first_2018, marinacci_first_2018, naiman_first_2018} and the TNG50-1 \citep{pillepich_first_2019, nelson_first_2019} simulation volumes which offer the highest satisfactory resolutions for our modelling purpose: TNG50 has a dark matter resolution of $m_{\text{DM}}=4.5\times10^{5}$ M$_{\odot}$ and mean baryonic gas mass resolution of $m_{\text{baryon}}=8.5\times10^{4}$ M$_{\odot}$, while TNG100 has resolutions of $m_{\text{DM}}=7.5\times10^{6}$ M$_{\odot}$ and $m_{\text{baryon}}=1.4\times10^{6}$ M$_{\odot}$.  In TNG50 the gravitational softening lengths are 290 pc at 
z = 0 for the stars and dark matter, while the gas softening is adaptive with a minimum of 74 comoving parsecs. In TNG100 the adaptive gas softening
length has the minimum comoving value $\epsilon_{gas,min} = 0.185$ kpc, while the stellar and dark matter softening length is $\epsilon_{DM,\star} = 0.74$ kpc in comoving units at $z = 0$.

\subsection{Classification of Elliptical Galaxies} %or maybe Selection of...
\label{sec: Classification of ETGs}

%In order to apply cosmographic methods to the elliptical galaxies in the IllustrisTNG simulations, we have to convert the three-dimensional particle representation (stars+dark matter) of galaxies into two-dimensional lensing convergence that is suitable for lens modelling in \texttt{Lenstronomy}.

As the TNG simulations do not directly classify the morphology of galaxies, we determine galaxies as elliptical in the simulation suite based on the following 3 criteria: (i) specific star formation rate, (ii) stellar dynamical property, and (iii) Sérsic index. We only consider galaxies with a minimum total stellar mass of $M^{\star}_{\text{tot}} = 10^{10.5}$ M$_{\odot}$ to implement our classification criteria.

The first criterion for selecting an elliptical galaxy is based on its specific star formation rate (sSFR) which is defined as the ratio of the star formation rate (SFR) and the stellar mass of the galaxy, i.e. sSFR $=$ SFR$/M^{\star}_{\text{tot}}$. The galaxy is likely to be an elliptical early-type galaxy (ETG) if it has an sSFR $<10^{-11}$ yr$^{-1}$, which follows from observational results and has been widely used in previous works separating star-forming from quiescent systems \citep{fontana_fraction_2009, donnari_star_2019, paspaliaris_star-forming_2023}.

The second criterion uses the dynamical properties of stars of the galaxy by measuring the fraction of kinetic energy invested in ordered rotation \citep{sales_origin_2012} given by
\begin{equation}
    k_{\text{rot}} = \frac{K_{\text{rot}}}{K} = \frac{1}{K} \sum \frac{1}{2} m_{i} \left( \frac{\boldsymbol{j}_{i} \cdot \hat{\boldsymbol{J}}}{|\boldsymbol{r}_{i}| \times \hat{\boldsymbol{J}}} \right)^2
\end{equation}
where $m_i$ is the mass of the $i$th stellar particle, $\boldsymbol{j}_{i} = \boldsymbol{r}_{i} \times \boldsymbol{v}_{i}$ is the specific angular momentum, $\hat{\boldsymbol{J}}$ denotes the direction of the total angular momentum, and $K=\sum \frac{1}{2} m_{i} |\boldsymbol{v}_{i}|^2$ is the total kinetic energy of the galaxy. In the calculation of $k_{\text{rot}}$, we only consider the stellar particles within the spherical radius $r_{90}$, which is the radius enclosing $90\%$ of the total stellar mass, in order to avoid possible bias due to satellites at the outskirts of the galaxy. For a perfectly rotating, completely disc-dominated galaxy $k_{\text{rot}}$ tends to 1, whereas for a bulge-dominated or elliptical galaxy $k_{\text{rot}}$ will tend to 1/3. We define galaxies with $k_{\text{rot}}<0.5$ as ``elliptical" \citep{penoyre_origin_2017}.

The final criterion is based on the morphological property of the galaxy where we fit a Sérsic profile to the two-dimensional projected light distribution of the galaxy given by 
\begin{equation}
    I(R) = I_{e} \exp\left( -b_{n}\left[ \left( \frac{R}{R_{e}} \right)^{1/n} - 1 \right] \right)
    \label{equ: Sersic light profile}
\end{equation} where $I_e$ is the intensity at half-light radius $R_e$, $n$ is the Sérsic index of the profile, $b_n=1.9992n-0.3271$, and the elliptic radius is
\begin{equation}
    R = \sqrt{ x^2+\frac{y^2}{q_{s}^2} }
\end{equation} with $q_s$ the axis ratio of the profile and $(x,y)$ are Cartesian coordinates centred on the galaxy light centre and rotated by the orientation angle of the ellipse %$\phi_s$
\citep{ciotti_analytical_1999}.We project the stellar distribution of a galaxy along the $X$, $Y$, and $Z$ axes of the simulation volume and use a constant mass-to-light ratio to produce 2D intensity maps of the galaxy. We fit the elliptic Sérsic profile to the stellar intensity within a radius enclosing $90\%$ of the stellar mass using a two-stage fitting procedure based on a global differential evolution (DE) search on a Cash statistic, followed by a local least-squares refinement on log-space residuals. %\rev{[need some citations]}. 
We then inspect the inferred best-fit value of the Sérsic index for all three projections: if $n_i \geq 2$ for $i=X,Y$ and $Z$, then the galaxy is classified as an ETG \citep{fisher_structure_2008, bruce_morphologies_2012, ferreira_catalogue_2025}. 

%\rev{Further details and results on the classification criteria can be found in Appendix \ref{Appendix A}.}

% There are $52\,049$ galaxy samples in TNG100-1 and $5382$ samples in TNG50-1 at $z=0$ with total stellar mass $M_{\text{tot}}^\star>10^{10.5}$ M$_{\odot}$ and sSFR $<10^{-11}$ yr$^{-1}$.
% \label{Section: Classification of ETGs}

\subsection{The Galaxy sample projection}

Using our set of criteria in Section \ref{sec: Classification of ETGs}, we select 512 elliptical galaxies from TNG100-1 and 83 from TNG50-1 at redshift $z=0$, the latter generating a smaller sample size owing to the much smaller cosmological volume. In order to analyse our samples so that they are comparable to observed strong lensing data, we artificially place the galaxies at redshift $z_{\text{lens}}=0.44$ which is the median redshift of lenses in galaxy surveys (Figure \ref{fig: Lens galaxy and source redshifts strong lenses in galaxy surveys}).
\\
We use the $z=0$ snapshot to characterise the angular structure of a well-assembled, late-time population of massive ETGs. For a given IllustrisTNG run, the fixed particle mass resolution means that massive galaxies can be sampled by large numbers of stellar and dark-matter particles, facilitating the measurement of relatively subtle $m=1,3$ and $4$ multipoles while mitigating particle-discreteness effects \citep{pillepich_first_2018}. The resulting sample provides a baseline for studying angular structure in mature massive ETGs. However, galaxy structural properties can evolve with redshift \citep{wang_early-type_2019}, and we do not assume that the multipole distributions are invariant with redshift; quantifying their evolution is beyond the scope of the present study.
Furthermore, our sample is selected to characterise the intrinsic structural properties of massive ETGs in IllustrisTNG, rather than to reproduce the population of observed strong-lens deflectors. Strong-lensing selection favours particular combinations of lens mass, velocity dispersion, projected structure and source–lens geometry \citep{turner_statistics_1984, collett_population_2015}.
%For example, in the singular isothermal sphere profile, the Einstein radius scales as ($\theta_{\rm E}\propto\sigma_{\rm v}^{2}$) at fixed lens and source redshifts, and the strong-lensing cross-section scales as ($\theta_{\rm E}^{2}$) \citep{turner_statistics_1984, collett_population_2015}. 
The selection of quadruply imaged systems can introduce additional dependencies on projected ellipticity, orientation and the caustic structure of the lens, while the surrounding environment can further affect lensing probabilities \citep{mandelbaum_galaxy_2009, sonnenfeld_strong_2023}. We do not apply strong-lensing or quad cross-section weighting to the simulated population. Accordingly, the multipole distributions presented here should be interpreted as an intrinsic demographic baseline for massive ETGs in IllustrisTNG, rather than as a direct prediction for the distribution of multipoles among observed strong-lens deflectors.

%Incorporating redshift evolution and the relevant observational selection functions will be necessary to establish the corresponding distributions for a lens-selected population.
%, which includes the Sloan Lens ACS (SLACS) Survey \citep{auger_sloan_2009}, BOSS Emission-Line Lens Survey (BELLS) \citep{brownstein_boss_2011}, Strong Lensing Legacy Survey (SL2S) \citep{sonnenfeld_sl2s_2013}, and ASTRO 3D Galaxy Evolution with Lenses (AGEL) Survey \citep{barone_agel_2025}. 

We project each three-dimensional galaxy data onto two-dimensional planes to produce 2D density maps along the simulation $X$, $Y$, and $Z$ directions. The star, gas, DM and Black Hole particles are binned over spatial position on a given projection plane, using a resolution of $0.03^{\prime \prime}$/pixel at $z_{\text{lens}}=0.44$. %and the total particle masses summed per pixel. 
We use an adaptive triangular shaped cloud (ATSC) implemented through the Python package \texttt{PMESH} to turn the particle distribution into a 2D density map. 
In the standard triangular shaped cloud method (TSC, \citealp{hockney_computer_1981}), the mass of a particle ($m_p$) at location $\mathbf{r}$ is distributed within a $3\times3$ square of grid cells centred around the nearest grid cell to the particle, with the cell at location $\mathbf{r+x}$ receiving a fraction of the particle's mass. The mass assignment kernel in real space is described by $W(\mathbf{x}) = \prod_i W(x_i)$, with
\begin{equation}
    W_{\text{TSC}} (x_i) = \begin{cases}
        0.75 - x_i^2 & |x_i|\leq 0.5 \\
        \frac{(1.5-|x_i|^2)}{2} & 0.5 < |x_i| \leq 1.5 \\
        0 & \text{otherwise}
    \end{cases}
\end{equation}
where $x_i$ is the $i$-th component of $\mathbf{x}$. So the mass of a particle deposited at the grid cell at location $\mathbf{r+x}$ is $m_i \times W(\mathbf{x})$.

In the ATSC method that we use, an adaptive smoothing length analogous to that in smoothed particle hydrodynamics (SPH) is calculated for each particle type (stars, DM, gas) based on the 3D distance to its 16th nearest neighbour, $r_{16}$. The smoothing length is related to this distance by $h=r_{16}/\Delta x$, where $\Delta x$ is the size of a pixel. The ATSC mass assignment kernel is defined by
\begin{equation}
    W_{\text{ATSC}} (x_i, h) = \frac{1}{h} W_{\text{TSC}} \left( \frac{x_i}{h} \right)
    \label{equ: ATSC mass assignment kernel}
\end{equation}
The minimum of $h$ is set to $h=1$, and the maximum  is set to $h=5$, since large $h$ results in the particle mass being split between many grid cells, which can be computationally expensive.

We make separate maps of the stars, DM, and gas using the ATSC method, and summed together with the black hole map to obtain the total projected 2D surface density, $\Sigma_{\text{gal}}$, along a particular line of sight -- the $X$, $Y$, and $Z$ axes of the simulation in our case. We scale $\Sigma_{\text{gal}}$ with the critical surface density to produce the lensing convergence map $\kappa$ as defined in Equation \ref{eqn : lensing kappa}. The critical lensing surface density $\Sigma_{\text{crit}}$ is calculated using Equation \ref{eqn: Sigma_crit}, by assuming a redshift of the background lensed source of $z_{\text{source}}=1.4$, the median source redshift of observed strong lensing systems (Figure \ref{fig: Lens galaxy and source redshifts strong lenses in galaxy surveys}). For the rest of this paper, we adopt the flat $\Lambda$CDM cosmology as our fiducial cosmology with $H_0=70$ km s$^{-1}$ Mpc$^{-1}$ and $\Omega_m=0.3$ for calculations and simulations. The choice of value for the fiducial $H_0$ is the average of the Planck CMB measurement \citep{Planck2018} and the SH0ES collaboration \citep{Riess2022} using Cepheids and SNe Ia. The choice of cosmology only affects the time-delay distances and we do not expect it to affect our analysis of the convergence profiles of the TNG galaxies.

% we select our galaxies based on properties of massive ellipticals catalogued in large-scale surveys, such as the SLACS survey, the Sloan Digital Sky Survey (SDSS, \citealp{collaboration_first_2003}), and ATLAS$^{\text{3D}}$ \citep{cappellari_atlas3d_2011}.

% \rev{Ellipticals also have high velocity dispersion of stars $\sigma$, and so we use an additional criteria of $\sigma\geq 160$ km s$^{-1}$ for filtering the subhaloes. This lower limit is motivated by observed lens sample from the SLACS survey. We select galaxies within the range $0<z<0.7$ where the upper limit is the approximate maximum redshift at which the BELLS survey has observed a lensing galaxy \citep{brownstein_boss_2012}.}

% Once a suitable elliptical galaxy has been identified according to our selection criteria, we bin the stars, gas and DM particles over spatial position on a given plane (e.g.\ $x-y$) and sum up their masses. The critical lensing surface density $\Sigma_{\text{crit}}$ is calculated using Equation \ref{eqn: Sigma_crit} \sout{by considering redshifts of the background source between $0.2 < z<3.0$ as found in the analysis of lens samples in \textsc{Project Dinos}.

\subsection{Einstein Radius Computation}

\begin{figure}
    \centering
    \includegraphics[width=1.0\linewidth]{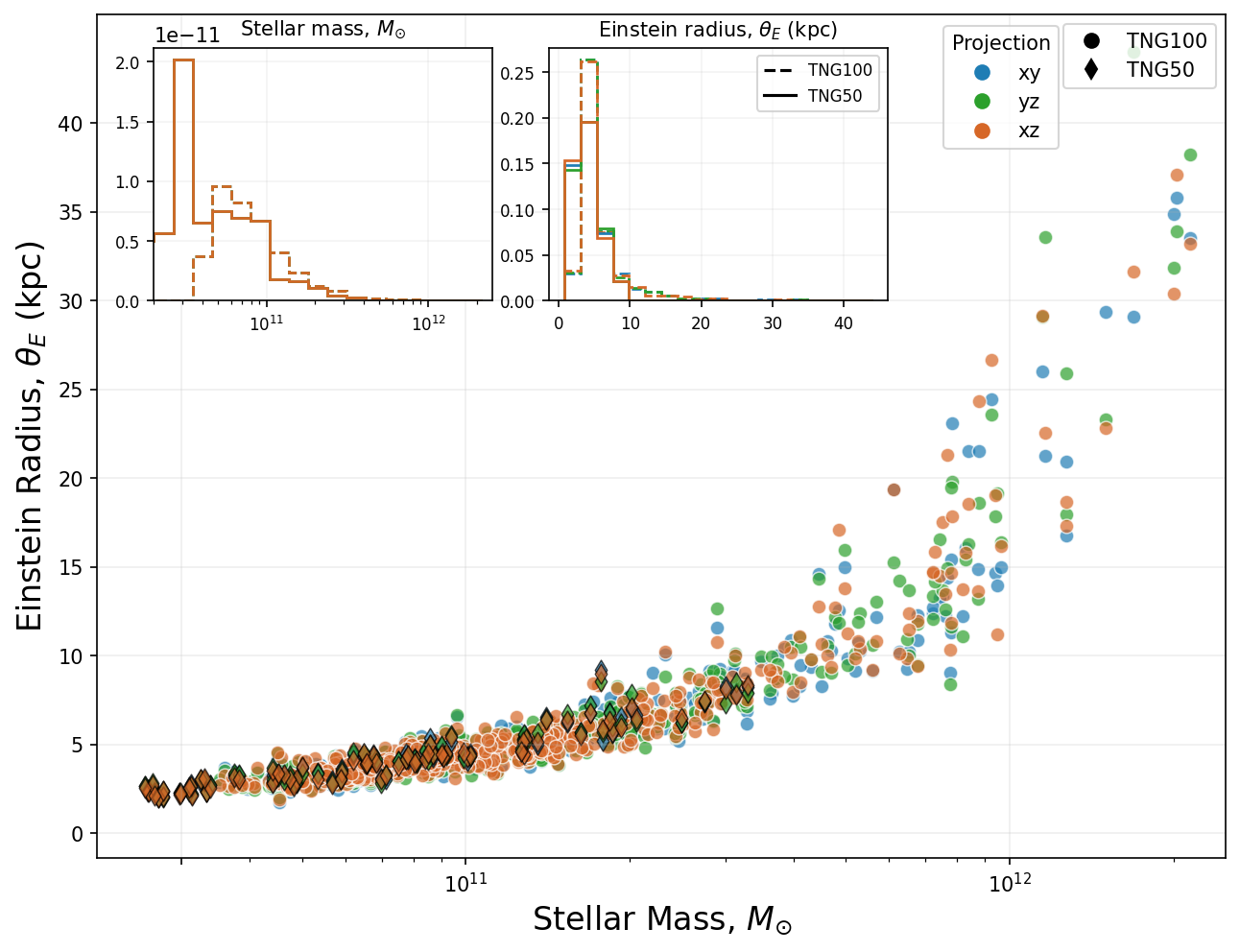}
    \caption{Distribution of effective Einstein radii against stellar masses obtained for the galaxy samples. The insets show the 1D histogram distributions for the total stellar mass and Einstein radius per projection. %The Einstein radii distributions are consistent across the three projections.
    }
    \label{fig: Einstein Radius vs. stellar mass}
\end{figure}

% \rev{The Einstein radius of a galaxy is a key quantity in determining its effectiveness in acting as a strong lens for a background source. In strong lensing statistics, the lensing optical depth, i.e. the probability for a lensing event, depends on the cross-section of a lens which in turn is directly proportional to the square of the Einstein radius, $\theta_E^2$ \add{(add citation: Gravitational Lensing: Strong, Weak and Micro; P Schneider book, What does Strong G Lensing; Robertson et al. 2020)}. The Einstein radius also provides a natural angular scale to describe the lensing geometry. In the case of multiple imaging, the typical angular separation of images is of order $2\theta_E$. Sources which are closer than about $\theta_E$ to the optic axis are significantly magnified, whereas sources which are located well outside the Einstein ring are magnified very little. In many lens models, the Einstein ring also represents roughly the boundary between source positions that are multiply-imaged and those that are only singly-imaged.} 
For an extended mass distribution that has circular symmetry in its projected surface mass density, the Einstein radius is given by 
\begin{equation}
    \theta_E = \sqrt{\frac{D_{LS}}{D_L D_S} \frac{4GM(\theta_E)}{c^2}}
\end{equation}
where $M(\theta_E)$ is the projected mass contained within a cylinder of radius $R=D_L\theta_E$. 
%Therefore, strong lensing is a powerful probe of the mass of the inner regions of galaxies.

Gravitational lensing preserves the surface brightness of background sources, but can magnify them by increasing their area. \textit{Critical lines} are regions of the lens plane where the magnification is formally infinite. The magnification is given by:
\begin{equation}
    \mu = \frac{1}{(1-\kappa-\gamma)(1-\kappa+\gamma)}
    \label{eqn: Magnification}
\end{equation}
where $\kappa$ is the convergence field and $\gamma$ is the gravitational shear \citep{saha_essentials_2024}. Similar to $\kappa$, the shear is given by the spatial derivatives of the projected lensing potential of a galaxy. 
%\rev{\sout{For the details of the derivation of the magnification and the lensing shear, see Appendix \ref{Appendix A}.}}

Equation \ref{eqn: Magnification} leads to two distinct types of critical curves: \textit{radial critical curves} appear where $1-\kappa+\gamma=0$, with images close to this line stretched perpendicularly to the direction of the line, while \textit{tangential critical curves} occur where $1-\kappa-\gamma=0$, at which images are stretched tangentially to the line. For axisymmetric lenses, the latter of these is a circle with a radius that is by definition the Einstein radius, $\theta_E$. We can extend this definition of $\theta_E$ to a general lens which does not have a circular tangential critical curve by using the \textit{effective} Einstein radius, $\theta_{E, \text{eff}}$. This is defined as:
\begin{equation}
    \theta_{E, \text{eff}} = \sqrt{\frac{A}{\pi}}
    \label{eqn: effective Einstein radius}
\end{equation} where $A$ is the area enclosed by the tangential critical curve \citep{redlich_strongest_2012}.

For each of our convergence maps, we solve the 2D Poisson relation in Equation \ref{eqn: lensing potential ODE} to obtain the lensing potential, $\psi$, using a FFT-based Green's function method implemented in \texttt{Lenstronomy}. The shear is then computed from the 2D potential map using a finite-difference method. Finally, we extract the tangential critical curves ($1-\kappa-\gamma=0$ contours) using a marching squares method. The area within the contour is calculated via a shoelace algorithm to obtain the effective (circularised) Einstein radius, $\theta_{E, \text{eff}}$, using Equation $\ref{eqn: effective Einstein radius}$. 
%Details on the computation method can be found in Appendix \ref{Appendix B}. 
Figure \ref{fig: Einstein Radius vs. stellar mass} shows the extracted Einstein radius distribution of our galaxies for each of the 3 line-of-sight projections.

\begin{figure*}
    \centering
    \begin{subfigure}[t]{0.33\textwidth}
        \centering
        \includegraphics[height=2.0in]{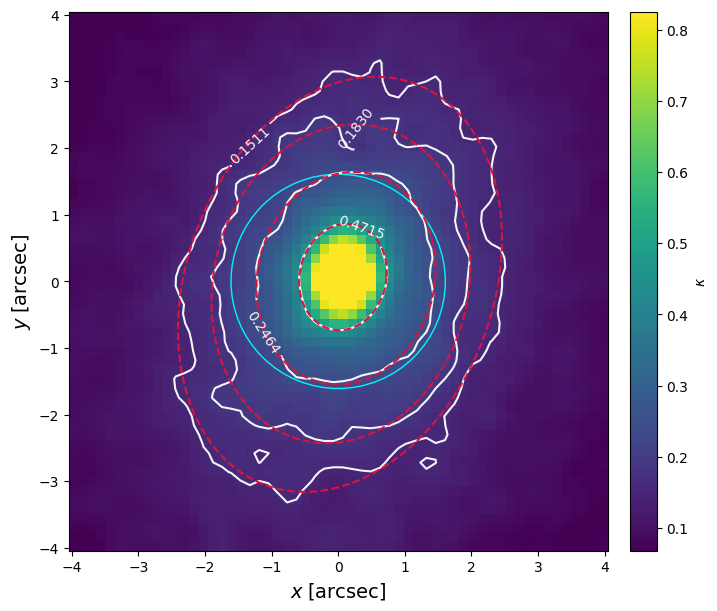}
        \caption{$XY$ plane}
    \end{subfigure}%
    ~ 
    \begin{subfigure}[t]{0.33\textwidth}
        \centering
        \includegraphics[height=2.0in]{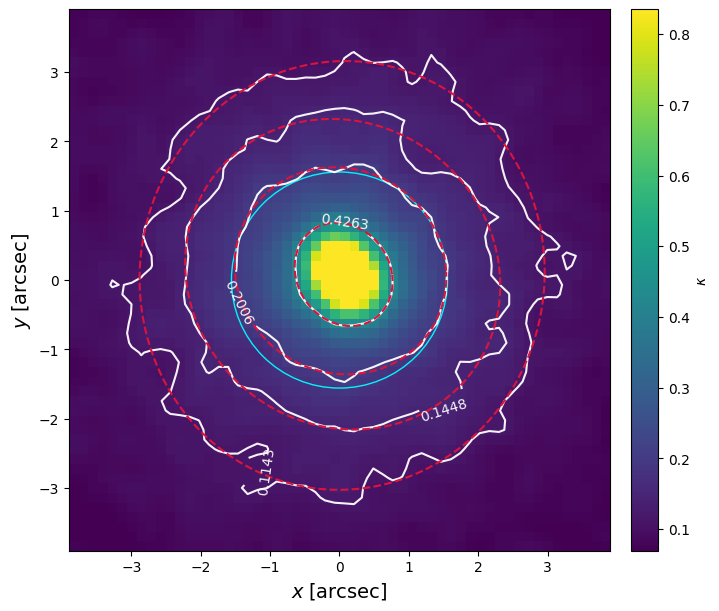}
        \caption{$YZ$ plane}
    \end{subfigure}%
    ~ 
    \begin{subfigure}[t]{0.33\textwidth}
        \centering
        \includegraphics[height=2.0in]{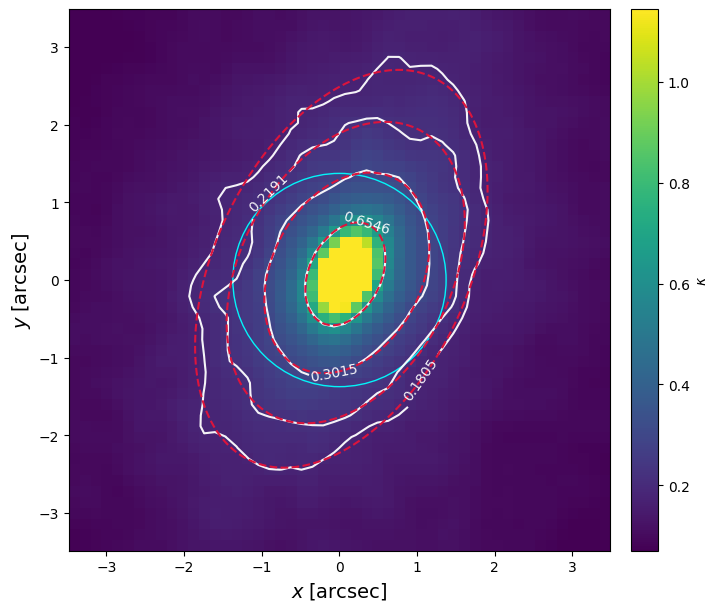}
        \caption{$XZ$ plane}
    \end{subfigure}
    \caption{Convergence maps of the three projections of an elliptical galaxy selected from IllustrisTNG, showing the iso-$\kappa$ contours (white solid line) and the best-fit ellipse (red dotted line) for four contour levels between 0.5$\theta_E$ and 2$\theta_E$ of the galaxy. This particular galaxy has a total stellar mass of $2.89\times10^{11} M_{\odot}$, sSFR $= 7.55\times10^{-12}$ yr$^{-1}$, $k_\text{rot} = 0.454$ and Sérsic indicies of 7.3, 6.3 and 3.8 for the $XY$, $YZ$, and $XZ$ projections respectively, thereby matching our criteria for being classified as an elliptical ETG. The circularised Einstein radius $\theta_{E, \text{eff}}$ of the galaxy is also traced in solid cyan.}
    \label{fig: convergence-map}
\end{figure*}

\subsection{Lensing Convergence Fitting}

\begin{figure}
    \centering
    \includegraphics[width=1.0\linewidth]{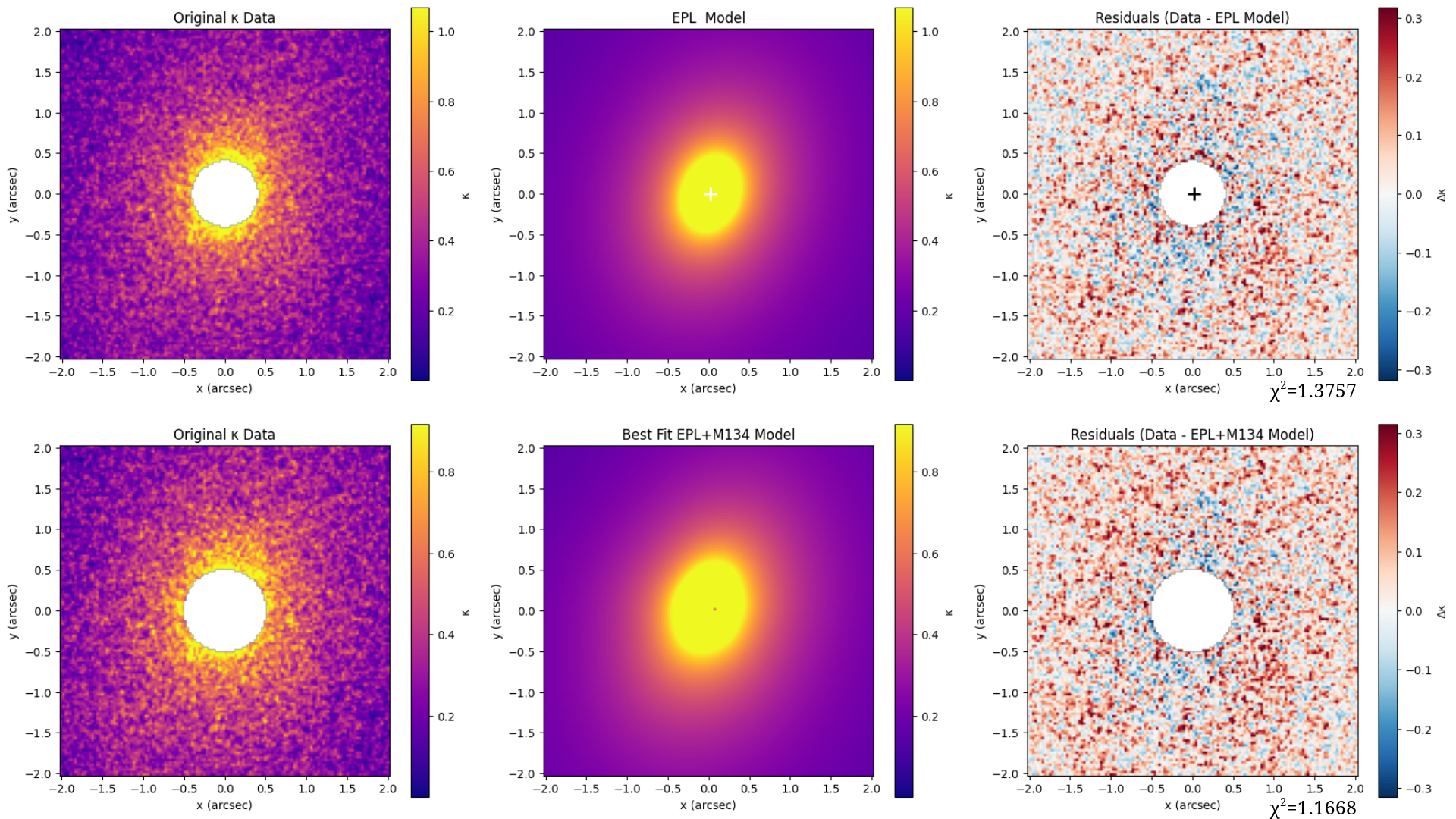}
    \caption{Convergence fitting result for a sample galaxy from TNG100, showing the best-fit convergence map and residuals of fitting with an EPL only and EPL+Multipole model. The reduced $\chi^2$ for the EPL+Multipole model is lower than the EPL model with a $\Delta$BIC = 770. So the EPL+Multipole model is a better description of the mass profile of the galaxy in the region between 0.5 and 2 $\theta_{E, \text{eff}}$}
    \label{fig:placeholder}
\end{figure}

To test whether the mass profiles of our galaxy samples deviate from a pure elliptic power law, we fit the 2D convergence maps with a %prescribed
lensing convergence model that takes into account higher-order multipolar perturbations. The analysis employs the following convergence model:

\begin{equation*}
    \kappa(x,y) = \kappa_{\text{EPL}}(x,y) + \sum_{m \in \{1,3,4\}} \kappa_{m}(x,y)
\end{equation*}
where $\kappa_{\text{EPL}}$ and $\kappa_{m}$ are the lensing convergences of the elliptical power law (EPL) model and multipolar perturbations defined in Equations \ref{eqn: EPL_convergence} and \ref{eqn: EPL-multipoles} respectively. Our model consists of 12 non-linear parameters in total to optimise:
\begin{itemize}
    \item EPL (6 parameters): Einstein radius ($\theta_E$), slope of the mass profile ($\gamma$), axis ratio ($q$), orientation angle ($\phi_{\text{EPL}}$), lens mass centre ($x_{\text{lens}}, y_{\text{lens}}$)

    \item Multipole components (6 parameters): Amplitudes ($a_m$) and orientation angles ($\phi_m$) for $m=1, 3, $ and $4$ multipole terms.
\end{itemize}
We also fit our convergence maps with only the EPL model to determine whether the addition of multipoles improves the fits.

For performing the 2D convergence fits, we choose a window between 0.5$\theta_{E,\text{eff}}$ and 2$\theta_{E,\text{eff}}$. This region typically comprises of the lensed features (arcs, multiple images, etc.) we are able to observe in galaxy-scale strong lenses, and is, therefore, where we are most interested in understanding the effect of multipole deformations on strong lensing observables. The pixel size is set at 0.05$^{\prime \prime}$ which is comparable to HST-like optical resolution scales.

The parameter optimisation is driven by minimisation of the following $\chi^2$ function
\begin{equation}
    \chi^2 = \sum_i \left( \frac{\kappa_i^{\text{data}}(\boldsymbol{x})- \kappa_i^{\text{model}}(\boldsymbol{x}) }{\sigma_i} \right)^2
    \label{equ: chi^2}
\end{equation}
where the uncertainties $\sigma_i$ in the data points are estimated using a Poisson-like model for the ATSC-based convergence. The error in each data point of the convergence map is computed as
\begin{equation}
    \sigma_{i} = \frac{1}{A_{bin} \cdot \Sigma_{crit}} \sqrt{\sum_{j=1}^{N_{i}} (W_{ij}^2 m_{j}^2)}
\end{equation} where $A_{\text{bin}}$ is the area of a two-dimensional bin for the data point, $m_j$ are the masses of all the particles (stars, gas, dark matter) in the bin, and $W_{ij}$ are the adaptive TSC weights scaled by the $k$-nearest neighbour distance ($k_{nn}=16$) in local density regions defined in Equation \ref{equ: ATSC mass assignment kernel}.

We use a combination of Differential Evolution (DE) and Markov Chain Monte Carlo (MCMC) to fit the convergence model to the lensing maps. The DE \citep{storn_differential_1997} is used to locate the best-fit parameter values, by setting the $\chi^2$ in Equation \ref{equ: chi^2} as the objective function to minimise through a global, gradient-free optimisation scheme. This approach is robust to multimodal likelihood landscapes and provides an excellent initialization point for the subsequent Bayesian analysis. We use 200 iterations with a population size of 135 for the EPL+Multipole model and 72 for the EPL-only model.

After the DE has converged, an MCMC is performed to sample the posterior probability distribution to obtain the model parameter estimates with uncertainties. We use the affine-invariant MCMC ensemble sampler implemented in \texttt{Python} called \texttt{emcee} \citep{goodman_ensemble_2010, foreman-mackey_emcee_2013}. 
%The MCMC samples the posterior distribution of the convergence model parameters. 
We use 25 walker initialised as a Gaussian centred around the DE parameter values with a standard deviation of 1\%, with 200 burn-in steps that are discarded and 1200 steps from which we obtain the final posterior on the parameters.

\begin{figure*}
    \centering
    \begin{subfigure}[t]{0.5\textwidth}
        \centering
        \includegraphics[height=3.55in]{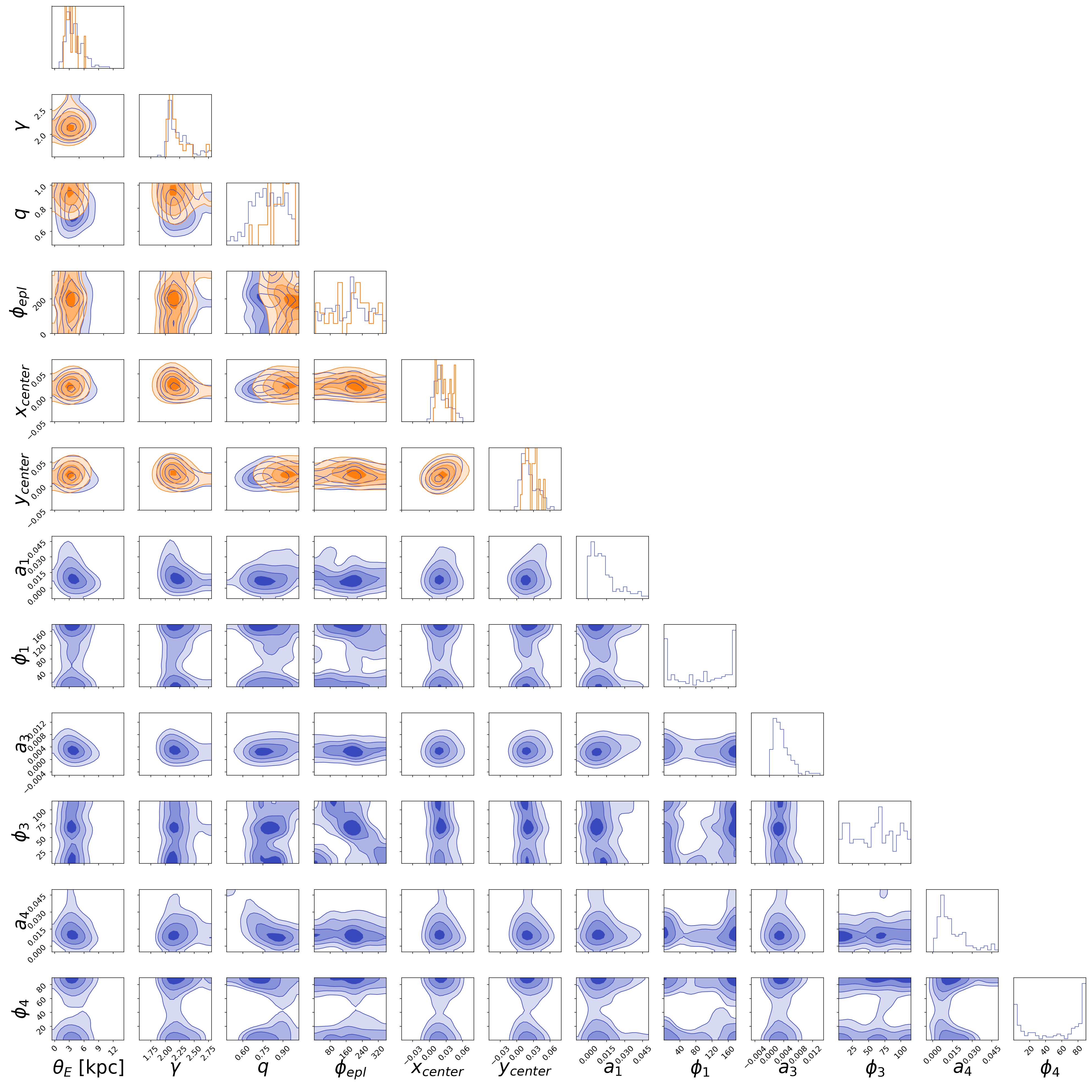}
    \caption{TNG50 distributions}
    \end{subfigure}%
    ~ 
    \begin{subfigure}[t]{0.5\textwidth}
        \centering
        \includegraphics[height=3.55in]{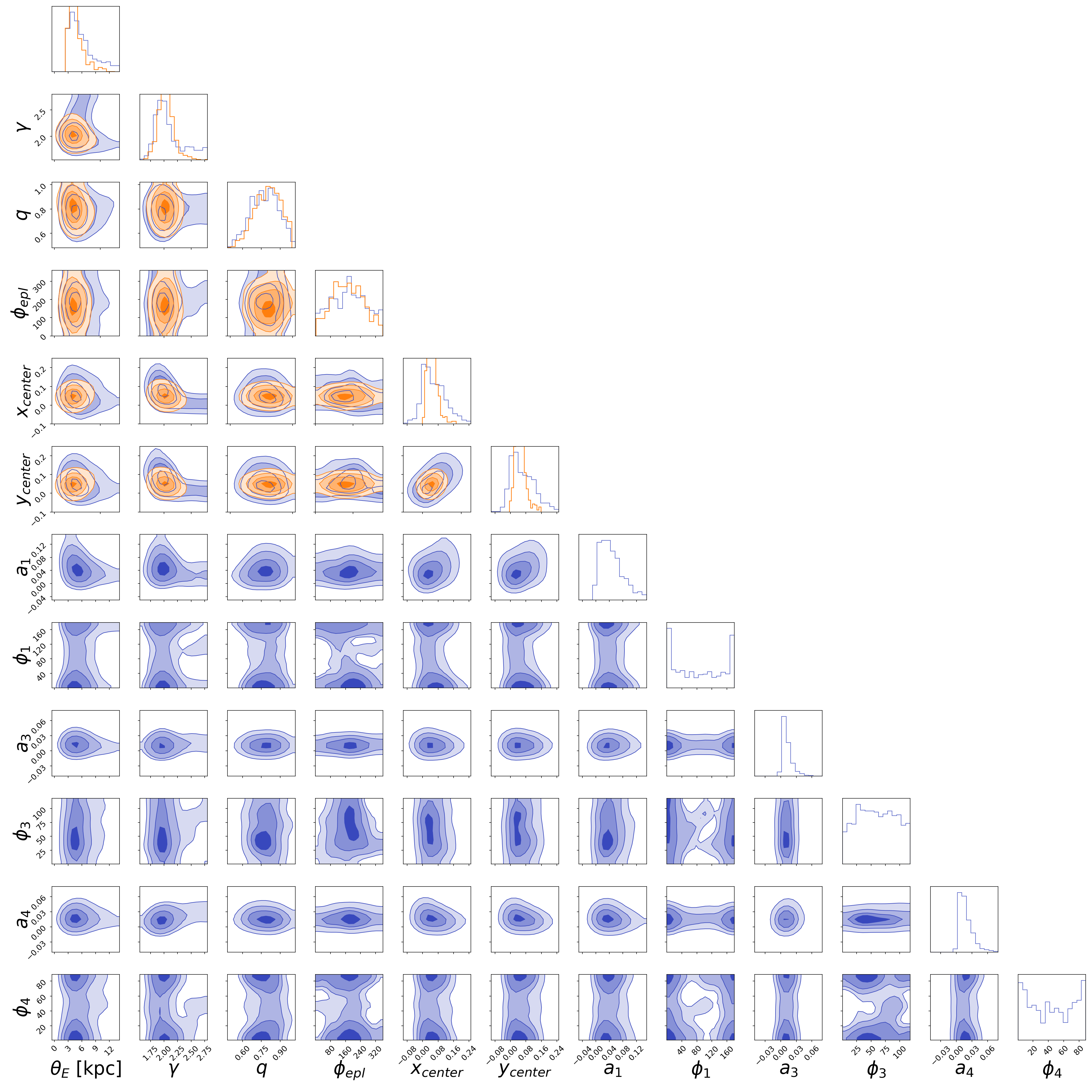}
    \caption{TNG100 distributions}
    \end{subfigure}
    \caption{Distribution of the power-law model parameters inferred for galaxies from the two TNG simulation boxes: Einstein radius $\theta_E$, slope $\gamma$, axis ratio $q$, orientation angle $\phi_{\text{EPL}}$ and mass centre $(x, y)$. The orange contour and histogram shows the best-fit parameter distributions for the EPL-only model while the blue contour and histogram shows the distributions for the EPL+Multipole model fit.}
    \label{fig: EPL and EPL+Multipole PL distribution}
\end{figure*}

\subsection{Galaxy Isodensity Fitting}

We further investigate the deviation of our galaxies from pure elliptical shapes by following a common method in the literature of characterising these deviations using isophotes of the light profile of the galaxy \citep{bender_isophote_1988, hao_isophotal_2006, penoyre_origin_2017}.
Instead, we consider the total mass density of a galaxy and develop a routine to fit an ellipse to an iso-convergence contour of the 2D lensing convergence map. The residuals of the fit are then expanded in terms of a Fourier series.
%Using the definition in \cite{hao_isophotal_2006}, T
The expansion of an iso-$\kappa$ contour in polar coordinates is given by
\begin{equation}
    R_{\kappa}(\phi) = R_0(\phi) + \sum_{m} A_m \cos(m\phi) + B_m\sin(m\phi)
\end{equation}
where $R_0(\phi)$ is the radius of the best-fit ellipse, and $A_m$ and $B_m$ are the amplitudes of the $m$th harmonic mode. The equation of the ellipse in polar coordinates, centred at the origin is
\begin{equation}
    R_0(\phi) = \frac{aq}{\sqrt{q^2\cos^2(\phi-\theta_{\text{PA}})+\sin^2(\phi-\theta_{\text{PA}})}}
\end{equation}where $a$ is the major axis of the ellipse, $q = b/a$ is the semi-minor axis ratio, and $\theta_{\text{PA}}$ is the relative orientation angle of the ellipse. We only consider multipole terms of order $m = 1, 3$, and $4$.
%as the $m = 2$ is already accounted for by the ellipticity parameter in the model.
Comparing to the lensing convergence definition in Equation \ref{eqn: EPL-multipoles}, the conversion between the two formalisms is given by:
\begin{equation}
    \begin{split}
        a_m &= \sqrt{A_m^2+B_m^2} \\
        \phi_m &= \frac{1}{m} \tan^{-1} (B_m/A_m)
    \end{split}
\end{equation}
so the results from isophote fitting can be directly compared to that from 2D convergence fits.

The iso-$\kappa$ fitting is done by a non-linear least-squares method using a Levenberg–Marquardt algorithm up to order $m=4$, giving equal weight to each residual, in order to find $R_0(\phi)$ as well as the sign and magnitude of the $A_m$ and $B_m$ coefficients. For each galaxy and its projections, we perform this analysis for four equally-spaced contours between 0.5$\theta_{E,\text{eff}}$ and 2$\theta_{E,\text{eff}}$ in order to take the average value of the multipole coefficients.

% \begin{itemize}
%     % \item Based on methods in \citep{hao_isophotal_2006} and \citep{penoyre_origin_2017} by fitting an ellipse with radius $R_{\text{ell}}(\phi)$ to an isophote with radius $R_{\text{iso}}(\phi)$ and express the residuals as a Fourier series $$\Delta(\phi) = R_{\text{iso}}(\phi) - R_{\text{ell}}(\phi) = \sum_{n=1}^{\infty} a_n \cos(n\phi) + b_n\sin(n\phi)$$
%     \item Expand the Fourier series of the residuals of a galaxy’s isodensity contours to find the sign and magnitude of the $a_1, a_3$ and $a_4$ coefficients
%     \item The fitting is done by 
%     \item For each galaxy, we perform this analysis for four nearly equally-spaced contours between 0.5 and 2.0 times the effective Einstein radius, $\theta_{\text{E, eff}}$, amd then average the results.
%     \item Figures 8 and 9
% \end{itemize}

% \begin{figure*}
%     \centering
%     \includegraphics[width=1.0\linewidth]{Figures/Isodensity fit and residuals.png}
%     \caption{Caption}
%     \label{fig:placeholder}
% \end{figure*}

% \begin{figure}
%     \centering
%     \includegraphics[width=1.0\linewidth]{Figures/a4 distribution with skewnorm fit.png}
%     \caption{Distribution of $a_4$ obtained by fitting Fourier series to residuals of best-fit ellipse of isodensity contours (See section 3.5)}
%     \label{fig:placeholder}
% \end{figure}

\section{Results}
\label{sec:results}

\begin{table}
\centering
\caption{Distribution of the difference in the Bayesian Information Criterion (BIC) between the two convergence models fitted to the selected TNG ellipticals, computed as $\Delta$BIC $=$ BIC(EPL) $-$ BIC(EPL+Multipole). The model with the lower BIC is more preferable, and if $\Delta$BIC$>2$ the fit prefers the EPL+Multipole model. 56\% of TNG100 and 85\% of TNG50 galaxies are likely to prefer the more complex mass model with multipoles.}
\renewcommand{\arraystretch}{1.5} % Adds padding to rows
\begin{tabular}{l c c}
\hline
 & \multicolumn{2}{c}{Number of samples}  \\ \hline
 & \textbf{TNG100} & \textbf{TNG50} \\ \hline
$ \Delta \text{BIC} < 2 $ & 707 & 38 \\ \hline
$2 \leq \Delta \text{BIC} < 6 $ & 36 & 4 \\ \hline
$6 \leq \Delta \text{BIC} < 10 $ & 38 & 2 \\ \hline
$\Delta \text{BIC} \geq 10 $ & 761 & 205 \\ \hline
\end{tabular}
\label{tab: Delta BIC}
\end{table}

\begin{table*}
\centering
\caption{Posterior median estimates and 68\% credible intervals of the elliptic power-law convergence model parameters fitted to selected elliptical galaxy projections from TNG100 and TNG50 with and without multipole perturbations added. We report the statistics of the distribution for galaxies which prefer the model with lower BIC.}
\renewcommand{\arraystretch}{1.8} % Adds padding to rows
\begin{tabular}{c l c c c c c c}
\hline
& \textbf{Power-Law Model Parameters} & $\theta_E$(kpc) & $\gamma$ & $q$ & $\phi_{\text{EPL}}(\degree)$ & $x_{\text{centre}} (^{\prime\prime})$ & $y_{\text{centre}} (^{\prime\prime})$  \\  \hline
TNG50 & EPL   & $3.127^{+0.970}_{-0.794}$ & $2.146^{+0.266}_{-0.071}$ & $0.908^{+0.060}_{-0.114}$ & $194.1^{+81.8}_{-114.4}$ & $0.023^{+0.017}_{-0.010}$ & $0.022^{+0.013}_{-0.008}$  \\  

& EPL + Multipole   & $3.750^{+2.193}_{-1.319}$ & $2.208^{+0.270}_{-0.134}$ & $0.789^{+0.131}_{-0.122}$ & $185.2^{+105.1}_{-115.6}$ & $0.019^{+0.019}_{-0.009}$ & $0.020^{+0.021}_{-0.010}$  \\ \hline

TNG100 & EPL   & $4.110^{+1.766}_{-0.831}$ & $2.024^{+0.123}_{-0.120}$ & $0.807^{+0.104}_{-0.113}$ & $171.9^{+85.7}_{-84.1}$ & $0.047^{+0.021}_{-0.018}$ & $0.048^{+0.020}_{-0.021}$  \\  
& EPL + Multipole   & $5.540^{+5.928}_{-2.018}$ & $2.030^{+0.541}_{-0.163}$ & $0.779^{+0.114}_{-0.114}$ & $178.6^{+104.3}_{-104.9}$ & $0.053^{+0.073}_{-0.051}$ & $0.049^{+0.078}_{-0.045}$  \\ \hline

\end{tabular}

% \begin{tabular}{c c c c c c c}
% \hline
%  \textbf{Multipole Parameters} & $a_1$ & $\phi_1$ & $a_3$ & $\phi_3$ & $a_4$ & $\phi_4$ \\ \hline
% TNG50   & $0.0098^{+0.0151}_{-0.0070}$ & $105.6^{+69.0}_{-100.6}$ & $0.0030^{+0.0030}_{-0.0017}$ & $62.1^{+36.9}_{-42.7}$ & $0.0119^{+0.0127}_{-0.0061}$ & $60.6^{+26.8}_{-56.2}$ \\

% TNG100   & $0.0433^{+0.0485}_{-0.0300}$ & $84.3^{+86.7}_{-76.8}$ & $0.0097^{+0.0112}_{-0.0056}$ & $58.3^{+38.4}_{-34.1}$ & $0.0155^{+0.0146}_{-0.0095}$ & $43.0^{+36.0}_{-33.4}$ \\ \hline

% \end{tabular}

\label{tab: convergence fit medians}

\end{table*}

\begin{figure*}
    \centering
    \begin{subfigure}[t]{0.33\textwidth}
        \centering
        \includegraphics[height=1.75in]{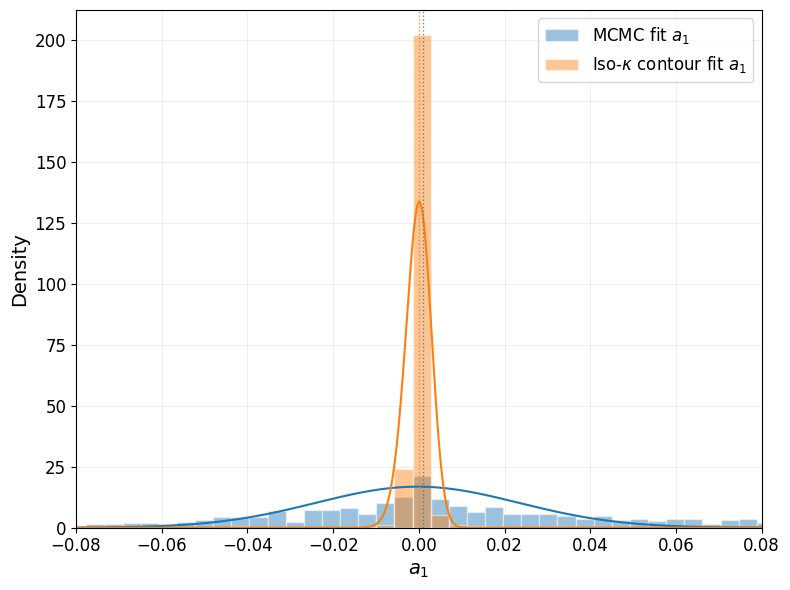}
        \caption{$a_1$}
    \end{subfigure}%
    ~ 
    \begin{subfigure}[t]{0.33\textwidth}
        \centering
        \includegraphics[height=1.75in]{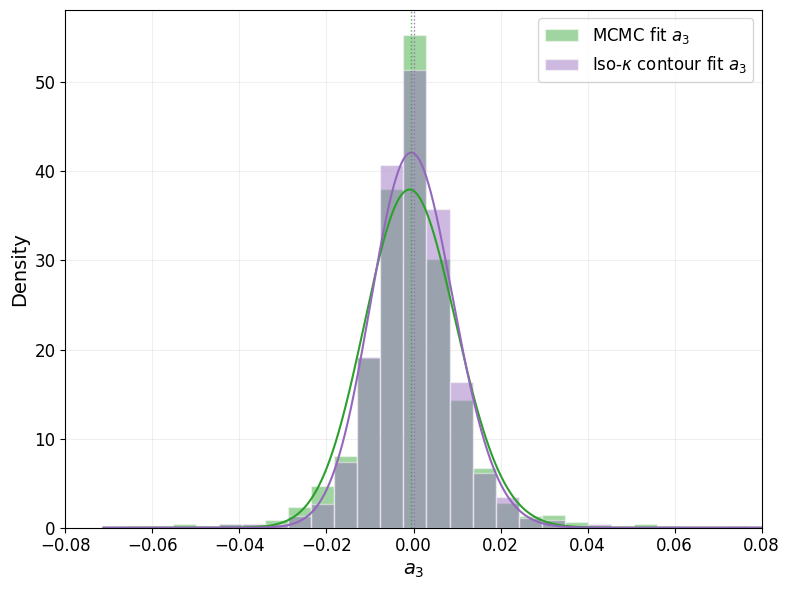}
        \caption{$a_3$}
    \end{subfigure}%
    ~ 
    \begin{subfigure}[t]{0.33\textwidth}
        \centering
        \includegraphics[height=1.75in]{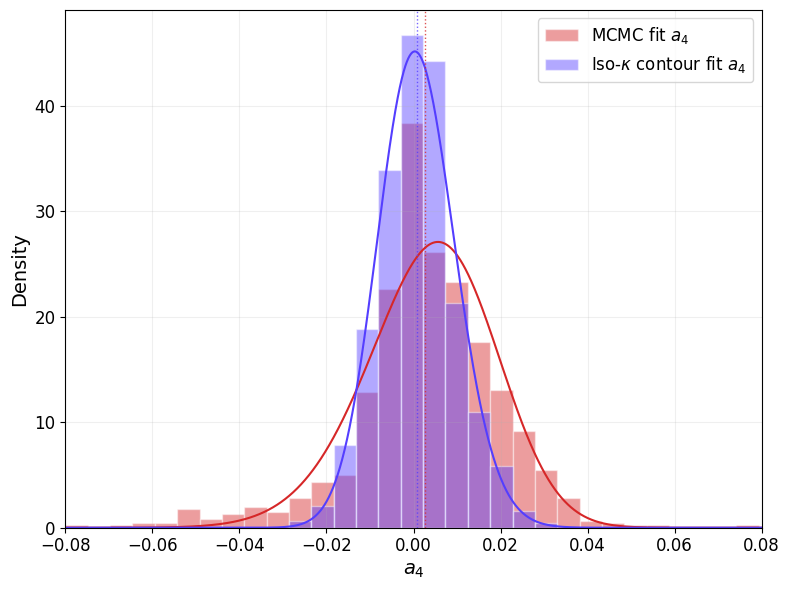}
        \caption{$a_4$}
    \end{subfigure}
    \caption{Distribution of the pure $a_m$ multipoles from the TNG100 samples, comparing the results of two different methods applied to infer the multipole pertubations. The physical multipole strengths are normalised by the semi-major axis for the iso-$\kappa$ contour fitting results, while those inferred from the 2D convergence model fits are normalised by the Einstein radius and ellipticity of the galaxy. A skew-normal curve is used for describing the distribution of the multipoles.}
    \label{fig: Multipole distributions TNG100}
\end{figure*}

\begin{figure*}
    \centering
    \begin{subfigure}[t]{0.33\textwidth}
        \centering
        \includegraphics[height=1.75in]{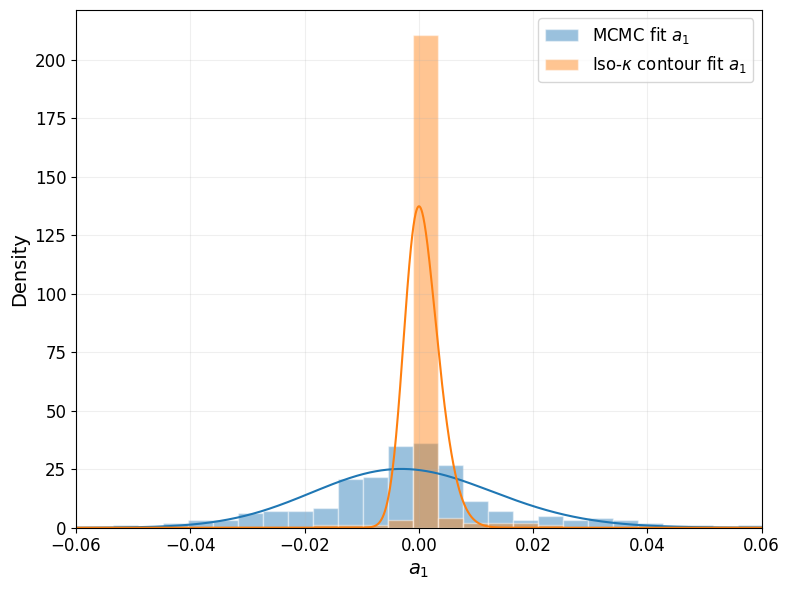}
        \caption{$a_1$}
    \end{subfigure}%
    ~ 
    \begin{subfigure}[t]{0.33\textwidth}
        \centering
        \includegraphics[height=1.75in]{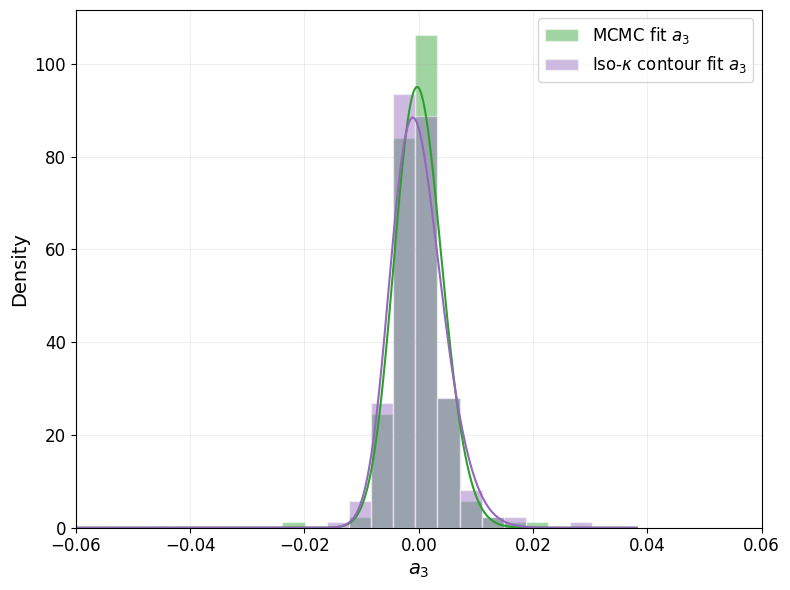}
        \caption{$a_3$}
    \end{subfigure}%
    ~ 
    \begin{subfigure}[t]{0.33\textwidth}
        \centering
        \includegraphics[height=1.75in]{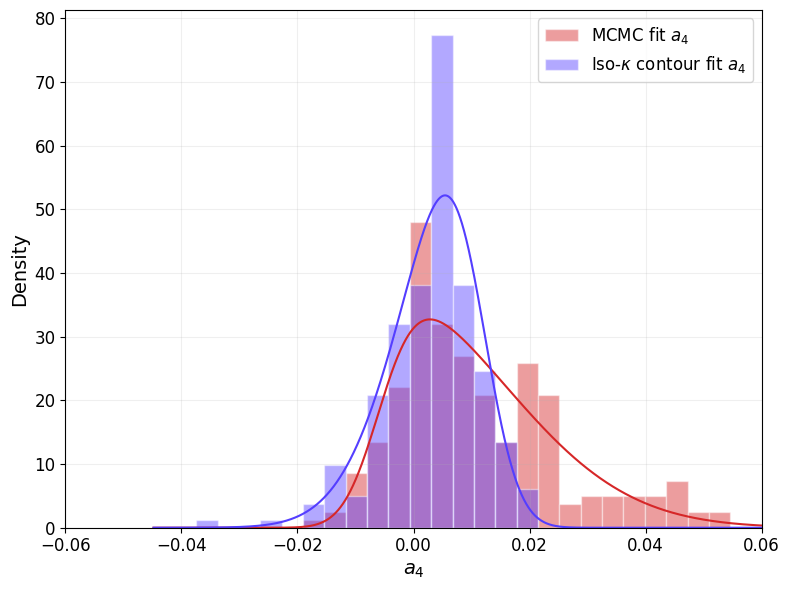}
        \caption{$a_4$}
    \end{subfigure}
    \caption{Distribution of the pure $a_m$ multipoles from the TNG50 samples obtained from iso-$\kappa$ contour and 2D convergence map fits, with skew-normal curves overlaid to describe each distribution.}
    \label{fig: Multipole Distributions TNG50}
\end{figure*}

\begin{figure*}
    \centering
    \begin{subfigure}[t]{1.\textwidth}
        \centering
        \includegraphics[height=1.75in]{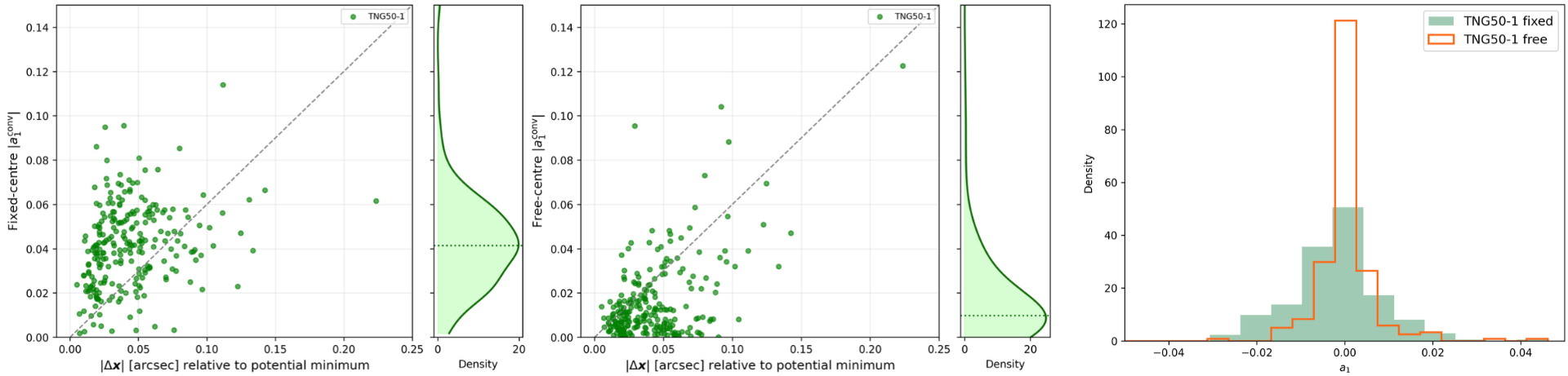}
    \caption{TNG50 refit: the median physical $m=1$ amplitude is $0.0098$ for the free-centre fits compared to $0.0414$ for the fixed-centre fits. The fixed-centre normalised $a_1$ distribution is approximately $43\%$ positive and $57\%$ negative with a median of $-0.0067$.}
    \end{subfigure}
    \vspace{6mm}
    ~ 

    \begin{subfigure}[t]{1.\textwidth}
        \centering
        \includegraphics[height=1.75in]{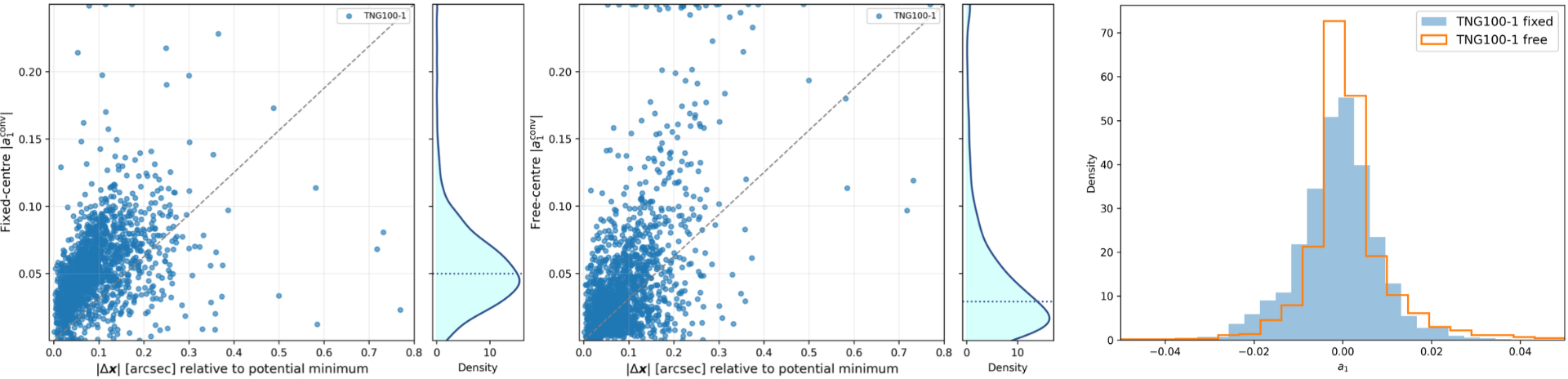}
    \caption{TNG100 refit: the median physical $m=1$ amplitude increases from $0.0289$ to $0.0414$ for the fixed-centre fits. The fixed-centre normalised $a_1$ distribution has approximately $47\%$ positive and $53\%$ negative normalised $m=1$ strengths with a median of $0.0014$.}
    \end{subfigure}
    \vspace{3mm}
    ~
    \caption{Comparison of the results of $m=1$ amplitude from refitting convergence map of the TNG50 and TNG100 galaxy projections with the EPL+Multipole model keeping the mass centre fixed to the minimum in the gravitational potential of the galaxy. \textbf{Left to right}: Physical $m=1$ amplitudes from the fixed-centre fit against the offset in the inferred free mass centre relative to reference fixed centre, physical $m=1$ amplitudes from the free-centre fit against the offset in the inferred free mass centre relative to reference fixed centre, and distributions of the scaled pure $a_1$ for both fits. Overall, the inferred $m=1$ strength increases with the fixed-centre runs with the fraction of positive and negative remaining nearly consistent.}

    \label{fig: fixed-centre a1 fit run}
\end{figure*}

\subsection{Lensing Convergence fitting}

\subsubsection{Model Comparison}

% \begin{figure}
%     \centering
%     \includegraphics[width=1.0\linewidth]{Figures/Model_Convergence_Fit_Results/reduced chi2 comparison.png}
%     \caption{Comparison of the reduced $\chi^2$ ($\chi^2_\nu$) from fitting EPL and EPL+Multipole convergence models to the ellipitical galaxies from TNG100 and TNG50 simulation boxes. Most points lie below the dashed line which means that most galaxies (around 98\%) prefer the more complex model. This could be attributed to the multipole convergence model having more flexible parametrisation, so the $\chi^2$ comparison should be interpreted cautiously.}
%     \label{fig: reduced chi^2 comparison}
% \end{figure}

We analysed all three projections of the  TNG100 and TNG50 galaxies. We treat each projection as an independent realisation of a possible elliptical galaxy and therefore have 1542 and 249 snapshots from TNG100 and TNG50 respectively.

We find that in over 98\% of samples, the EPL+Multipole model leads to a lower reduced $\chi^2$ in the data fits which would suggest that the more complex model is preferred. However, to check that the lower $\chi^2$ is not simply due to the multipole model having an increased number of free parameters to fit,
 %In order to compare the results of fitting our two models -- an EPL profile versus a more complex EPL with Multipole components -- to the galaxy data, 
 we additionally employ the Bayesian Information Criterion (BIC). The BIC is used not as a means to assess global model preference, but to determine which of the two models is a better description of the mass distribution of a galaxy, where both fits for each galaxy sample share the same data set and likelihood function. A model is favourable if it has a lower BIC. The BIC is defined as:
\begin{equation}
    \text{BIC} = k \ln n - 2 \ln \hat{\mathcal{L}}
    \label{eqn: BIC definition}
\end{equation} where $k$ is the number of parameters, $n$ is the number of data points and $\hat{\mathcal{L}}$ is the maximum likelihood of the MCMC fit.

The BIC comparison assumes that the number of independent data points, $n$, is given by the number of fitted pixels of the convergence map. However, neighbouring pixels in our projected mass maps are not strictly independent, owing to the particle-to-grid assignment and adaptive smoothing used in map construction. The resulting BIC values should therefore be interpreted with this limitation in mind, rather than as exact measures of model-selection evidence. We assess the impact of these correlations through null tests in which synthetic maps generated from smooth EPL profiles are processed and fitted using the same pipeline (Appendix \ref{app:epl_null_tests}). 

Table \ref{tab: Delta BIC} shows the distribution of the computed BIC difference of the two fitted models. We find that of the 1791 total projection samples in both simulation boxes, only 835 in TNG100 and 211 in TNG50 have $\Delta$BIC greater than 2, and would therefore prefer the EPL+Multipole mass model. We further test the sensitivity of our results to the map resolution and smoothing length by repeating the model comparison under alternative configurations (Appendix \ref{app:resolution_tests}). The fraction  and subset of galaxies for which multipole components are favoured remains approximately stable across these tests, suggesting that our principal population-level conclusions are not significantly affected by the adopted pixel scale or smoothing prescription.

%\rev{comment on why different rates of multipoles in the two volumes}

\begin{table*}
\caption{Statistical summary of the distribution of the normalised pure $a_m$ multipole perturbation amplitudes for different fitting methods and simulation boxes. The distribution of $a_3$ and $a_4$ obtained by \protect\cite{hao_isophotal_2006} from analysis of stellar isophotes of ellipticals are reported for comparison.}
\renewcommand{\arraystretch}{1.8} % Adds padding to rows

\begin{tabular}{l | c c c c c}
\hline
& \multicolumn{4}{c}{$\boldsymbol{a_1}$}  \\ \hline
\textbf{Sample} & Mean + Std. Dev. & Median + 68\% credible interval & Positive fraction (\%)  & Negative fraction (\%)  \\ \hline

TNG50 (Iso-$\kappa$ fit)  & $0.00072 \pm 0.00305$ & $0.0000^{+0.0004}_{-0.0005}$ & 50.0 & 50.0  \\

TNG100 (Iso-$\kappa$ fit)  & $-0.0002 \pm 0.0031$ & $-0.0001_{-0.0010}^{+0.0011}$ & 48.5 & 51.5  \\

TNG50  (Convergence Model fit) & $-4.4\times10^{-5} \pm 0.0167$ & $-0.0014^{+0.0153}_{-0.0137}$ & 43.0 & 57.0 \\

TNG100 (Convergence Model fit)  & 0.0005 $\pm$ 0.0024 & $0.0014^{+0.0546}_{-0.0390}$ & 52.1 & 47.9  \\ \hline

\end{tabular}

\begin{tabular}{l | c c c c c}
\hline
& \multicolumn{4}{c}{$\boldsymbol{a_3}$}  \\ \hline
\textbf{Sample} & Mean + Std. Dev. & Median + 68\% credible interval & Positive fraction (\%)  & Negative fraction (\%)  \\ \hline

Hao06 & 0.00041 $\pm 0.00440$ & $0.00025^{+0.00355}_{-0.00300}$ & 53.4 & 46.6 \\

TNG50 (Iso-$\kappa$ fit)  & $-4.3\times10^{-5} \pm 0.0047$ & $-0.00048^{+0.00376}_{-0.00355}$ & 46.3 & 53.7  \\

TNG100 (Iso-$\kappa$ fit)  & $-7.5 \times 10^{-5} \pm 0.0096$ & $0.0000_{-0.0082}^{+0.0080}$ & 49.7 & 50.3  \\

TNG50  (Convergence Model fit) & 0.00023 $\pm$ 0.00434 & $-0.00040^{+0.00363}_{-0.00305}$ & 44.4 & 55.6 \\

TNG100 (Convergence Model fit)  & -0.00041 $\pm$ 0.01107 & $-0.00084^{+0.00880}_{-0.00820}$ & 44.5 & 55.5  \\ \hline

\end{tabular}

\begin{tabular}{l | c c c c c}
\hline
 & \multicolumn{4}{c}{$\boldsymbol{a_4}$}  \\ \hline
\textbf{Sample} & Mean + Std. Dev. & Median + 68\% credible interval & Positive fraction (\%)  & Negative fraction (\%)  \\ \hline

Hao06 & 0.0062 $\pm 0.0137$ & $0.0027^{+0.0119}_{-0.0055}$ & 65.6 & 34.4 \\

TNG50 (Iso-$\kappa$ fit)  & 0.00316 $\pm$ 0.00805 & $0.0043^{+0.0067}_{-0.0078}$ & 71.0 & 29.0  \\

TNG100 (Iso-$\kappa$ fit)  & 0.0009 $\pm$ 0.00891 & $0.00076^{+0.00820}_{-0.00839}$ & 55.1 & 44.9  \\

TNG50  (Convergence Model fit) & 0.0110 $\pm$ 0.0136 & $0.0085^{+0.0153}_{-0.0100}$ & 79.0 & 21.0 \\

TNG100 (Convergence Model fit)  & 0.0036 $\pm$ 0.0152 & $0.0029^{+0.0160}_{-0.0127}$ & 58.2 & 41.8  \\ \hline

\end{tabular}

\label{Tab: Multipole comparison with Hao06}

\end{table*}

\subsection{Mass Parameter and Multipole Model Distributions}

\label{sec: mass parameter distributions}

We report the distribution and posterior median values of both fitted model parameters to the lensing convergence for the TNG100 and TNG50 galaxy samples which have a reduced $\chi^2 < 2.0$ and lower BIC among the two models in Figure \ref{fig: EPL and EPL+Multipole PL distribution} and Table \ref{tab: convergence fit medians}.

The galaxies selected at redshift $z=0$ have total power-law slopes that are close to isothermal ($\gamma=2.0$) when fitted with either an EPL or EPL+Multipole model to the 2D convergence maps. Combining the results from both simulation boxes, we obtain mean values for the slope of $2.13 \pm 0.27$ for the EPL model and $2.21 \pm 0.41$ for the EPL+Multipole model.
This is slightly greater than the mean slope of $\braket{\gamma} = 2.011 \pm 0.007$ of $z=0$ early-type galaxies in the IllustrisTNG suite obtained by \cite{wang_early-type_2020} from 1D fits to the total mass density between 0.4 and 4 times the stellar half-light radius. Our values for the slope are statistically consistent with that obtained from analyses of both simulated and real galaxies in surveys (see \citealt{ferreira_catalogue_2025} for a summary of other works). 
% We note, however, that the 2D fits here have a relatively large scatter in the inferred slope with a non-trivial fraction of galaxy sample projections having a slope much greater than 2.
% Figure \ref{fig: Multipole distributions} shows the distribution of multipoles from the EPL+Multipole model fitting to the convergence maps.

% \subsubsection{Multipole Distributions}
% \label{sec: multipole distributions}

% \begin{figure*}
%     \centering
%     \includegraphics[width=0.88\linewidth]{Figures/Parameter_Inference_and_Time-Delay_Cosmography/Quad_configurations.png}
%     \caption{Three basic lensing configurations used in this study. The inner caustic in the image plane (green) and the associated critical curve in the source plane (red) are drawn for an elliptic power law with external shear model.  The source positions are shown with star symbols and the corresponding images are shown with diamonds. \textbf{Left:} A source located near the centre of the caustic produces a “cross” configuration \textbf{Centre:} A source located near the caustic and between two adjacent cusps produces a “fold” configuration. \textbf{Right:} A ``cusp" configuration is produced for a source near a cusp of the inner caustic.}
%     \label{fig: Quad Configurations}
% \end{figure*}

Figure \ref{fig: Multipole distributions TNG100} and \ref{fig: Multipole Distributions TNG50} show the the distributions of the $a_1$, $a_3$, and $a_4$ multipoles of the galaxies that were analysed using the 1D iso-$\kappa$ contour fitting and Bayesian model fits to the 2D spatial convergence profile. In order to compare the results from both analyses, we express the inferred multipole strengths as fractional deviations from the elliptical profile. The iso-$\kappa$ contour amplitudes are normalised by the semi-major axis as $a_m^{\text{iso}} \mapsto a_m^{\text{iso}}/a_{\text{major}}$. The physical multipole amplitudes inferred from the 2D convergence Bayesian fits are rescaled as $a_m^{\text{conv}} \mapsto a_m^{\text{conv}} \frac{\sqrt{q}}{\theta_E} \cos(m \phi_m)$, additionally accounting for the multipole orientation with respect to the orientational angle of the elliptic model. This ``normalised" $a_m$ convention allows us to compare our results with the multipole distribution obtained by \cite{hao_isophotal_2006} from analysing 847 nearby ETGs in the Sloan Digital Sky Survey (SDSS) Data Release 4 (Table \ref{Tab: Multipole comparison with Hao06}). Similar to our method of expanding the iso-$\kappa$ contours of galaxies as a Fourier series, the multipole strength of ETGs in their paper were obtained by fitting ellipses to isophotes of the stellar light distribution of the galaxies. 
We find a greater fraction of galaxies having ``discy" ($a_4>0$) than ``boxy" ($a_4<0$) multipoles in both simulation boxes in agreement with the findings of \citet{hao_isophotal_2006}. The fraction of galaxies with positive $a_3$, however, is less than those with negative $a_3$ in slight contrast with the distribution of \citet{hao_isophotal_2006} which found a relatively higher share of galaxies with $a_3>0$. In our TNG samples, around $40\%$ of galaxies have $a_4$ perturbations between 1 and 3\%, and $10\%$ have perturbations exceeding 3\%. In the case of the $a_3$ multipole, around 22\% of ellipticals have perturbations exceeding 1\%. Perturbations of order $m=1$ for the convergence fit results appear to exceed 1\% in around 70\% of galaxies, with about $50\%$ exceeding a 2\%, but is substantially lower for the iso-$\kappa$ contour fits.

We note that we do not find a strong correlation between the multipole strengths inferred for each galaxy using the iso-$\kappa$ and convergence fitting methods, though the overall distribution of $a_3$ and $a_4$ modes is comparable to previous works. The $a_1$ mode has the most obvious difference in the two distributions, and is likely attributed to the fact that lopsidedness manifests as a form of  displacement of the mass distribution from the lens centre (see Figure \ref{fig:Multipole perturbations}). Since we do not explicitly fix the lens centre of the best-fit ellipse in the iso-$\kappa$ contour fitting, the inferred $a_1$ is close to zero. 
% We also perform a very simple strong lensing cross-section selection filter on the galaxies for a pixel resolution of $0.05$ arcsecond. The gravitational lensing cross-section is the angular area on the source plane in which a source will produce lensed images, given by $\pi\theta_E^2$ for an SIS mass profile. We find that all multipole-preferring galaxies have a large enough cross-section to satisfy the strong lensing property.

\subsubsection{Fixed-centre $m=1$ multipole convergence fitting}
\label{sec:fixed-centre-summary}

To assess the degeneracy between lens-centre position and dipole-like $m=1$ mass multipole, we refitted the TNG100-1 and TNG50-1 galaxy projections with the centre fixed at the IllustrisTNG \texttt{SubhaloPos} position, corresponding to the particle with minimum gravitational potential energy. 
% We retained the production map-making prescription, using adaptive triangular-shaped-cloud smoothing with $k_{\rm nn}=16$, a pixel scale of $0.05\,{\rm arcsec}$, and the same uncertainty maps and annular fitting region, $0.5\,\theta_{\rm E,eff}\leq r\leq2\,\theta_{\rm E,eff}$. Each projected convergence map was fitted with an EPL model plus a single $m=1$ convergence multipole, with the centre fixed in both components. The six free parameters were optimised using differential evolution and subsequently sampled with \texttt{emcee}. The inferred amplitudes were compared with those from the existing free-centre EPL+Multipole fits. We also examined the signed component $a_{1,x}=a_1\cos\phi_1$ to distinguish changes in the absolute dipole amplitude from a population-wide directional preference.

The inferred $a_1$ for the multipole-preferring galaxy projections is higher in general for both simulation boxes (Figure \ref{fig: fixed-centre a1 fit run}). In TNG50, the median absolute physical $m=1$ amplitude increases from $0.0098$ for the free-centre fits to $0.0414$ for the fixed-centre fits. Despite the increase in absolute physical amplitude, the fixed-centre signed distribution remained approximately balanced, with $47\%$ and $53\%$ of projections having positive and negative normalised pure $a_1$ strengths, respectively. For TNG100, the fixed-centre fits yielded a median absolute amplitude of $a_1=0.0498$ compared with $0.0289$ in the
matched free-centre results. The fraction of normalised $a_1$ strengths is approximately $43\%$ positive and $57\%$ negative.

Overall, both simulations indicate that the inferred absolute $m=1$ amplitude is sensitive to the adopted centre convention. Fixing the centre at the potential minimum generally yields larger dipole amplitudes, consistent with a free-centre model absorbing some dipole-like structure through shifts in its fitted centroid. However, the signed distributions show no strong common directional preference. The fixed-centre amplitudes should be interpreted as residual dipole structure relative to an independently defined reference centre.

\begin{figure*}
    \centering
    \begin{subfigure}[t]{0.51\textwidth}
        \centering
        \includegraphics[height=1.23in]{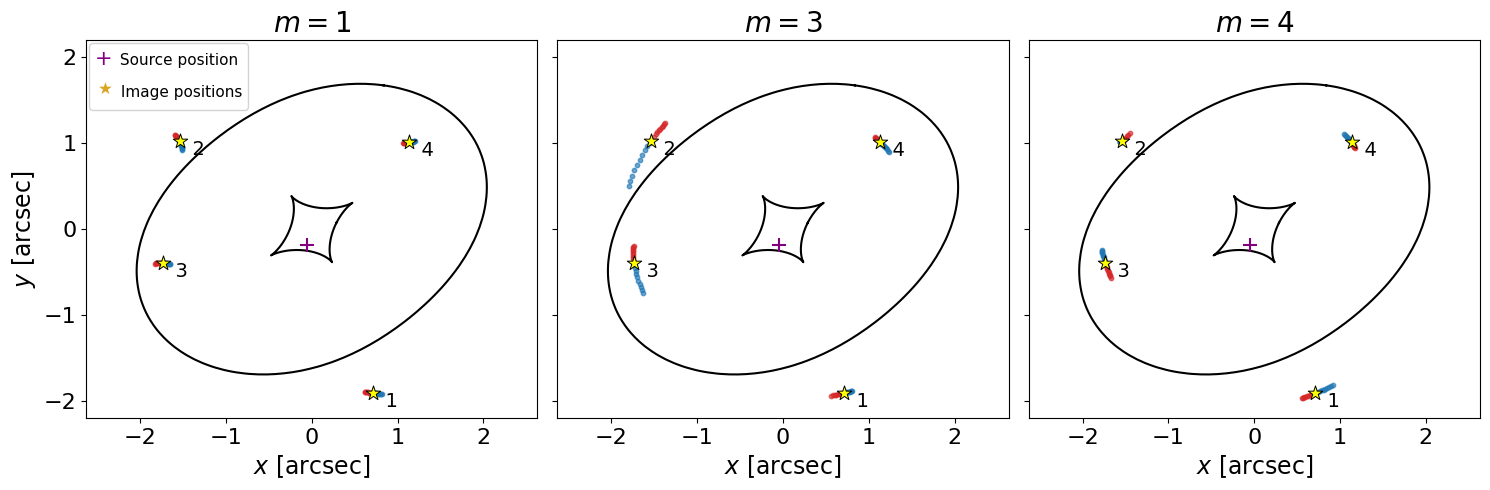}
    \end{subfigure}%
    ~
    \begin{subfigure}[t]{0.51\textwidth}
        \centering
        \includegraphics[height=1.23in]{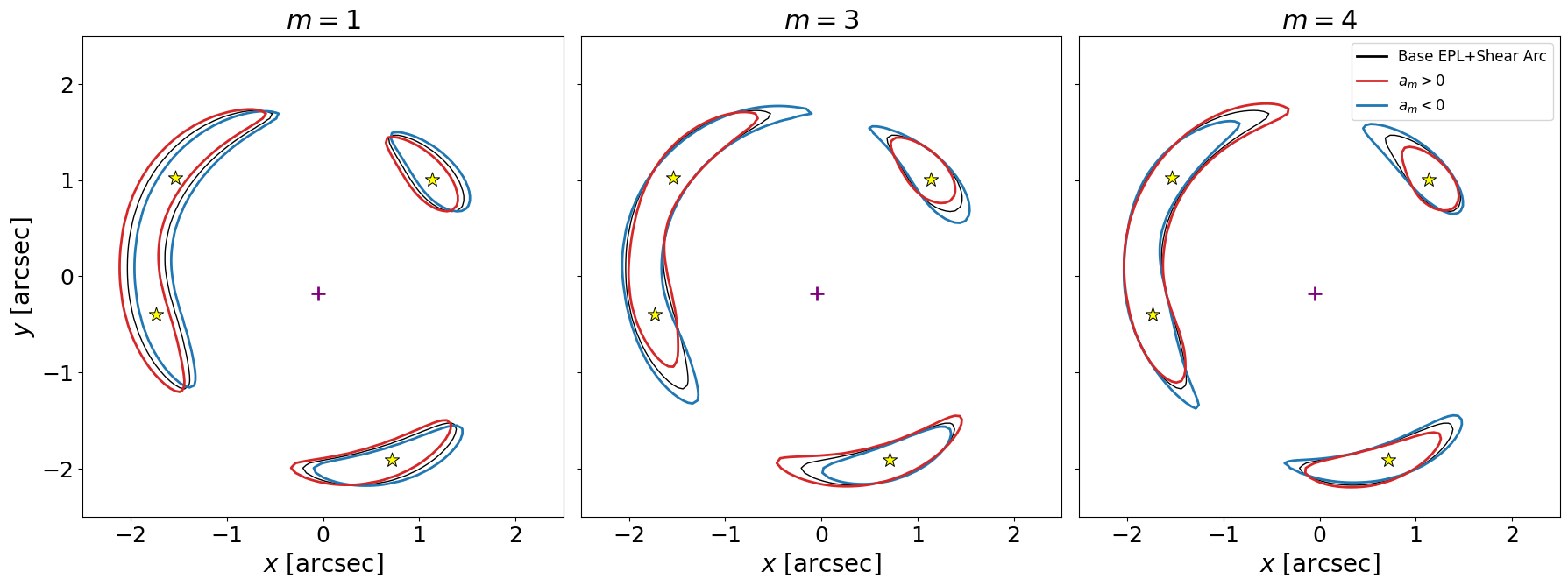}
    \end{subfigure}
    \begin{subfigure}[t]{1\textwidth}
        \centering
        \caption{Fold lens with source position $(x,y) = (-0.05, 0.018)^{\prime\prime}$. Mass Parameters: $\theta_E = 1.8^{\prime\prime}$, $\gamma=2.1$, $q=0.7$, $\phi_{\text{EPL}}=23 \degree$, $(\gamma_1, \gamma_2) = (0.016, -0.05)$}
    \end{subfigure}
    ~
    ~
    \begin{subfigure}[t]{0.51\textwidth}
        \centering
        \includegraphics[height=1.23in]{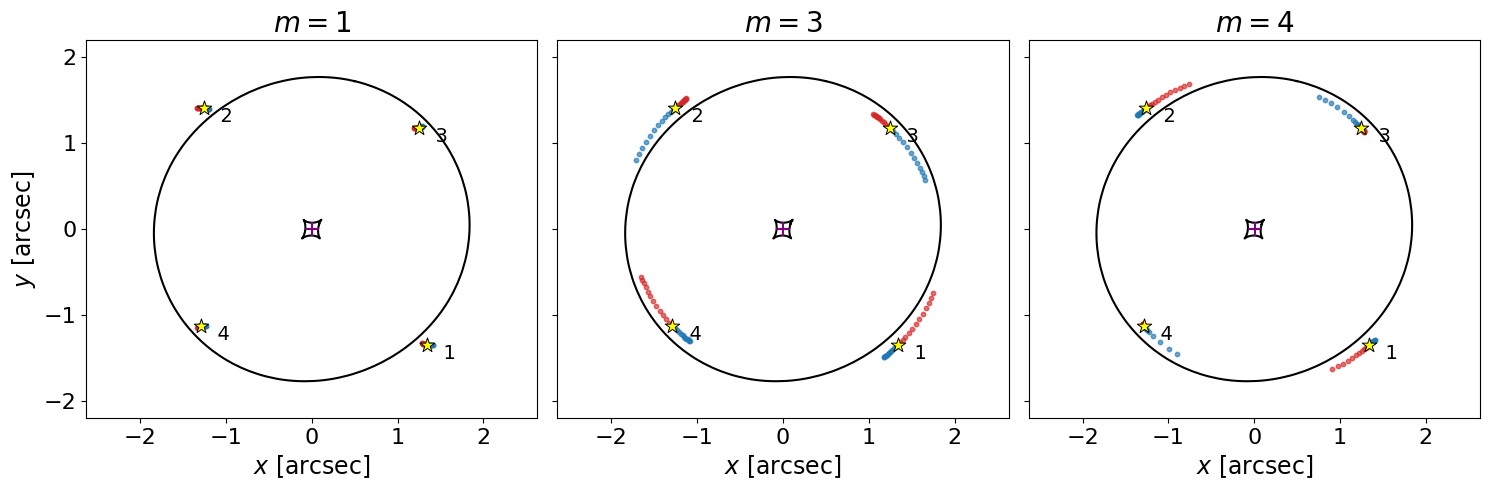}
    \end{subfigure}%
    ~
    \begin{subfigure}[t]{0.51\textwidth}
        \centering
        \includegraphics[height=1.23in]{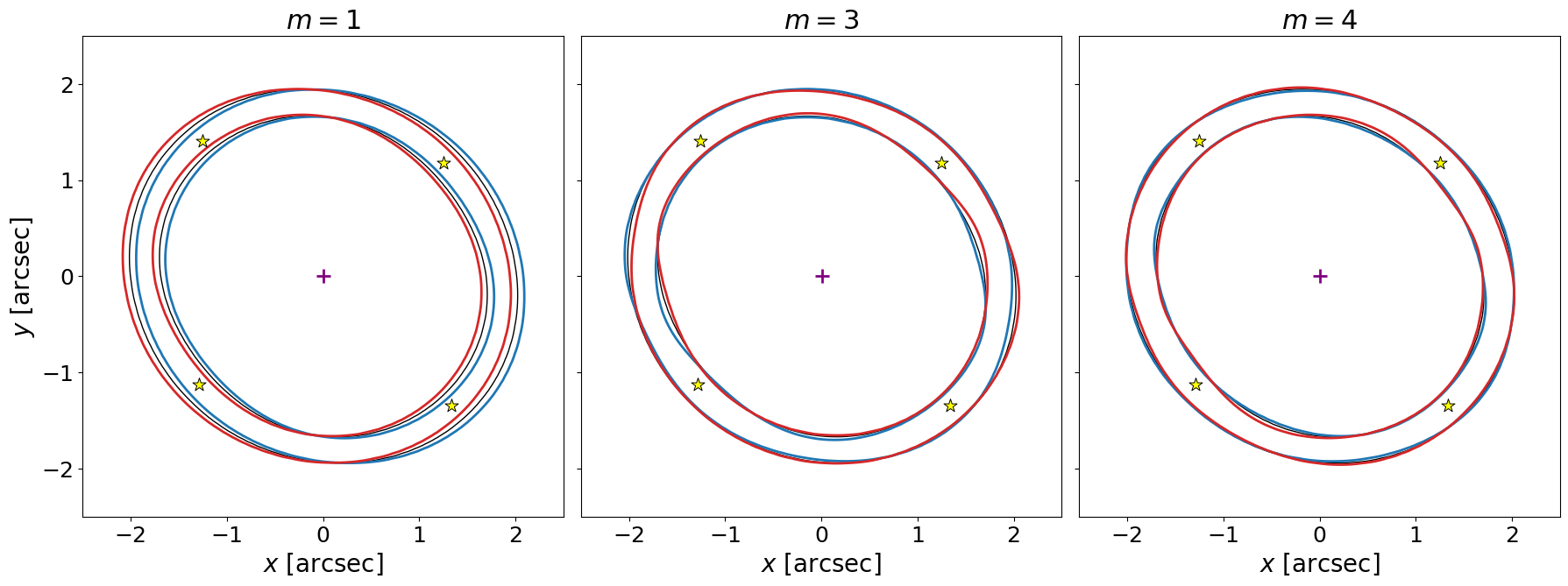}
    \end{subfigure}
    \begin{subfigure}[t]{1.\textwidth}
        \centering
        \caption{Cross lens with source position $(x,y) = (0.005, 0.001)^{\prime\prime}$. Mass Parameters: $\theta_E = 1.8^{\prime\prime}$, $\gamma=2.0$, $q=0.88$, $\phi_{\text{EPL}}=-23 \degree$, $(\gamma_1, \gamma_2) = (0.03, -0.075)$}
    \end{subfigure}
    ~
    \begin{subfigure}[t]{0.51\textwidth}
        \centering
        \includegraphics[height=1.23in]{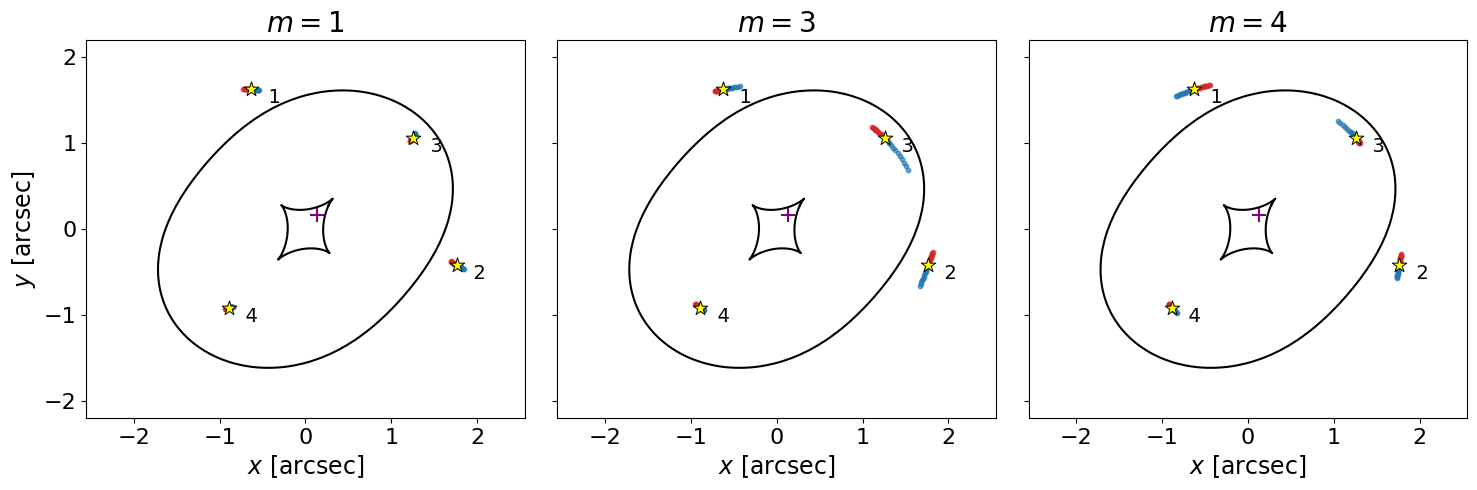}
    \end{subfigure}%
    ~
    \begin{subfigure}[t]{0.51\textwidth}
        \centering
        \includegraphics[height=1.23in]{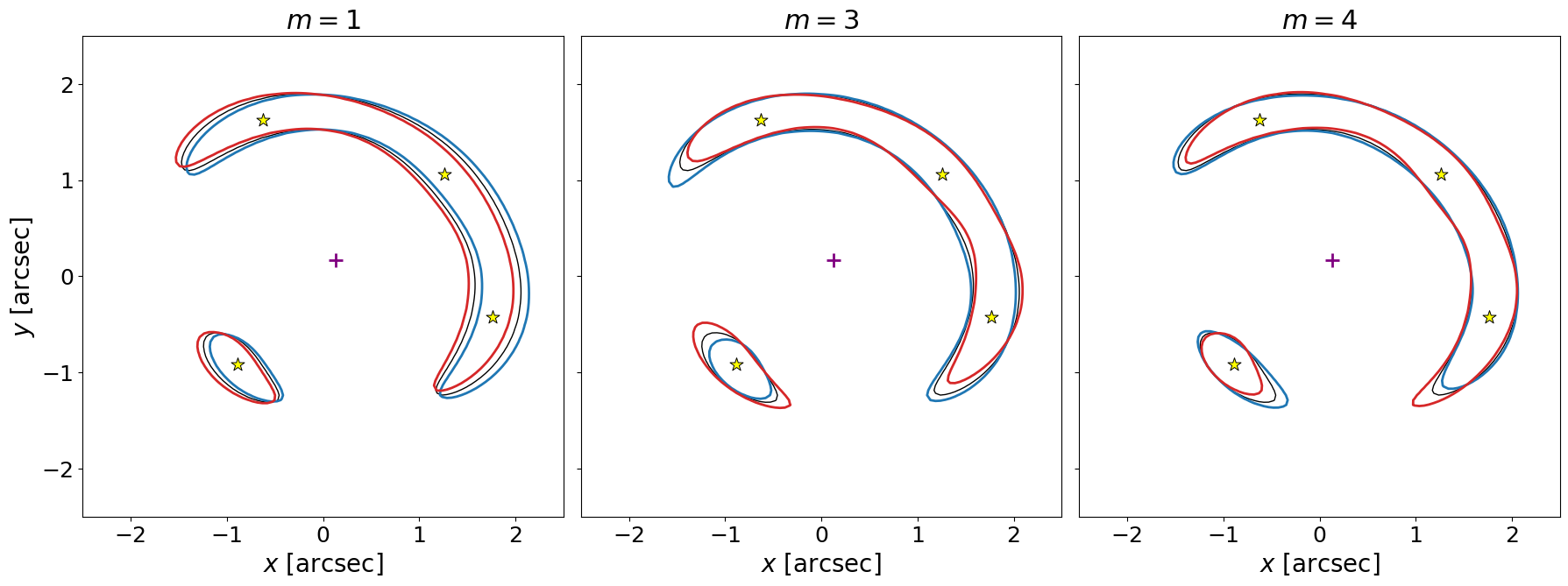}
    \end{subfigure}
    \begin{subfigure}[t]{1.\textwidth}
        \centering
        \caption{Cusp lens with source position $(x,y) = (0.13, 0.17)^{\prime\prime}$. Mass Parameters: $\theta_E = 1.6^{\prime\prime}$, $\gamma=2.1$, $q=0.75$, $\phi_{\text{EPL}}=23 \degree$, $(\gamma_1, \gamma_2) = (0.07, -0.07)$}
    \end{subfigure}
    ~ 
    \caption{Change in the image positions and lensed arcs of the source galaxy's flux for multipole perturbations of order $m=1,3$ and $4$ added to an EPL with external shear model. The left panel of each subplot shows the variation in image positions for multipole magnitudes up to 15\%, with blue circles representing the change in image position due to positive amplitude and red circles corresponding to negative amplitude. The right panel shows how the lensed arcs are changed for multipole deviation of $\pm 15$\% where the red line shows the perturbation for positive and blue line shows the perturbation for negative amplitudes. The amplitudes are much larger than would be expected for real systems for illustrative purposes.}
    \label{fig:}
\end{figure*}

\section{Parameter Inference and Time-Delay Cosmography}
\label{sec:impact_tdcosmo}

In this section, we investigate the impact of multipoles in the mass model of deflectors on the inferred $H_0$ in time-delay cosmography by simulating mock strong lensing systems. For the modelling and fitting of simulated mock lens images, we use the open-source software package \texttt{Lenstronomy} \citep{birrer_lenstronomy_2018, birrer_lenstronomy_2021}.

\subsection{Mock Image Construction}

We simulate our mock images using $HST$-like observation conditions, using a pixel size of 0.05 arcsec, a background rms noise of 0.005 photons/sec per pixel, and a maximum exposure time of 5000 sec. We choose a simple Gaussian point spread function (PSF) model with a full width at half maximum of 0.10 arcsec to avoid PSF-related complications in the lens model fitting. In real observations, the PSF is usually more complex and has to be reconstructed from the data.

We only focus on simulating quad configurations, i.e. quadruply-lensed sources. This is for two reasons: (i) cosmographic studies prioritise quad systems as they provide multiple time delays for better cosmological precision, and (ii) dark matter investigations based on flux ratio anomalies currently depend on such samples \citep{vyvere_tdcosmo_2022, cohen_general_2024}. We position the source at different locations with respect to the inner caustic to produce three different image configurations of quad lenses: cross, fold, and cusp. % (Figure \ref{fig: Quad Configurations}).

\subsection{Lens Modelling}

We briefly describe the model choices and parameters used for the mock systems in this section.

\subsubsection{Source and Lens Light Profile}

The host galaxy light model of the mock systems consists of a lensed point source (e.g. quasar or supernova) and its host galaxy which is modelled by the elliptic Sérsic light profile defined in Equation \ref{equ: Sersic light profile}. The source light profile is centred at the position of the point source, with Sérsic index $n=1$, the Sérsic half-light radius $R_e$ sampled between 0.15 and 0.2 arcsec and axis ratio between $q=0.7$ and $q=0.9$. The host is brighter than the point source by around 1.5 mag.

The light profile of the main deflector galaxy is modelled as a Sérsic profile with  Sérsic index $n=4$ and half-light radius between $0.4 \leq R \leq 0.5$ arcsec.

\subsubsection{Lens Mass Profile}
\label{sec: 5.2. Lens Mass Profile}

The mass profile of the deflector is modelled by an elliptical power-law (EPL) model with elliptical multipoles as described by Equations \ref{eqn: EPL_convergence} -- \ref{eqn: EPL-multipoles}. The slope is set between $1.9 \leq \gamma \leq 2.1$ (close to isothermal) and ellipticities $ 0.7 \leq q \leq 0.9$ are used.

An external shear component is also added to the model to simulate tidal distortions caused by surrounding galaxies and matter along the line of sight. The shear is defined by the lensing potential
\begin{equation}
    \psi_{\text{shear}}(x,y) = \frac{1}{2} (\gamma_{1}x^2+2\gamma_{2}xy - \gamma_{1}y^2)
    \label{eqn: EPL-shear}
\end{equation}
where the shear strength is $|\gamma_{\text{ext}}| = \sqrt{ \gamma_{1}^2+\gamma_{2}^2 }$ and its orientation $\phi_{\text{ext}} = \frac{1}{2}\arctan(\gamma_{2}/\gamma_{1})$. Shear magnitudes up to 0.1 are used.

The lensed images of the point source in the plane of the lens is obtained by solving the lensing equations for the specified mass profile.

\subsection{Time-delay and $H_0$ inference}
\label{sec: time-delay and H0 inference}
\subsubsection{Fitting Methodology}

We use \texttt{Lenstronomy} to simultaneously reconstruct the lens mass distribution, image positions, and the light distributions of the lens and host galaxy. The inference is performed at the pixel level of the mock images. The models we use for fitting have the following %29 
non-linear free parameters:
\begin{itemize}
    \item \textbf{Lens galaxy mass model }%(EPL+Multipole+Shear)
    : Einstein radius ($\theta_{\text{E}}$), power-law slope of the mass profile ($\gamma$), axis ratio ($q_{\text{lens}}$), orientation angle ($\phi_{\text{lens}}$), first- ($a_1$), third- ($a_3$), and fourth-order ($a_4$) multipole deviation strengths, lens galaxy mass centre ($x_{\text{lens}}, y_{\text{lens}}$), shear strength along $x$-axis ($\gamma_\text{shear,1}$), shear strength along $x=y$ direction ($\gamma_\text{shear,2}$). 
    
    \item \textbf{Lens galaxy light model} %(Sérsic)
    : Sérsic radius ($R_{\textrm{s,lens}}$), Sérsic index ($n_{\textrm{s,lens}}$), axis ratio ($q_{\textrm{s,lens}}$), orientation angle ($\phi_{\textrm{s,lens}}$), lens galaxy light centre ($x_{\textrm{s,lens}}, y_{\textrm{s,lens}}$).
    
    \item \textbf{Host galaxy light model} %(Sérsic)
    : Sérsic radius ($R_{\textrm{s,host}}$), Sérsic index ($n_{\textrm{s,host}}$), axis ratio ($q_{\textrm{s,host}}$), orientation angle ($\phi_{\textrm{s,host}}$), lens galaxy light centre ($x_{\textrm{s,host}}, y_{\textrm{s,host}}$).
    
    \item \textbf{SN images:} positions of four point sources ($x_{\textrm{A}}, \, y_{\textrm{A}}, \, x_{\textrm{B}}, \, y_{\textrm{B}}, \, x_{\textrm{C}}, \, y_{\textrm{C}}, \, x_{\textrm{D}}, \, y_{\textrm{D}}$).
\end{itemize} We adopt Gaussian priors for these, in combination with conservative upper and lower bounds to ensure good convergence. The time-delay distance ($D_{\Delta t}$) is also set as a free parameter which is simultaneously sampled in the inference.

We use Particle Swarm Optimisation (PSO; \citealp{kennedy_particle_1995, shi_modified_1998}) to find the model parameters that maximise the likelihood. The particles are initialised with a $10\%$ Gaussian prior around their starting positions in the multi-dimensional parameter space, and then the PSO is run for 250 iterations with 200 particles. The likelihood contains terms associated with the image residuals and measured time delays. We simulate the measured relative time delays for each image pair by drawing from a normal distribution with standard deviation of about $0.5-2$ days ($<5\%$ time-delay uncertainty). This is usually the accuracy with which we can measure the relative time delays between SN light curves. The mean of the normal distribution is the predicted time delay obtained by solving the lens equations for the model parameters used to create the mock images. 

After the PSO has converged, an MCMC sampling is performed using the \texttt{emcee} package within \texttt{Lenstronomy} to explore the parameter space to obtain the lens model and time-delay parameter estimates with uncertainties. We use 460 walkers with 200 burn-in steps that are discarded and 1200 steps from which we obtain the final posterior on the model parameters. We initialise the walkers as a Gaussian centred around the PSO parameter values with a standard deviation of 1\%.

\begin{figure*}
    \centering
    \begin{subfigure}[t]{1\textwidth}
        \centering
        \includegraphics[height=2.73in]{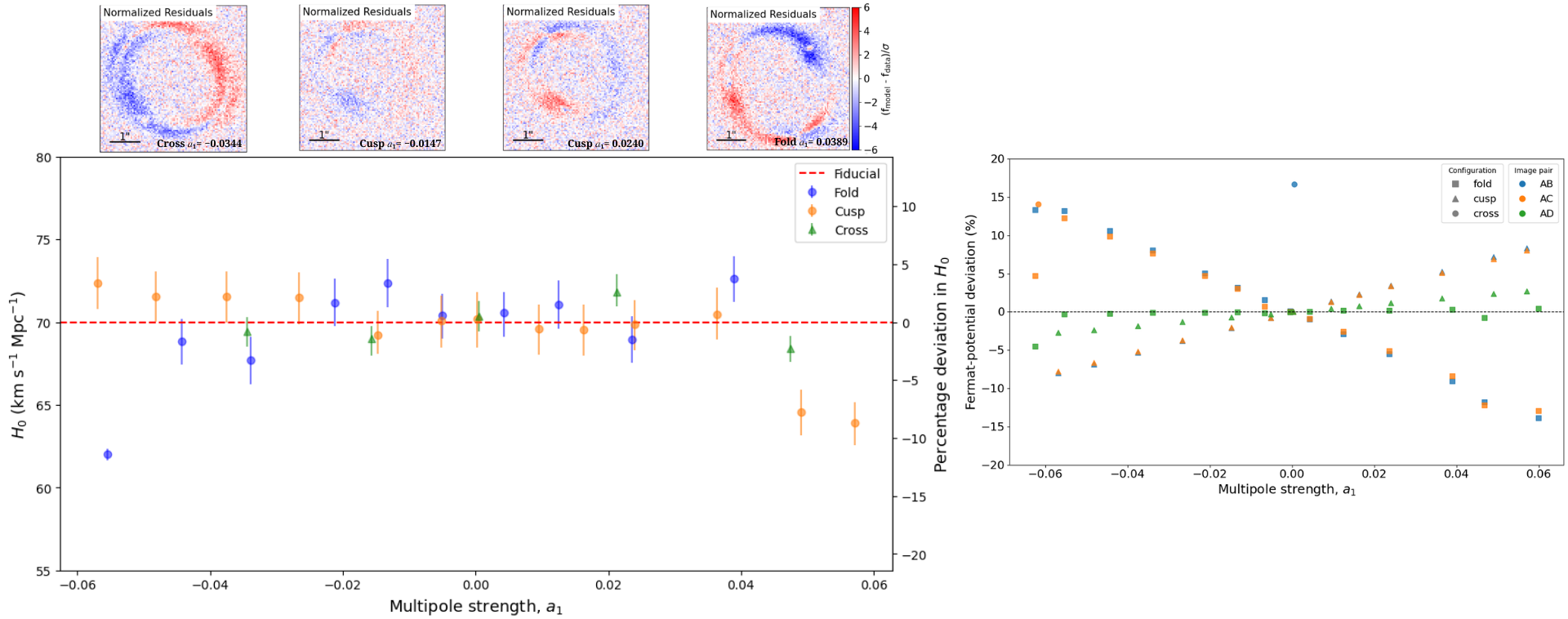}
    \caption{$m=1$ multipole}
    \end{subfigure}
    ~ 
    \begin{subfigure}[t]{1\textwidth}
        \centering
        \includegraphics[height=2.73in]{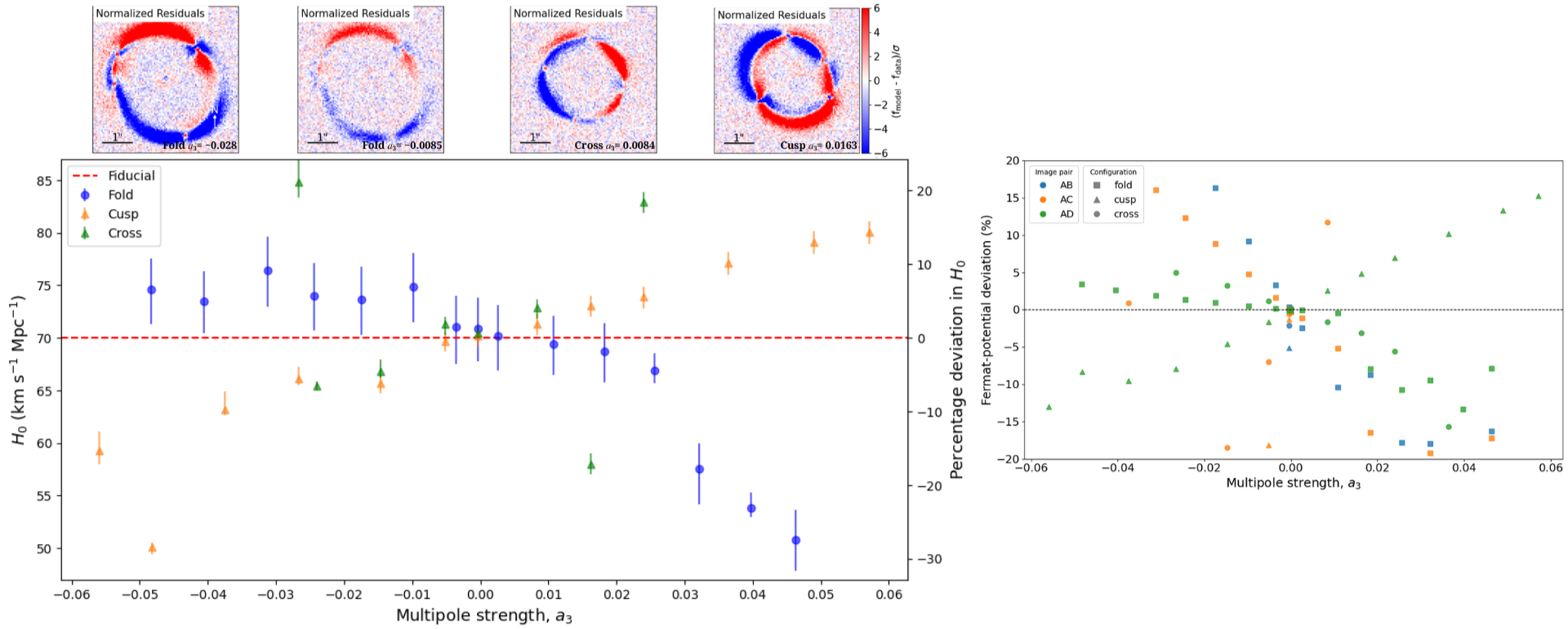}
    \caption{$m=3$ multipole}
    \end{subfigure}
    ~ 
    \begin{subfigure}[t]{1\textwidth}
        \centering
        \includegraphics[height=2.73in]{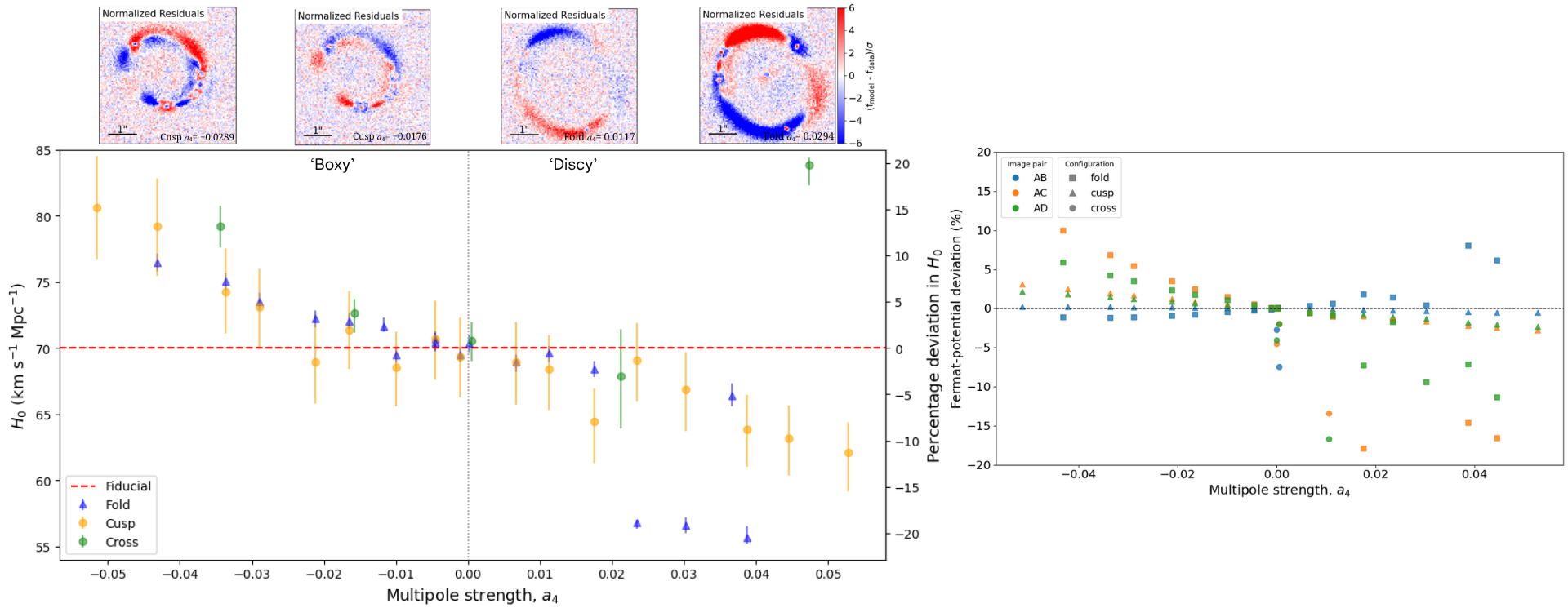}
    \caption{$m=4$ multipole}
    \end{subfigure}
    \caption{Hubble Constant ($H_0$) from time-delay cosmographic inference of mock lensing systems of different quad configurations simulated with different multipole strengths. The simulations are performed at two different redshifts of the lens and source: $z_{\text{lens}} = 0.44, z_{\text{source}} = 1.4$ (circular markers) and $z_{\text{lens}} = 0.85, z_{\text{source}} = 2.0$ (triangular markers). The lensed images are created with a power-law mass model with multipole perturbations of order 1, 3 or 4 at a time, and fitted with only a power-law model. The residual imprint pattern of the multipole from fitting are also shown for each type of perturbation. The $a_4$ multipole has a more clear directional bias in $H_0$ than the $a_3$ and $a_1$ multipoles.}
    \label{fig: Model mismatch Impact on H0}
\end{figure*}

\begin{figure*}
    \centering
    \begin{subfigure}[t]{0.33\textwidth}
        \centering
        \includegraphics[height=2.3in]{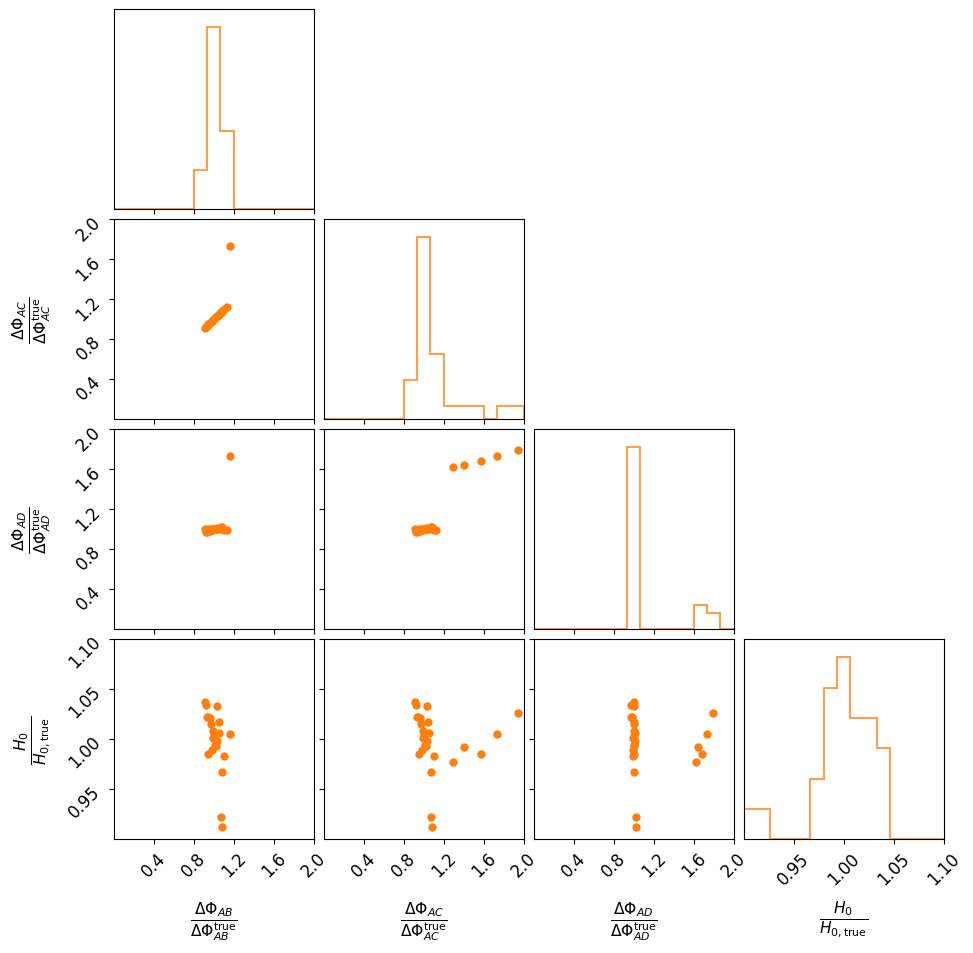}
        \caption{$a_1$}
    \end{subfigure}%
    ~ 
    \begin{subfigure}[t]{0.33\textwidth}
        \centering
        \includegraphics[height=2.3in]{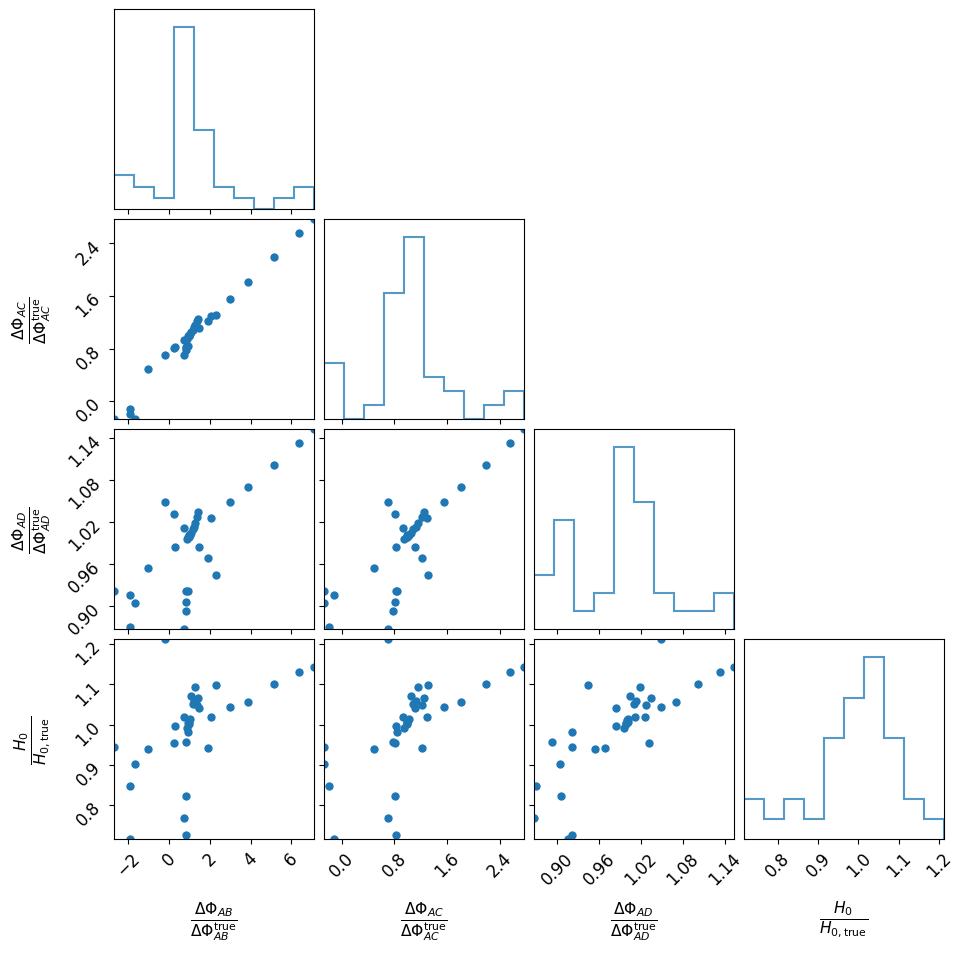}
        \caption{$a_3$}
    \end{subfigure}%
    ~ 
    \begin{subfigure}[t]{0.33\textwidth}
        \centering
        \includegraphics[height=2.3in]{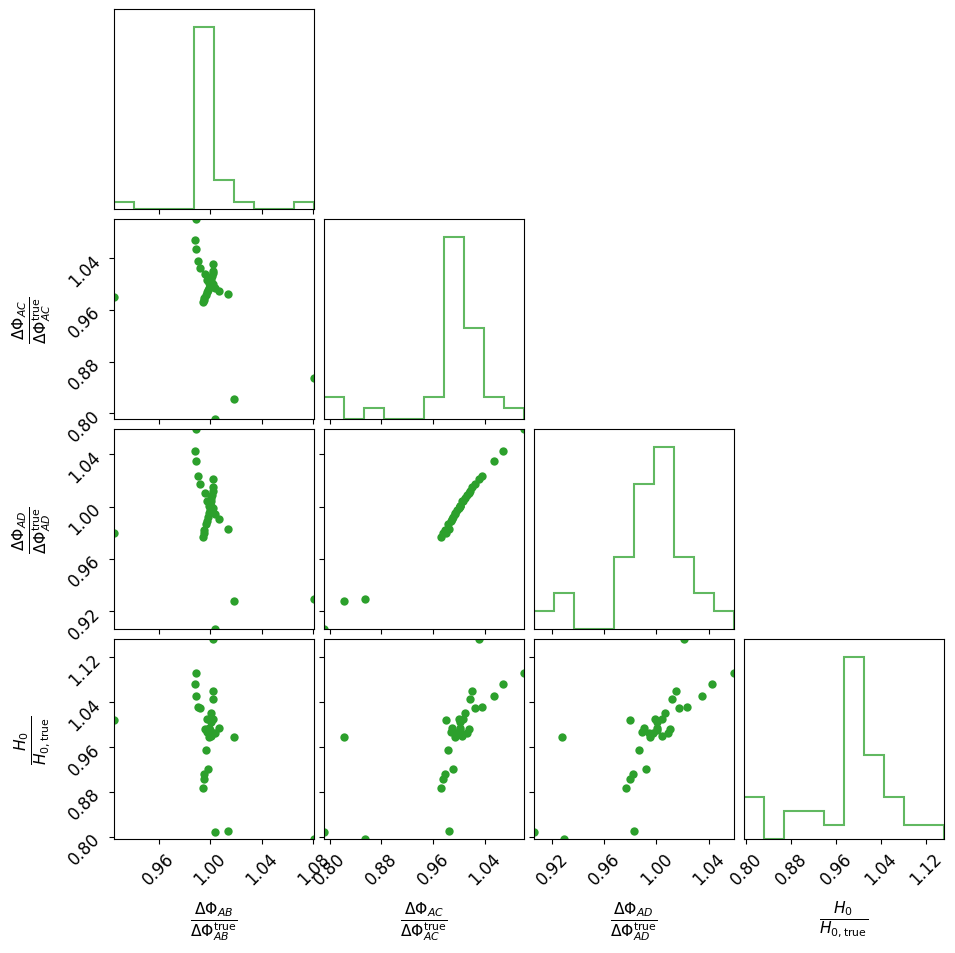}
        \caption{$a_4$}
    \end{subfigure}
    \caption{Corner plot distribution of the inferred  Fermat potential difference for the three image pairs of the quad lens and corresponding $H_0$ from the model mismatch fits in Figure \ref{fig: Model mismatch Impact on H0}. The larger deviations from 1.0 correspond to larger amplitude strengths generally of the underlying model, resulting in the inferred quantities to be further biased from the fiducial values.}
    \label{fig: Fermat-H0 corner plot}
\end{figure*}

\begin{table*}
\renewcommand{\arraystretch}{1.5}
\centering
\caption{Representative $H_0$ values for $a_{1}$. Blank entries indicate no sampled amplitude within $\pm0.004$ of the column value. Errors are asymmetric.}
\label{tab:h0_a1_summary}
\begin{tabular}{lcccccccc}
\hline
Configuration & $a_{1}=-0.050$ & $a_{1}=-0.038$ & $a_{1}=-0.025$ & $a_{1}=-0.012$ & $a_{1}=+0.012$ & $a_{1}=+0.025$ & $a_{1}=+0.038$ & $a_{1}=+0.050$ \\
\hline
Fold &  & $67.70^{+1.40}_{-1.44}$ & $71.19^{+1.42}_{-1.44}$ & $72.35^{+1.49}_{-1.47}$ & $71.05^{+1.45}_{-1.43}$ & $68.96^{+1.38}_{-1.40}$ & $72.61^{+1.36}_{-1.38}$ &  \\
Cusp & $71.56^{+1.49}_{-1.56}$ & $71.54^{+1.51}_{-1.55}$ & $71.48^{+1.51}_{-1.62}$ & $69.22^{+1.47}_{-1.15}$ & $69.60^{+1.44}_{-1.58}$ & $69.85^{+1.45}_{-1.57}$ & $70.47^{+1.59}_{-1.55}$ & $64.57^{+1.34}_{-1.42}$ \\
Cross &  & $69.45^{+0.84}_{-0.93}$ &  & $68.99^{+0.78}_{-1.01}$ &  & $71.81^{+1.09}_{-0.85}$ &  & $68.38^{+0.80}_{-0.81}$ \\
\hline
\end{tabular}
\end{table*}

\begin{table*}
\renewcommand{\arraystretch}{1.5}
\centering
\caption{Representative $H_0$ values for $a_{3}$. Blank entries indicate no sampled amplitude within $\pm0.004$ of the column value. Errors are asymmetric.}
\label{tab:h0_a3_summary}
\begin{tabular}{lcccccccc}
\hline
Configuration & $a_{3}=-0.050$ & $a_{3}=-0.038$ & $a_{3}=-0.025$ & $a_{3}=-0.012$ & $a_{3}=+0.012$ & $a_{3}=+0.025$ & $a_{3}=+0.038$ & $a_{3}=+0.050$ \\
\hline
Fold & $74.57^{+2.98}_{-3.26}$ & $73.46^{+2.82}_{-3.05}$ & $74.00^{+3.07}_{-3.31}$ & $74.89^{+3.18}_{-3.43}$ & $69.39^{+2.70}_{-2.96}$ & $66.91^{+1.64}_{-1.25}$ & $53.82^{+1.44}_{-0.88}$ & $50.74^{+2.88}_{-2.92}$ \\
Cusp & $50.07^{+0.42}_{-0.70}$ & $63.14^{+1.78}_{-0.50}$ & $66.15^{+1.08}_{-0.62}$ & $65.69^{+0.73}_{-1.01}$ & $73.01^{+1.00}_{-1.03}$ & $73.87^{+0.95}_{-1.07}$ & $77.08^{+1.03}_{-1.07}$ & $79.07^{+1.06}_{-1.10}$ \\
Cross &  &  & $65.45^{+0.45}_{-0.45}$ & $66.83^{+1.08}_{-0.62}$ & $66.96^{+1.02}_{-0.96}$ & $78.87^{+1.00}_{-1.03}$ &  &  \\
\hline
\end{tabular}
\end{table*}

\begin{table*}
\renewcommand{\arraystretch}{1.5}
\centering
\caption{Representative $H_0$ values for $a_{4}$. Blank entries indicate no sampled amplitude within $\pm0.004$ of the column value. Errors are asymmetric.}
\label{tab:h0_a4_summary}
\begin{tabular}{lcccccccc}
\hline
Configuration & $a_{4}=-0.050$ & $a_{4}=-0.038$ & $a_{4}=-0.025$ & $a_{4}=-0.012$ & $a_{4}=+0.012$ & $a_{4}=+0.025$ & $a_{4}=+0.038$ & $a_{4}=+0.050$ \\
\hline
Fold &  & $75.04^{+0.65}_{-0.66}$ & $72.25^{+0.56}_{-0.68}$ & $69.46^{+0.53}_{-0.58}$ & $69.60^{+0.55}_{-0.63}$ & $56.77^{+0.22}_{-0.41}$ & $55.67^{+0.88}_{-0.48}$ &  \\
Cusp & $80.64^{+3.90}_{-3.93}$ & $74.23^{+3.28}_{-3.10}$ & $68.97^{+3.01}_{-3.19}$ & $68.51^{+2.71}_{-2.91}$ & $68.39^{+2.55}_{-3.10}$ & $69.09^{+2.82}_{-3.12}$ & $63.81^{+2.62}_{-2.81}$ & $62.10^{+2.27}_{-2.94}$ \\
Cross &  & $79.19^{+1.55}_{-1.57}$ &  & $72.60^{+1.13}_{-1.43}$ &  & $67.90^{+3.53}_{-3.97}$ &  & $83.85^{+0.59}_{-1.53}$ \\
\hline
\end{tabular}
\end{table*}

% \begin{figure*}
%     \centering
%     \includegraphics[width=0.9\linewidth]{Figures/Parameter_Inference_and_Time-Delay_Cosmography/a3+4-H0-impact-mismatch.png}
%     \caption{}
%     \label{fig:}
% \end{figure*}

\begin{figure}
    \centering
    \begin{subfigure}[t]{0.45\textwidth}
        \centering
        \includegraphics[height=2.2in]{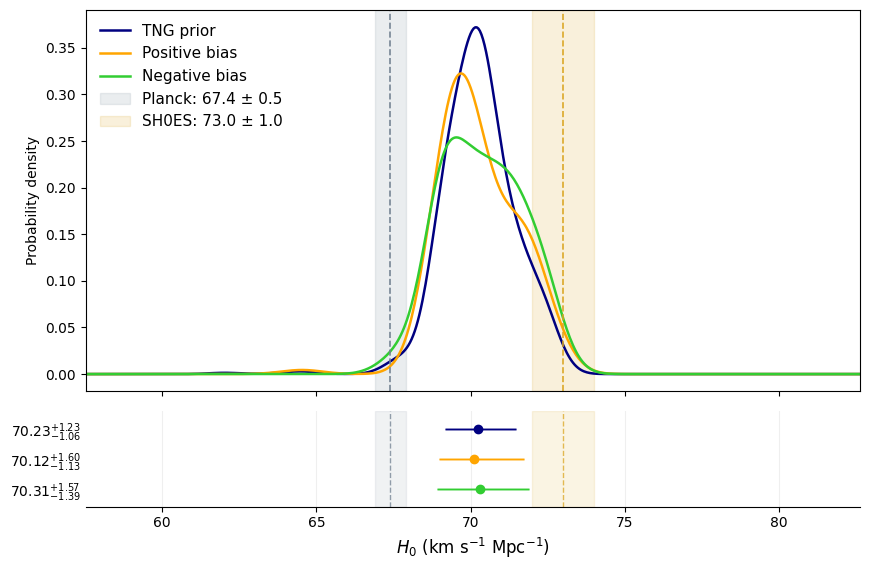}
    \caption{$H_0$ distributions weighted by the distribution of $a_1$ multipole perturbation. The distribution remains centred close to the fiducial value and imposing asymmetric priors produces minimal shifts in the median.}
    \end{subfigure}
    \vspace{6mm}
    ~ 

    \begin{subfigure}[t]{0.45\textwidth}
        \centering
        \includegraphics[height=2.2in]{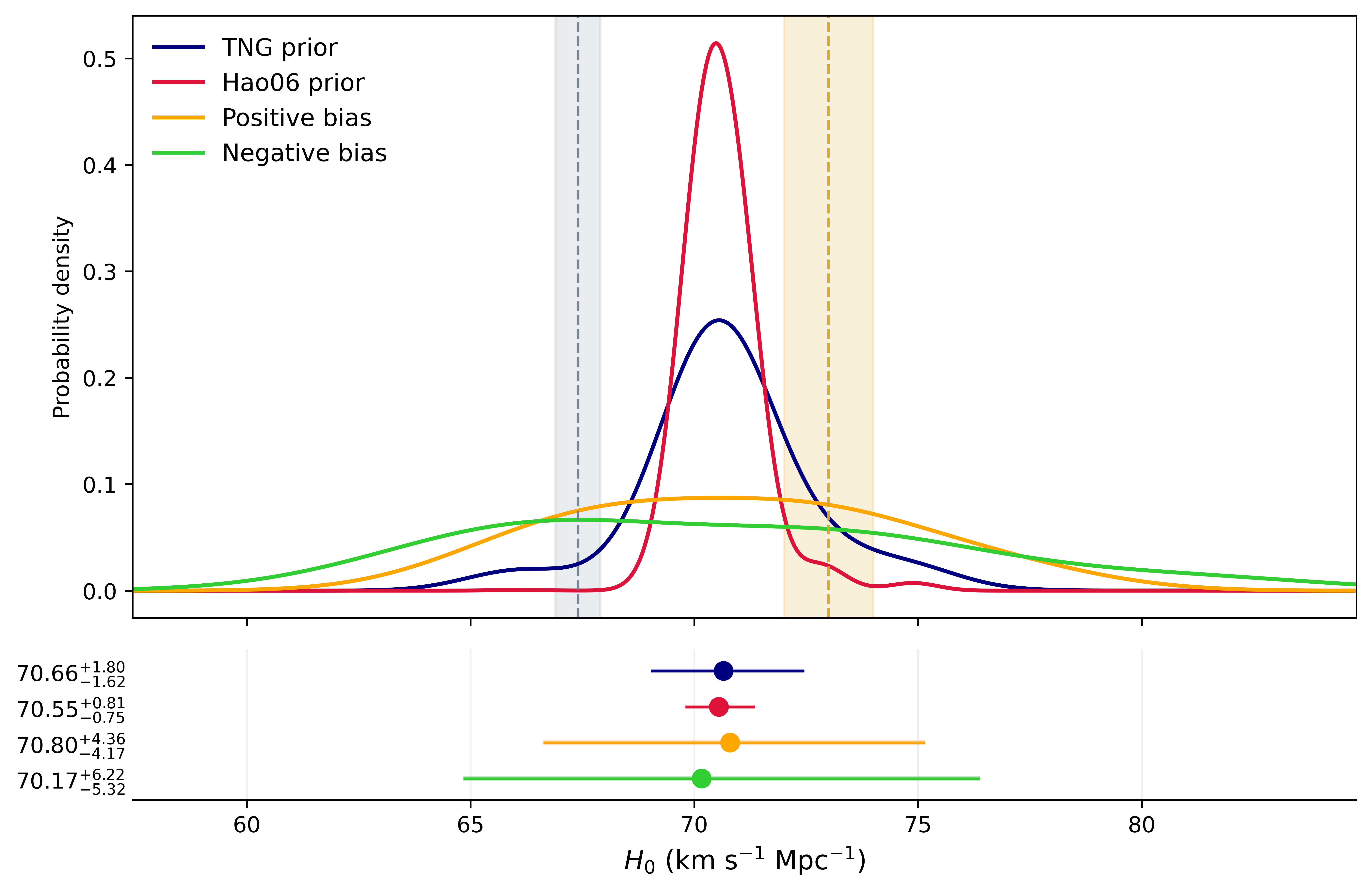}
    \caption{$H_0$ distributions weighted by the distribution of $a_3$ multipole perturbation. Weighting by the narrow Hao06 isophote prior yields a tightly constrained median compared to the broader TNG mass distribution. Under asymmetric priors, the dispersion increases noticeably due to the larger differences in the biased $H_0$ from fold and cusp lenses of the same multipole strength.}
    \end{subfigure}
    \vspace{3mm}
    ~
    \caption{Distribution of the $H_0$ from our model mismatch cosmographic tests for the $m=1$ and $3$ perturbations, with the fiducial value taken to be $70$ km s$^{-1}$ Mpc$^{-1}$. The distribution is weighted by the occurrence of the multipoles from our TNG galaxy samples and the catalogue obtained by Hao06, as well as if the assumed multipole distribution is positively or negatively skewed (median of 2\% on either side with the same spread as the TNG distribution). The $H_0$ values and uncertainties from the Planck CMB measurement \citep{Planck2018} and SH0ES collaboration \citep{Riess2022} are shown by the shaded bands for comparison with the current state of the Hubble tension.}

    \label{fig: weighted H0 PDF 1}
\end{figure}

\begin{figure}
    \centering
        \centering
        \includegraphics[height=2.2in]{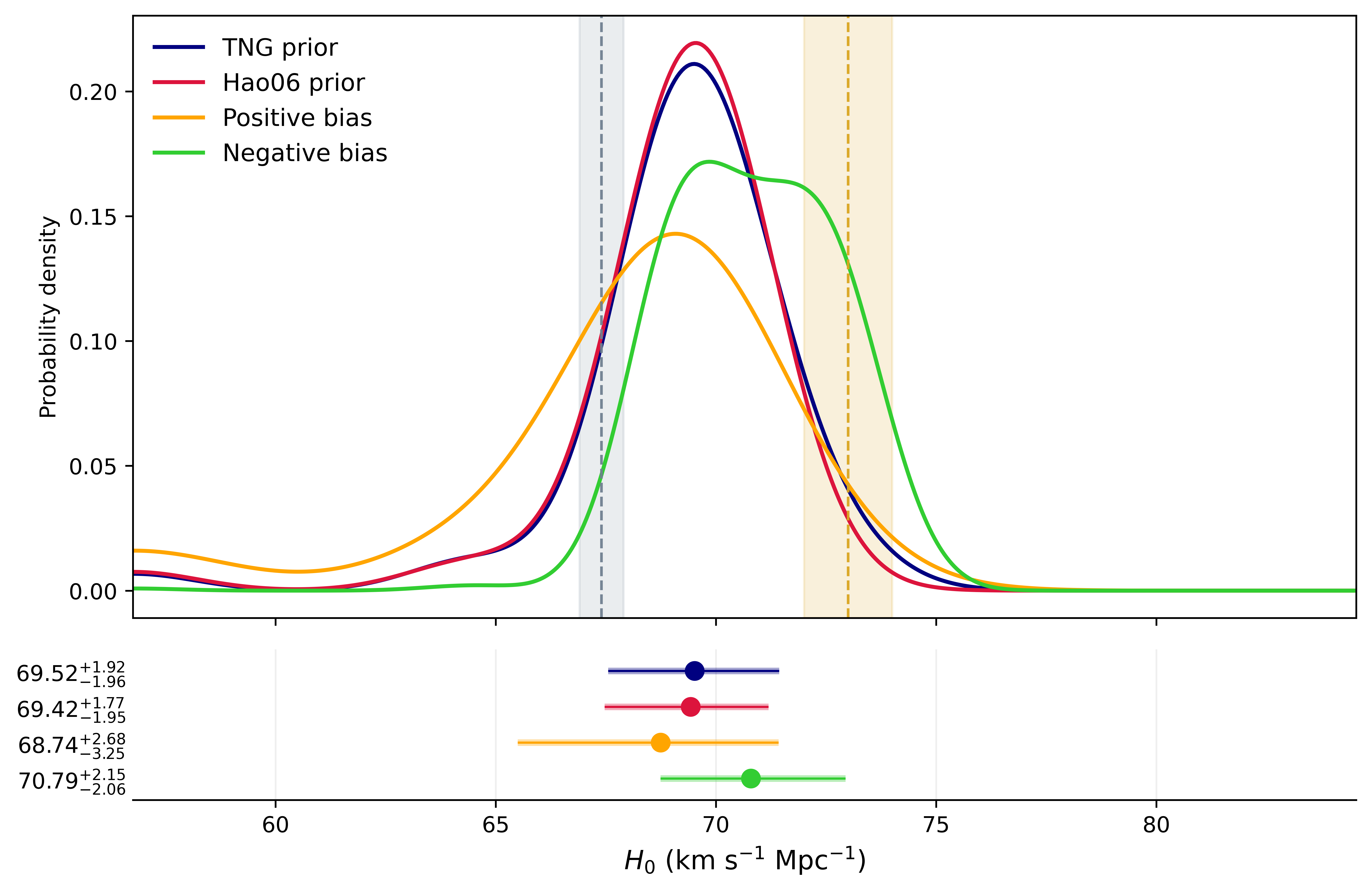}
    \caption{Distribution of the $H_0$ from our model mismatch cosmographic tests for the $m=4$ perturbations, weighted by different $a_4$ priors. A clear systematic shift in the population median is expected due to the directionality of the $H_0$ bias with amplitude strength. The median shifts downward under a positively skewed prior (discy profiles) ($-1.8\%$ offset), whereas a negatively skewed prior (boxy profiles) shifts the median upward ($+1.1\%$ offset) from the fiducial.}
    \label{fig: weighted H0 PDF m=4}
\end{figure}
\subsubsection{Impact of Model Mismatch on $H_0$}
\label{sec: results of impact of multipole mismatch}
%\rev{histogram to show uncertainty reduction}

\begin{figure*}
    \centering
    \includegraphics[width=1.0\linewidth]{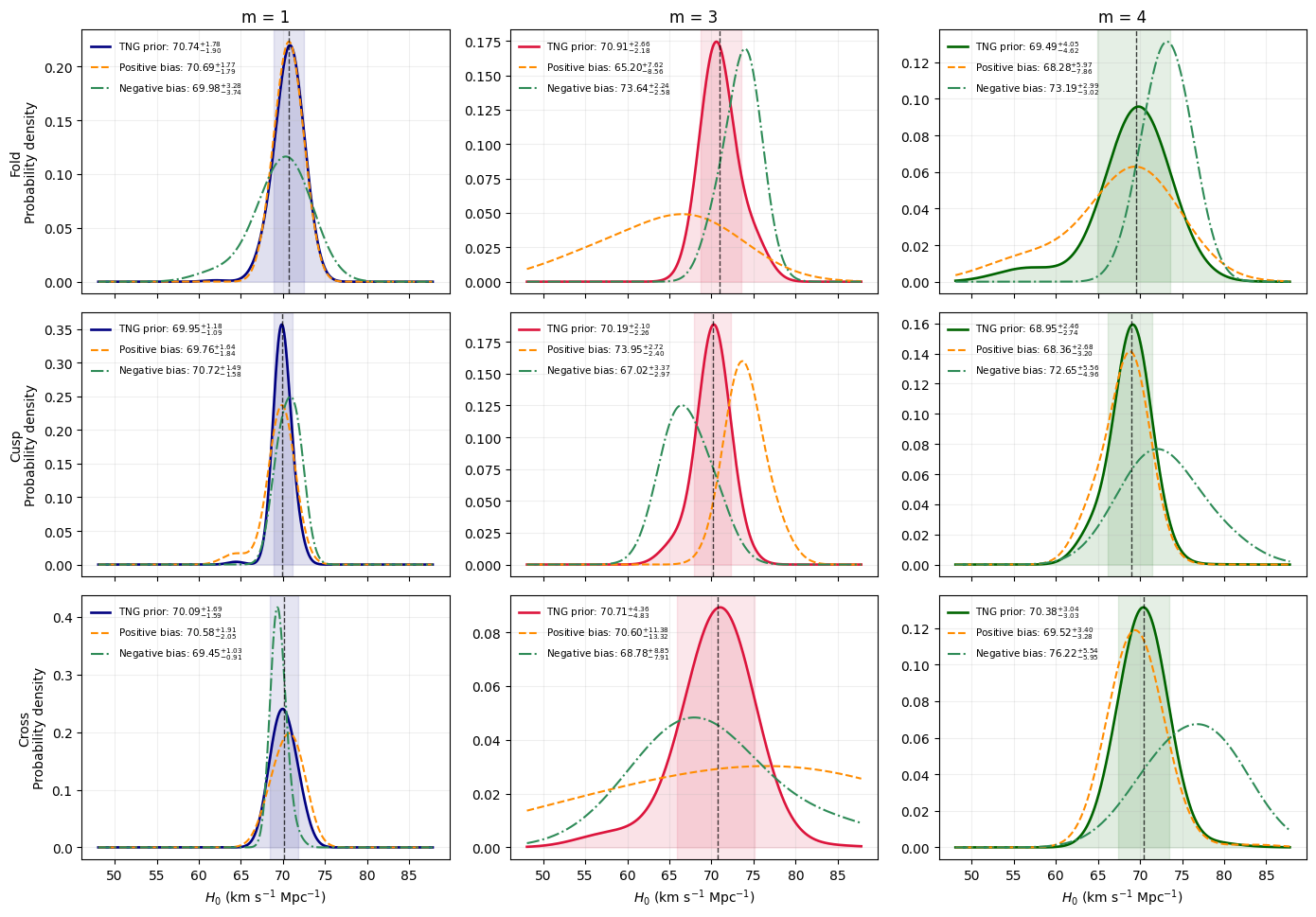}
    \caption{Multipole-weighted probability density distributions of the inferred Hubble constant $H_0$ separated by image configuration (rows: fold, cusp, and cross) and multipole order (columns: $m = 1$, $3$, and $4$). Each panel displays the resulting $H_0$ distribution obtained by convolving the deterministic model-mismatch response curves with different parent prior distributions: the empirical mass multipole distribution from our IllustrisTNG sample (solid curve, with the shaded region marking the $68\%$ credible interval), a positively biased prior ($a_m > 0$; orange dashed curve), and a negatively biased prior ($a_m < 0$; green dash-dotted curve). The true fiducial value ($H_0 = 70\,{\rm km\,s^{-1}\,Mpc^{-1}}$) is marked by the vertical black dashed line in each panel. While $m = 1$ remains stable across all configurations, $m = 3$ induces substantial variance and exhibits configuration-dependent median shifts in opposite directions for folds and cusps. For $m = 4$, negative perturbations ($a_4 < 0$, boxy) consistently overestimate $H_0$, with the most pronounced shift occurring in cross configurations.}
    \label{fig: am_weighted_H0_distributions_by_configuration}
\end{figure*}

We perform our time-delay cosmography inference by generating mock images using an EPL mass model that includes multipole perturbations ($a_1,a_3,a_4$) drawn from the distributions inferred from the TNG galaxies. However, assuming that we do not know to check for the presence of multipoles, we perform the fitting using only an EPL profile with external shear for the lens mass model to quantify the systematic bias in $H_0$ for different orders of multipole perturbations. Figure \ref{fig: Model mismatch Impact on H0} shows how $H_0$ is biased for varying strengths and order of multipole perturbations for different lensing quad configurations. In strong gravitational lensing, the inferred $H_0$ scales directly with the Fermat potential difference $\Delta\phi$ and inversely with the arrival time difference $\Delta t$ (Equations \ref{equ: time-delay equation} and \ref{equ: Hubble-time delay disctance scaling}):
\begin{equation}
    H_0 \propto \frac{\Delta\phi_{AB}}{\Delta t_{AB}}.
    \label{eq:H0_proportionality}
\end{equation}
Because the simulated time delays incorporate an observational measurement uncertainty ($\sigma_{\Delta t} \sim 0.5\text{--}2$~days), the total fractional variance of the inferred $H_0$ contains contributions from both the observational precision and the model-induced Fermat potential recovery:
\begin{equation}
    \left(\frac{\sigma_{H_0}}{H_0}\right)^2 \approx \left(\frac{\sigma_{\Delta \phi}}{\Delta \phi}\right)^2 + \left(\frac{\sigma_{\Delta t}}{\Delta t}\right)^2.
    \label{eq:H0_variance_split}
\end{equation}
%For the inference of individual galaxies, the bias in $H_0$ is of the order of several percent for multipole deviations of a few percent.

% Our results for the fourth-order multipole are in strong qualitative agreement with the findings of \cite{vyvere_tdcosmo_2022}. 
As illustrated in Figure \ref{fig: Model mismatch Impact on H0}c, we recover a clear directional bias for fourth-order multipoles: boxy lens galaxies ($a_4<0$) lead to a systematic overestimation of $H_0$, while discy galaxies ($a_4>0$) generally bias $H_0$ toward lower values. For the range of $a_4$ amplitudes emergent in our TNG sample (typically $|a_4|\lesssim 0.02$), the individual lens bias reaches $\sim3-5\%$.
This confirms that the $m=4$ perturbations alter the Fermat potential difference $\Delta\phi$ in a predictable way by changing the "sharpness" of the potential at the image positions. If a population of lenses has an equal distribution of boxy and discy structures, this bias might cancel out; however, as noted in Section 4, the TNG population shows a relatively greater preference for discy configurations, which could leave a residual sub-percent systematic in large-scale surveys like LSST.

The most significant departure from the work of \citet{vyvere_tdcosmo_2022} arises from our inclusion of the $m=1$ (lopsidedness) and $m=3$ (triangularity) modes. As shown in Figures 12a and 12b, the odd-order multipoles do not exhibit the same monotonic directional bias as the boxy/discy mode, and the bias depends on the particular quad configuration. Instead, $a_1$ and $a_3$ introduce a significant stochastic scatter in the inferred $H_0$. Lopsidedness, in particular, can be interpreted as a displacement between the center of the light and the center of the mass. When we force an EPL model to fit a lopsided TNG galaxy, the model attempts to compensate by shifting the lens center or adjusting the external shear magnitude. 
%This results in $H_0$ deviations of up to $\sim$[X]\%, even when the image residuals remain below the noise level (see Section 6).

\subsubsection{Prior-weighted probability distributions of $H_0$}

To evaluate how unmodelled angular complexity manifests across an ensemble of systems, we construct multipole-weighted probability density functions (PDFs) of $H_0$ following \citet{vyvere_tdcosmo_2022}. Rather than performing full MCMC sampling across a massive synthetic catalogue, we map the response of the inferred Hubble constant, $H_0(a_m)$, over the discrete grid of multipole perturbation amplitudes shown in Figure \ref{fig: Model mismatch Impact on H0}.

For each multipole order ($m=1,3,4$) and for each quad configuration (fold, cusp, and cross), we interpolate the point estimates of $H_0(a_m)$, excluding individual MCMC measurement errors to isolate the purely systematic bias induced by model mismatch, and weight them by parent prior distributions, $P(a_m)$. We use four distinct weighting priors:
\begin{enumerate}
\item The mass multipole distributions extracted from our IllustrisTNG sample (Section~\ref{sec: mass parameter distributions});
\item The observed stellar isophote distribution from the SDSS ETG sample of \citet{hao_isophotal_2006} (for $m = 3$ and $m = 4$)
\item A positively biased skew-normal distribution prior ($\mu= + 0.02$) representing a population dominated by positive multipole amplitudes ($a_m>0$); and
\item A positively biased skew-normal distribution prior ($\mu= - 0.02$) representing an excess of negative amplitudes ($a_m<0$).
\end{enumerate}
The resulting combined population PDFs are shown in Figures \ref{fig: weighted H0 PDF 1} and \ref{fig: weighted H0 PDF m=4} For the lopsided $m=1$ mode, the median remains stable across all weighting schemes, exhibiting minimal directional shift broadening in the PDF. The triangular $m=3$ mode preserves a consistent median near the fiducial of the simulations but induces large variance inflation under skewed priors, where the $68\%$ credible interval expands to $70.80^{+4.36}_{-4.17}\,{\rm km\,s^{-1}\,Mpc^{-1}}$ and $70.17^{+6.22}_{-5.32}\,{\rm km\,s^{-1}\,Mpc^{-1}}$ for positive and negative biases, respectively, encompassing the entire span between the \textit{Planck} 
($67.4\pm 0.5\,{\rm km\,s^{-1}\,Mpc^{-1}}$) and SH0ES 
($73.0\pm 1.0\,{\rm km\,s^{-1}\,Mpc^{-1}}$) benchmarks. Meanwhile, the $m=4$ mode (Fig.~\ref{fig: weighted H0 PDF m=4}) introduces a clear 
directional systematic shift. Both the TNG and Hao06 priors shift the 
medians downward from the fiducial driven by the demographic excess of discy ($a_4 > 0$) profiles, while positive and negative biased distributions shift the medians to $68.74^{+2.68}_{-3.25}\,{\rm km\,s^{-1}\,Mpc^{-1}}$ ($-1.8\%$) and $70.79^{+2.15}_{-2.06}\,{\rm km\,s^{-1}\,Mpc^{-1}}$ ($+1.1\%$), respectively.

The configuration-resolved distributions are detailed in the $3\times3$ grid of Figure \ref{fig: am_weighted_H0_distributions_by_configuration}. In fold configurations, positive $a_{3}$ perturbations bias $H_{0}$ strongly downward to a median of $65.20_{-8.56}^{+7.62}$ km s$^{-1}$ Mpc$^{-1}$, while negative perturbations bias $H_0$ upward to $73.64_{-2.58}^{+2.34}$ km s$^{-1}$ Mpc$^{-1}$. Conversely, for cusp configurations, this trend reverses: positive $a_3$ pushes the median up to $73.95_{-2.40}^{+2.72}$ km s$^{-1}$ Mpc$^{-1}$, while negative $a_{3}$ pulls it down to $67.02_{-2.97}^{+3.37}$ km s$^{-1}$ Mpc$^{-1}$. In cross configurations, the three-fold symmetry produces an enormous spread.
%, with uncertainties reaching $\sigma_{H_{0}}\approx11-13$ km s$^{-1}$ Mpc$^{-1}$. 
Because the images in fold, cusp, and cross configurations probe different azimuthal chords of the potential relative to the triangular lobes of the $m=3$ multipole, the resulting Fermat potential differences $\Delta\phi$ shift in opposite directions depending on the position of the source relative to the caustic cusps. While folds and cusps under negative $a_4$ (boxy) perturbations yield median $H_0$ estimates of $73.19^{+2.99}_{-3.02}\text{ km s}^{-1}\text{ Mpc}^{-1}$ and $72.65^{+5.56}_{-4.90}\text{ km s}^{-1}\text{ Mpc}^{-1}$, respectively, cross configurations exhibit notable vulnerability, shifting to $H_0 = 76.22^{+3.54}_{-5.95}\text{ km s}^{-1}\text{ Mpc}^{-1}$ -- an overestimation of $\sim+9\%$. Cross configurations occur when the source is close to the centre of the astroid caustic, producing four images at comparable radial distances across the four quadrants. When a boxy perturbation pushes mass into the $\sim45^\circ$ diagonal corners, an EPL model attempting to fit all four images simultaneously undergoes an extreme over-adjustment of its Einstein radius and shear, significantly affecting the relative Fermat potential.

While \citet{vyvere_tdcosmo_2022} concluded that multipoles are unlikely to be a source of substantial systematic bias at the population level ($<1$\%) which agrees with our combined TNG prior-weighted $H_0$, it is still possible that bias in $H_0$ inferences could exceed the sub-percent level if the multipole perturbations in elliptical galaxies in the universe are very skewed in one direction as well as depending on the configuration of quad lenses in the cosmographic sample. Consequently, assuming that multipole perturbations cleanly average out to sub-percent precision across a population of lenses is fundamentally dependent on the true, currently unconstrained demographic balance of galaxy shapes in the Universe.

\begin{figure}
    \centering
    \begin{subfigure}[t]{0.45\textwidth}
        \centering
        \includegraphics[height=2.2in]{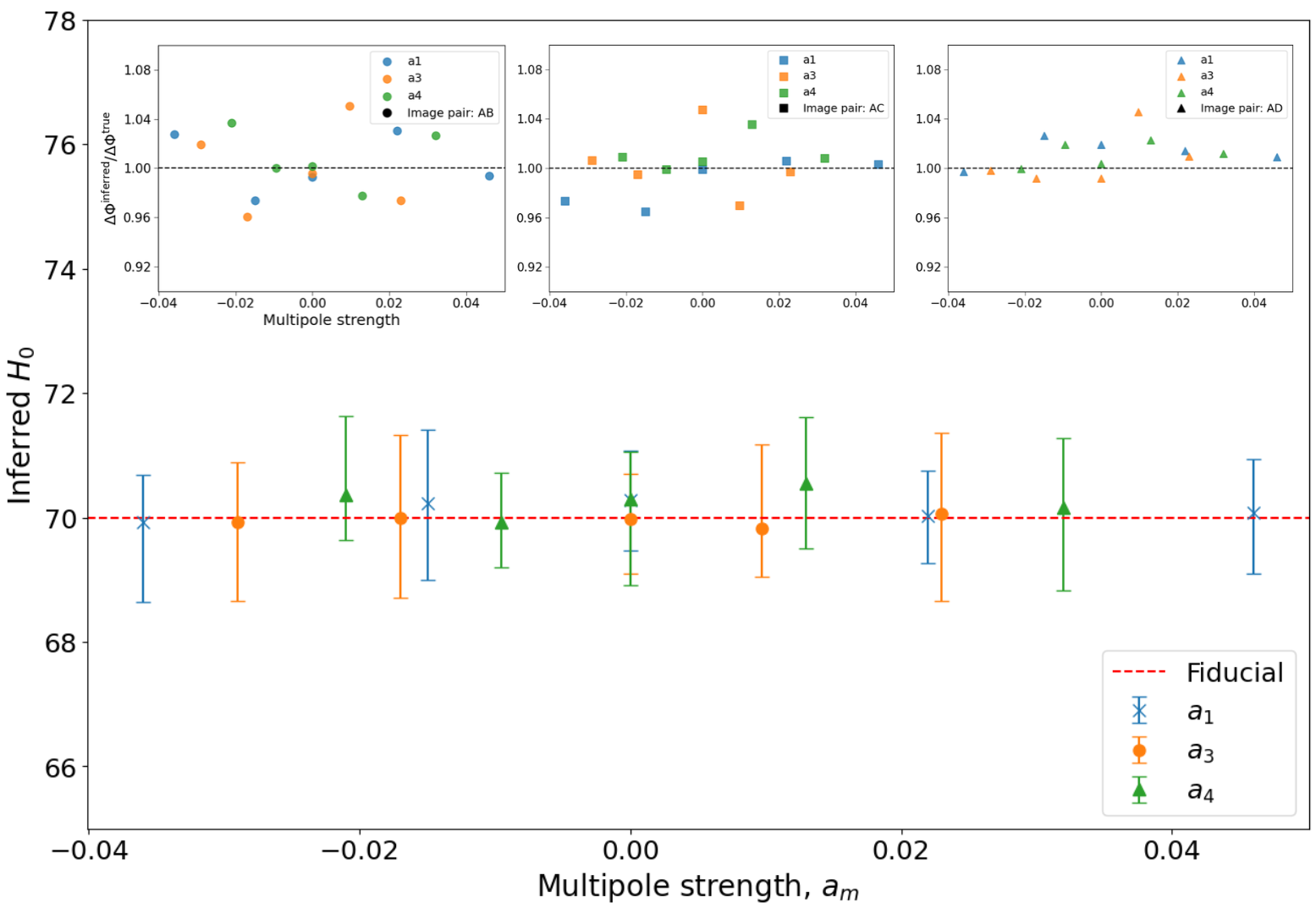}
    \caption{Recovered $H_0$ for different multipole strengths for each order $m=1,3,4$ for simulated quad lens. The insets show the corresponding inferred Fermat potential differences, $\Delta\phi$, for each image pair AB, AC, and AD as fractions of the expected $\Delta\phi$ values.}
    \end{subfigure}
    \vspace{6mm}
    ~ 

    \begin{subfigure}[t]{0.45\textwidth}
        \centering
        \includegraphics[height=1.8in]{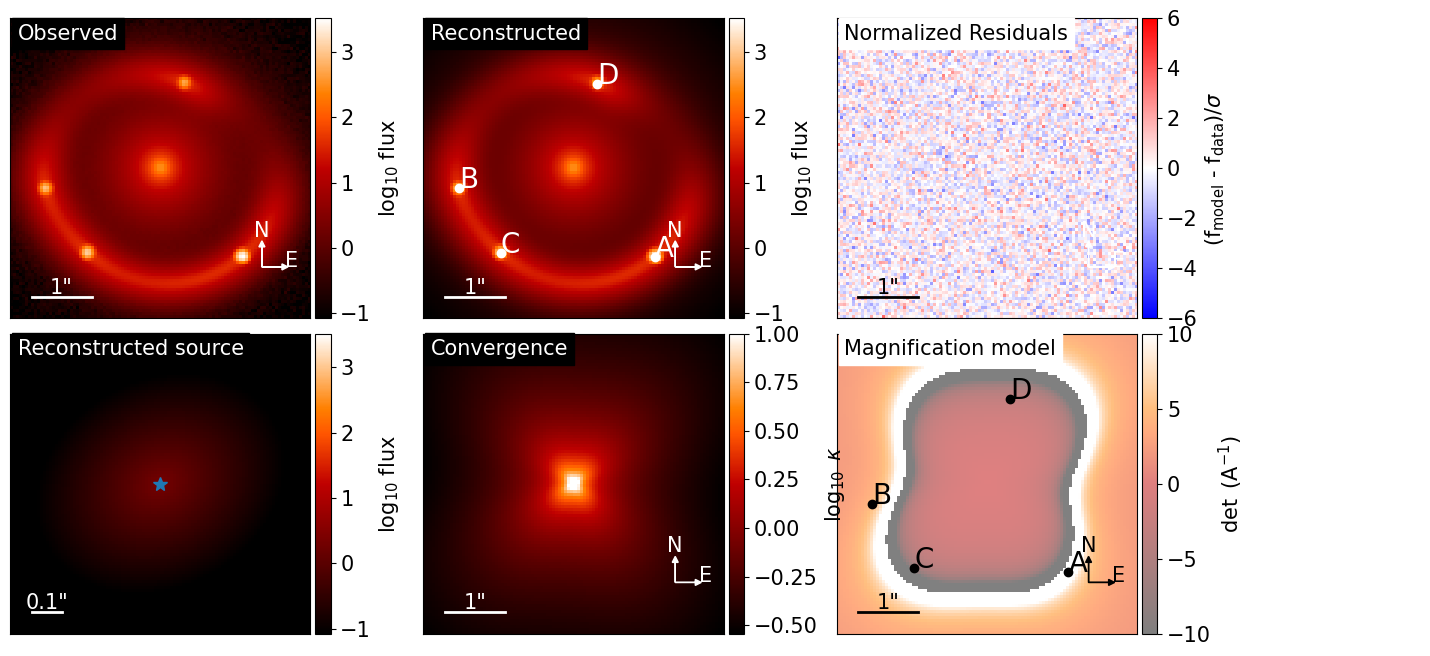}
    \caption{\textit{Top panel from left to right}: The simulated data with noise similar to a high-resolution observed image of a strongly lensed system, the best-fit model EPL+Multipole reconstruction, and the normalised residuals between the input and recreated data from model fitting. \textit{Bottom panel from left to right}: the reconstructed source light intensity, convergence map of the best fit EPL+Multipole model, and magnification model with the four lensed source image positions (A--D). We use an idealised Gaussian PSF with FWHM = 0$^{\prime\prime}$.1 in both the simulation and fitting.}
    \end{subfigure}
    \vspace{3mm}
    ~
    \caption{Model recovery validation tests where an EPL+Multipole model is fitted to mock data simulated with the same mass model.}

    \label{fig: model recovery}
\end{figure}

\subsection{Effect of lensed arc observation}
\label{sec: effect of lensed arc observation}

In this section, we briefly describe the detectability limit of angular complexities in the lens mass distribution.
The detectability of these perturbations is limited by the signal-to-noise ratio (S/N) of the lensed images. As a proxy for low S/N observations we perform a series of lens fitting and cosmography tests where the background galaxy brightness amplitude is very low to check whether unmodelled multipoles can bias $H_0$ inference in cases of poorly resolved or absent lensed arcs of the source galaxy (e.g. a bright supernova source far from the host galaxy centre).

In our low host galaxy brightness scenario, we find that realistic multipole amplitudes can remain undetected in the image residuals. As summarised in Table \ref{tab: effect of source galaxy brightness}, models with multipole deviations as high as 3.3\% can yield a reduced $\chi^2$ near unity, appearing statistically indistinguishable from a pure EPL profile. Crucially, even when the image residuals remain consistent with white noise, the inferred $H_0$ can be biased by several percent (e.g., the $a_4 = 0.022$ case). This confirms that a satisfactory fit to the imaging data is an insufficient criterion for cosmological accuracy, as the standard EPL + external Shear model possesses the parametric flexibility to stochastically absorb complex mass features at the cost of the correct Fermat potential difference. Unlike the $m=3$ and $m=4$ mode, which produces a distortion that is somewhat easier to spot in residuals, the $m=1$ perturbation is more easily mimicked by a slight shift in the lens centre or a perturbation in the external shear.

% (e.g. in the case where the lensed point source such as a supernova is far from the host galaxy centre).

% Table \ref{tab: effect of source galaxy brightness} shows the residuals and inferred $H_0$ for some typical multipole pertubations fractions. We find that in the absence of clearly resolved or prominent lensed arcs, the multipoles can remain undetected and can still lead to hidden biases in the constrained $H_0$.

\subsection{External shear compensating multipole perturbation}

\begin{figure*}
    \centering
    \begin{subfigure}{0.9\textwidth}
        \centering
        \includegraphics[height=1.8in]{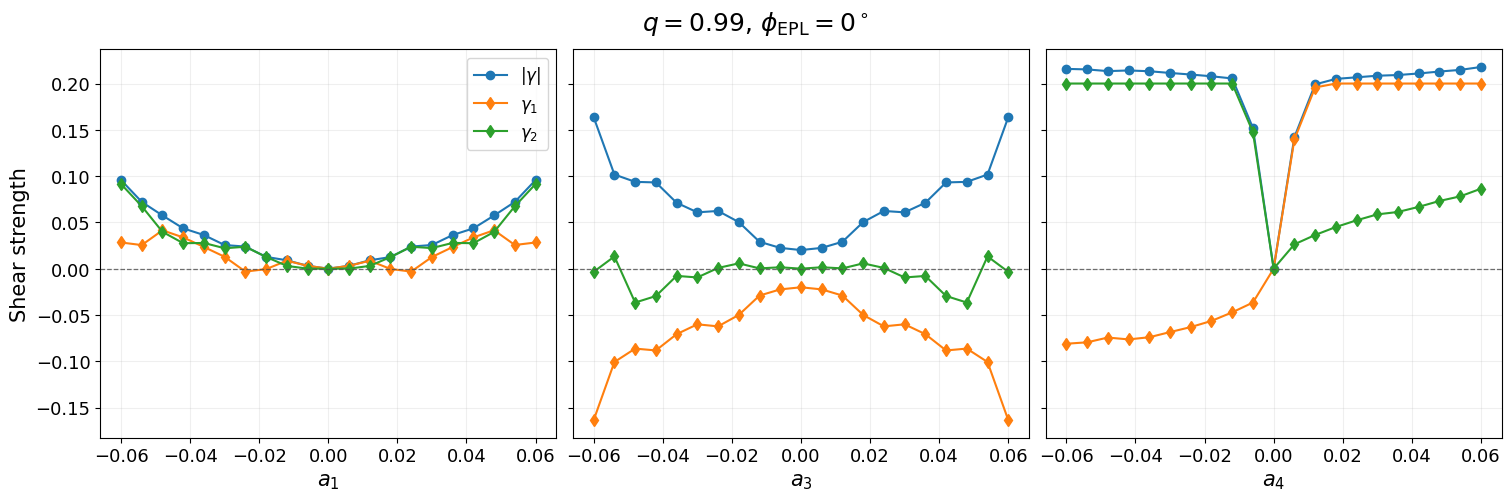}
    \end{subfigure}
    ~ 
    \begin{subfigure}{0.9\textwidth}
        \centering
        \includegraphics[height=1.8in]{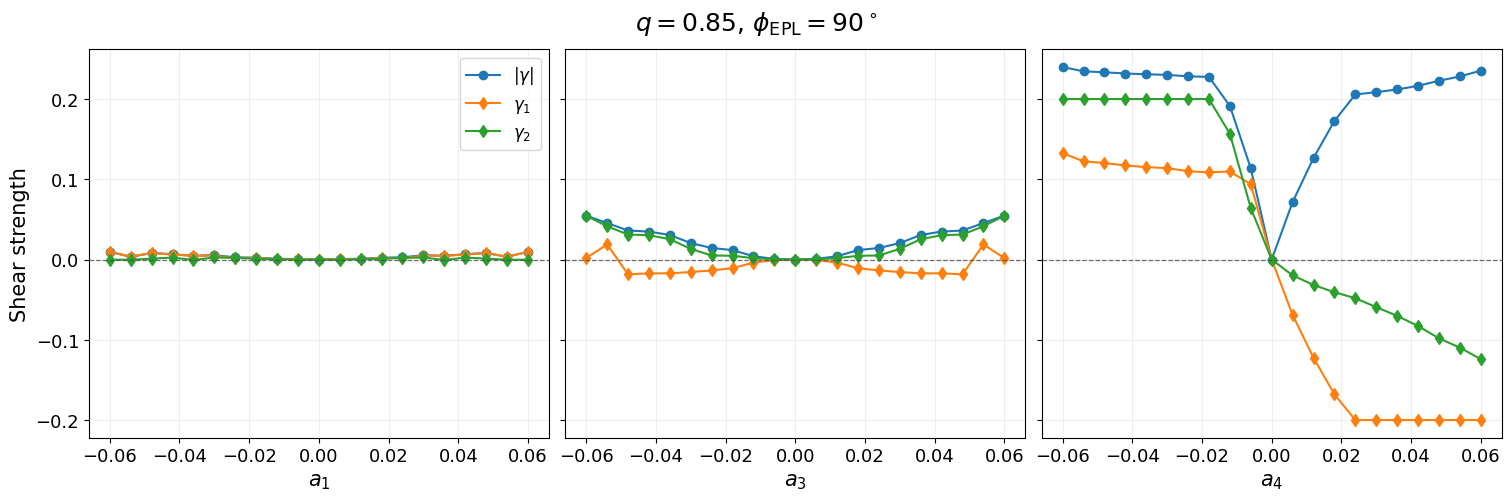}
    \end{subfigure}
    ~
    \begin{subfigure}{0.9\textwidth}
        \centering
        \includegraphics[height=1.8in]{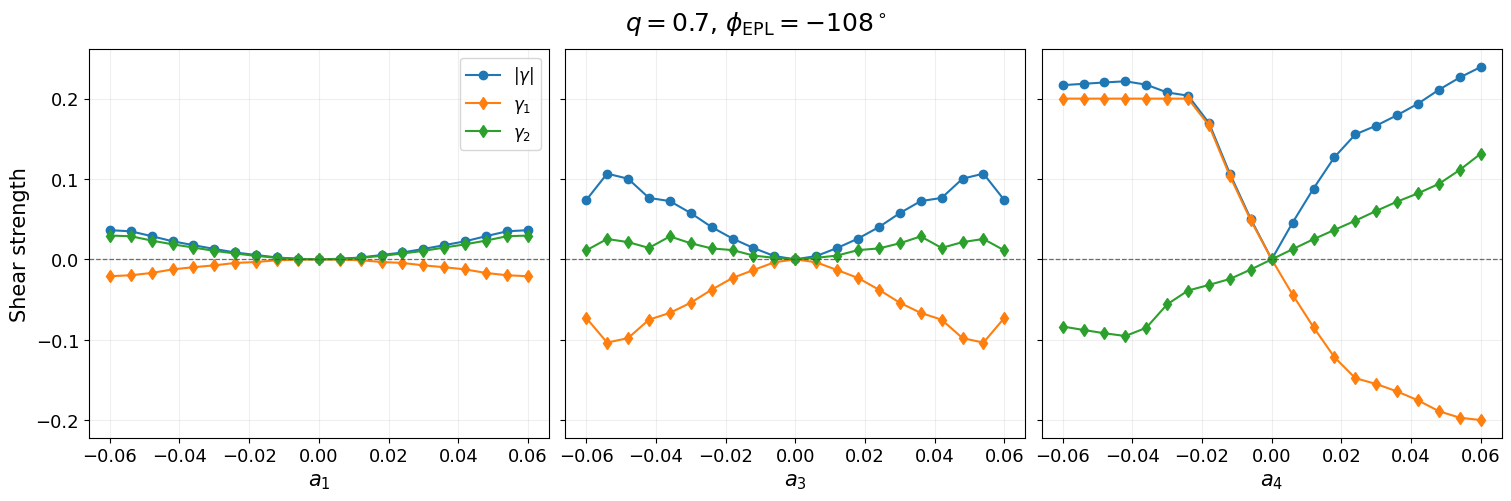}
    \end{subfigure}
    
    \caption{Plot of the inferred shear parameters against multipole strength for each of $m=1,3$ and $4$ from analysis of the critical-curve region of mock strong lensing galaxies for different ellipticity configurations and slope $\gamma=2.0$. An EPL+external shear model with no multipoles is fitted to the magnification map of fiducial models simulated with multipoles but no external shear to demonstrate how the external shear is degenerate with angular structures of the mass model. The $a_4$ appears to generate the largest shear response compared to $a_3$ and $a_1$, whilst the shear has a more symmetric variation with the odd-order multipoles for different signs of the amplitude.}
    \label{fig: Shear-compensating-multipole}
\end{figure*}

To demonstrate that low-order multipole perturbations in the lens potential can be absorbed by an effective external shear, we design a simple controlled mock-lensing experiment in which the fiducial model is an elliptical power-law lens augmented by $m=1$, $m=3$, and $m=4$ multipoles representative of the lenses analysed in this work. From this model we compute the determinant of the lens Jacobian,
\begin{equation}
\det A = \frac{1}{\mu},
\end{equation}
on a regular grid. 
%spanning $2.5$ arcsec
The critical curve is traced by the $\det A=0$ contour, and a region of interest is defined adaptively around the curve by selecting pixels with sufficiently small $|\det A|$. This focuses the comparison on the near-critical morphology of the lensing galaxy, where the degeneracy between shear and higher-order angular structure is expected to be strongest.

The mock determinant map is then fitted with a simpler EPL plus external-shear model using bounded least-squares minimization. The residuals are formed from the clipped determinant map within the near-critical region, so that the fit is driven by the shape of the critical-curve neighbourhood rather than by the full field. In practice, this recovers the best-fitting shear amplitude and orientation that compensate for the omitted multipole terms. 
% For the fiducial realisation, the inferred shear is non-zero even though the truth model contains no external tidal field, demonstrating that the simplified model absorbs part of the non-elliptical structure into the shear term. The residual maps remain small and localized, indicating that the EPL+shear model reproduces the dominant critical-curve geometry well despite being physically incomplete.
Figure \ref{fig: Shear-compensating-multipole} shows the results for a few different ellipticities and orientation of the EPL model. We see that the inferred shear is non-zero even though the true model contains no external tidal field, demonstrating that the simplified model absorbs part of the non-elliptical structure into the shear term. To isolate the effect of each multipole order, we repeat the experiment while varying one amplitude at a time. For each realization we refit the EPL+Shear model and record the recovered shear amplitude $|\gamma_{
\text{ext}}|$ (defined in Section \ref{sec: 5.2. Lens Mass Profile}).

We find that the recovered shear increases with the strength of the imposed multipole perturbation for all three cases, but notably the largest shear is obtained for the $a_3$ and $a_4$ deviations. The direction of the recovered shear also changes with the phase of the $a_4$ perturbation, showing that the degeneracy is vectorial in the shear plane rather than a simple scalar amplitude trade-off. This phase dependence is particularly important because it means that the same multipole strength can project onto different effective shear vectors depending on the orientation of the perturbation relative to the principal axes of the lens.

\begin{table*}
\centering
\caption{The residuals and inferred $H_0$ comparison for some typical multipole pertubation fractions for low host galaxy brightness. In the absence of clearly resolved or prominent lensed arcs, the multipoles can remain undetected and can lead to hidden biases in the constrained $H_0$.} %Comment on the choice of a_m parameters
\label{tab:}
\begin{tabular}{|c|c|c|c|}
\hline
\makecell{Multipole \\ deviation } & $a_4$ & $a_3$  & $a_1$ \\
\hline
$0.024$ & \makecell{ \includegraphics[width=0.15\linewidth]{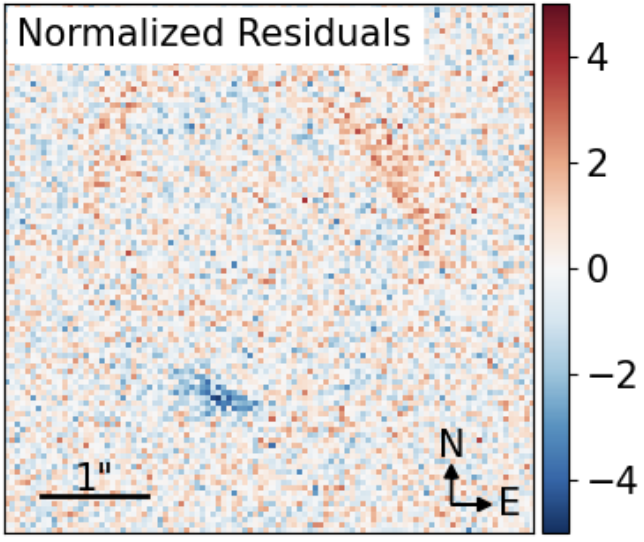} \\  $\chi^2_\nu = 1.090$ \\ $H_0 = 66.78$ \\ $\Delta H_0 = -4.59\%$ } & \makecell{ \includegraphics[width=0.15\linewidth]{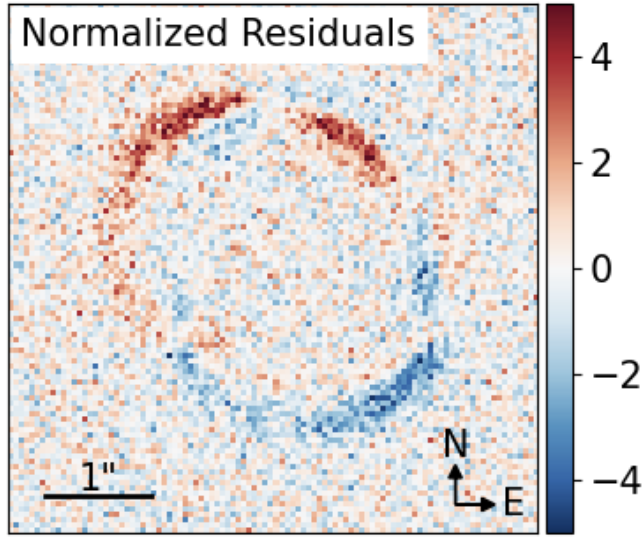} \\ $\chi^2_\nu=0.972$ \\ $H_0 = 76.40$ \\ $\Delta H_0 = +9.14\%$} & \makecell{ \includegraphics[height=0.9in]{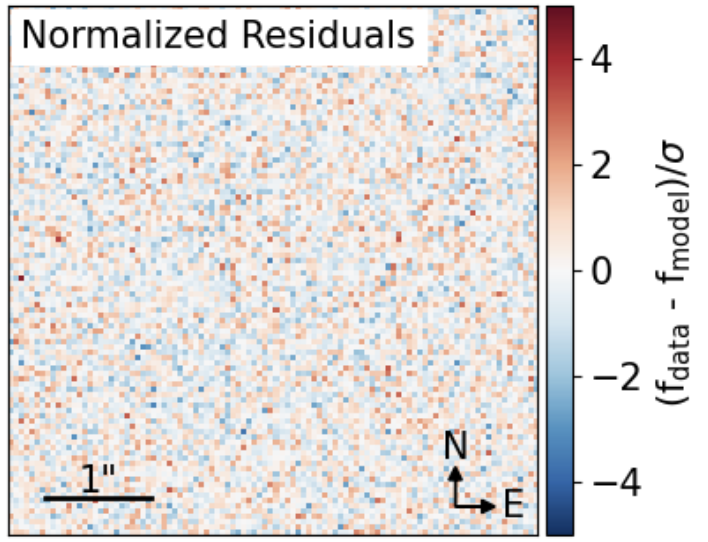} \\ $\chi^2_\nu=1.016$ \\ $H_0 = 72.24$ \\ $\Delta H_0 = +3.20\%$} \\

\hline

$0.011$ & \makecell{ \includegraphics[width=0.15\linewidth]{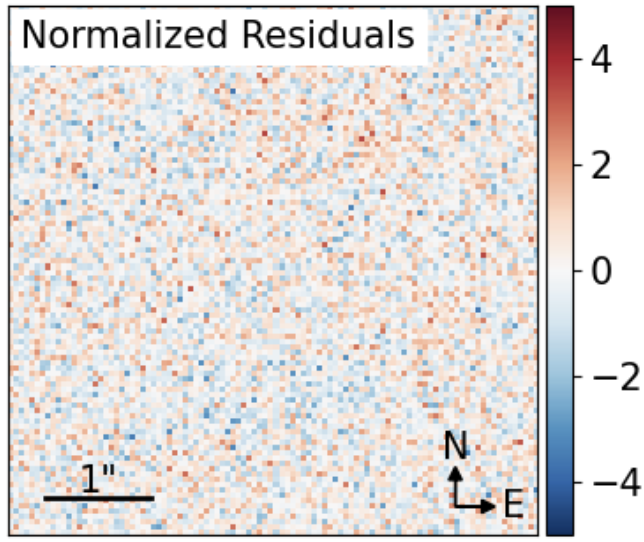} \\ $\chi^2_\nu = 1.022$ \\ $H_0 = 70.80$ \\ $\Delta H_0 = +1.14\%$ } & \makecell{ \includegraphics[width=0.15\linewidth]{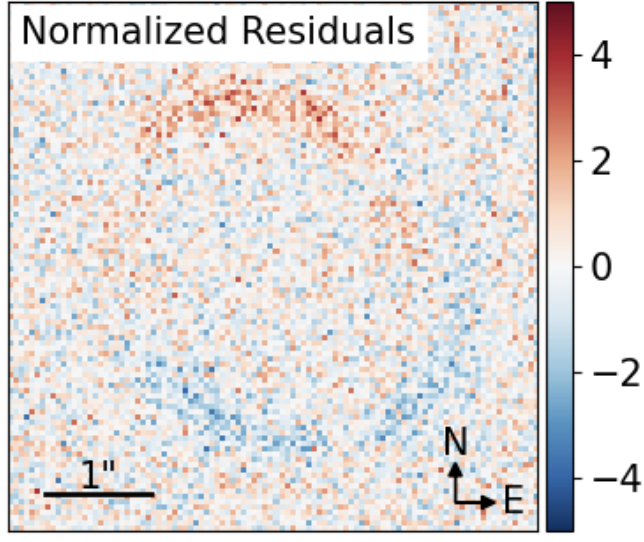} \\ $\chi^2_\nu = 1.078$ \\ $H_0 = 68.93$ \\ $\Delta H_0 = -1.51\%$ }  & \makecell{ \includegraphics[height=0.9in]{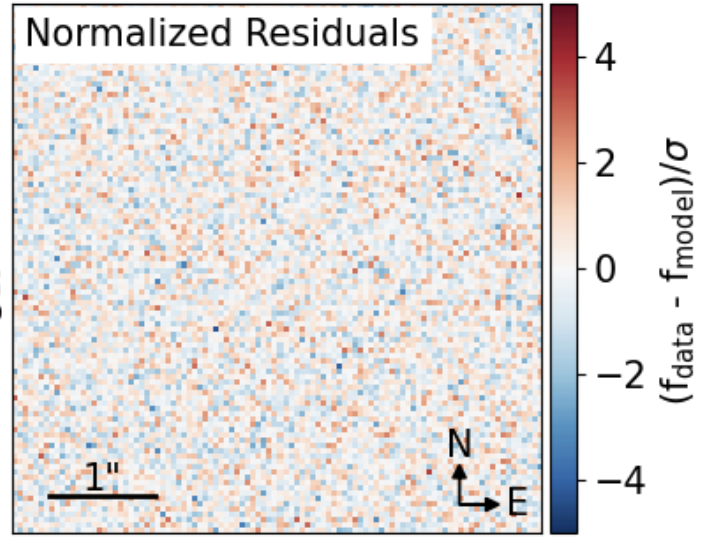} \\ $\chi^2_\nu = 0.972$ \\ $H_0 = 70.85$ \\ $\Delta H_0 = +1.2\%$ } \\ 
\hline

$-0.015$ & \makecell{ \includegraphics[width=0.15\linewidth]{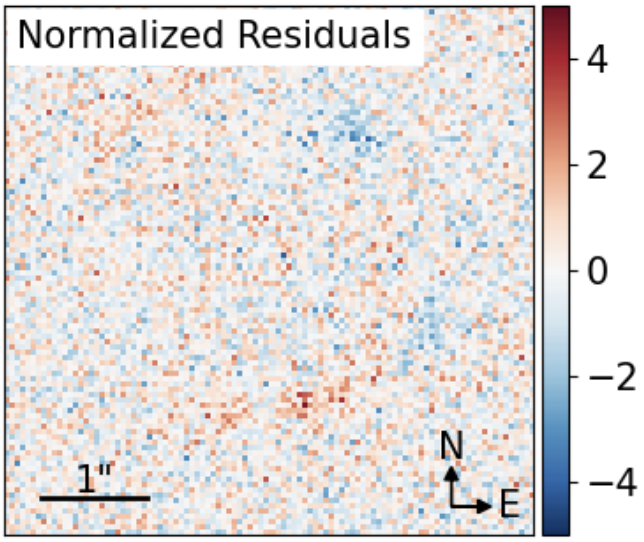} \\ $\chi^2_\nu = 1.047$ \\ $H_0 = 71.46$ \\ $\Delta H_0 = +2.09\%$ } & \makecell{ \includegraphics[width=0.15\linewidth]{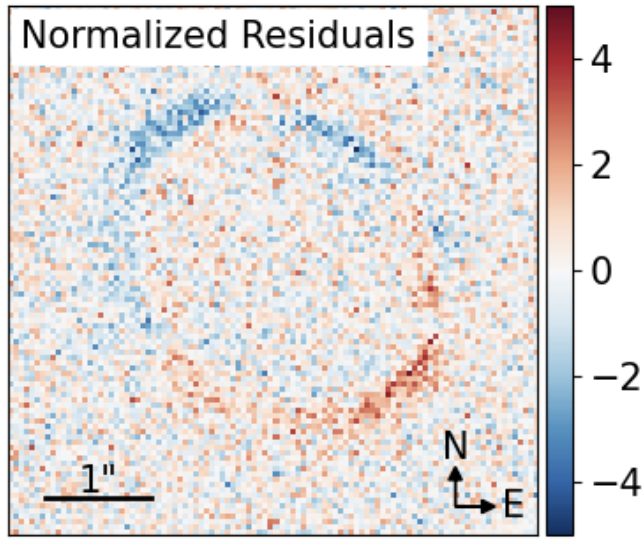} \\ $\chi^2_\nu = 1.156$ \\ $H_0 = 67.9361$ \\ $\Delta H_0 = -2.95\%$ } & \makecell{ \includegraphics[height=0.9in]{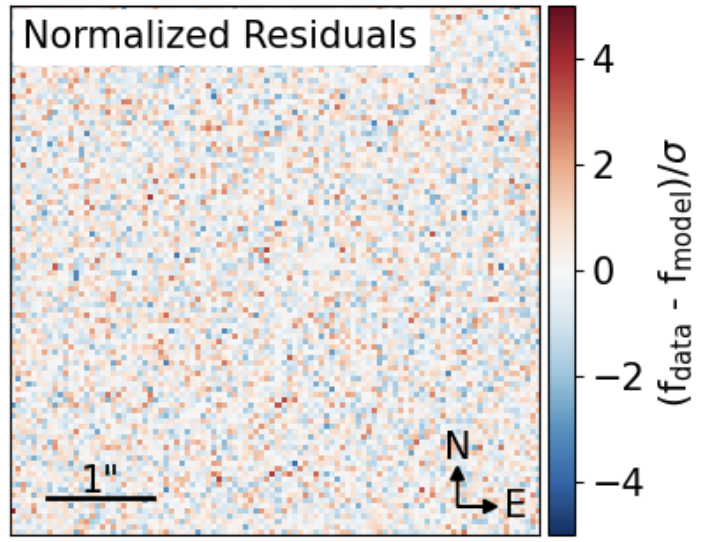} \\ $\chi^2_\nu=1.013$ \\ $H_0 = 68.64$ \\ $\Delta H_0 = -1.94\%$} \\
\hline

$-0.022$ 
& \makecell{ \includegraphics[width=0.15\linewidth]{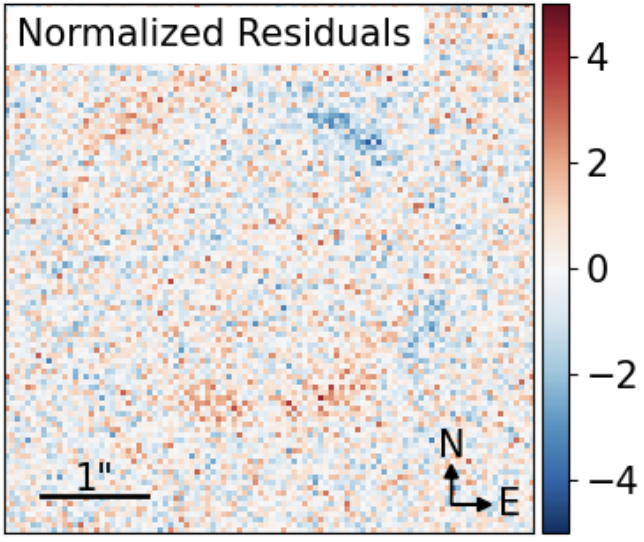} \\ $\chi^2_\nu = 1.052$ \\ $H_0 = 71.50$ \\ $\Delta H_0 = +2.16\%$} &
\makecell{ \includegraphics[width=0.15\linewidth]{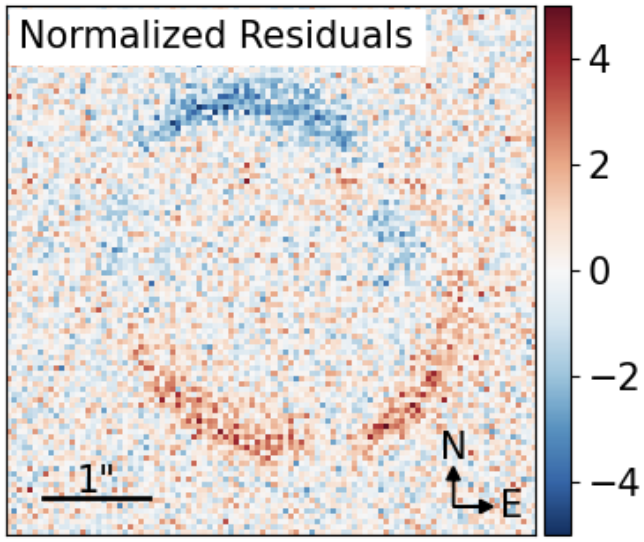} \\ $\chi^2_\nu = 1.301$ \\ $H_0 = 74.11$ \\ $\Delta H_0 = +5.87\%$ } & \makecell{ \includegraphics[height=0.9in]{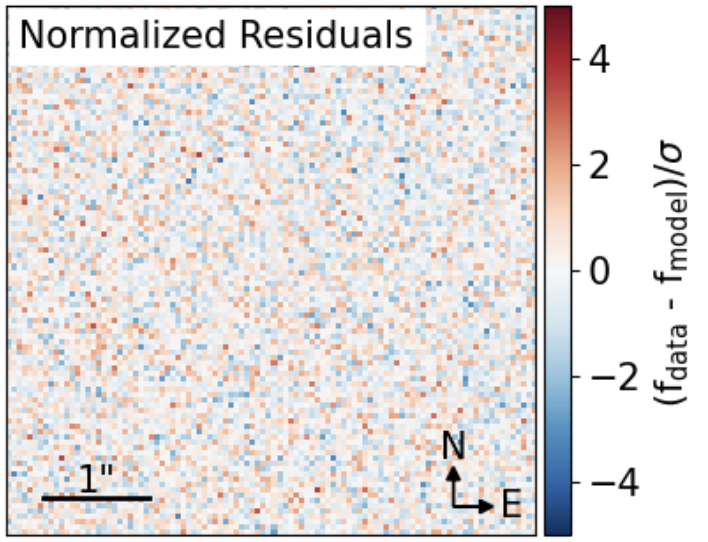} \\ $\chi^2_\nu=1.001$ \\ $H_0 = 68.17$ \\ $\Delta H_0 = -2.6\%$} \\
\hline
$-0.033$ & \makecell{ \includegraphics[width=0.15\linewidth]{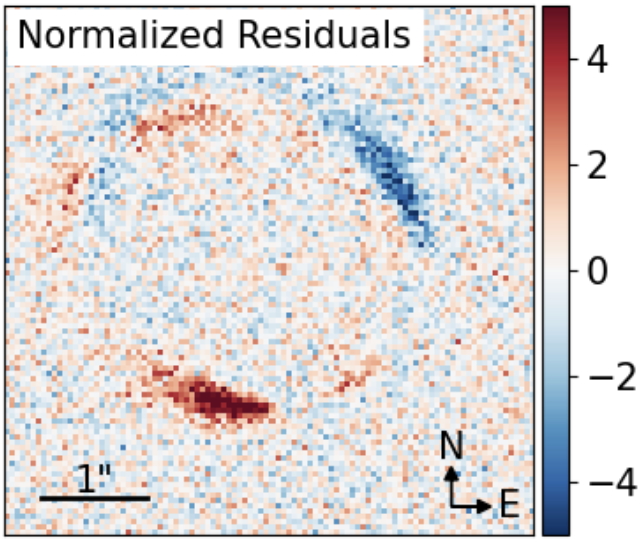} \\ $\chi^2_\nu = 1.443$ \\ $H_0 = 74.44$ \\ $\Delta H_0 = +6.34\%$ } & \makecell{ \includegraphics[width=0.15\linewidth]{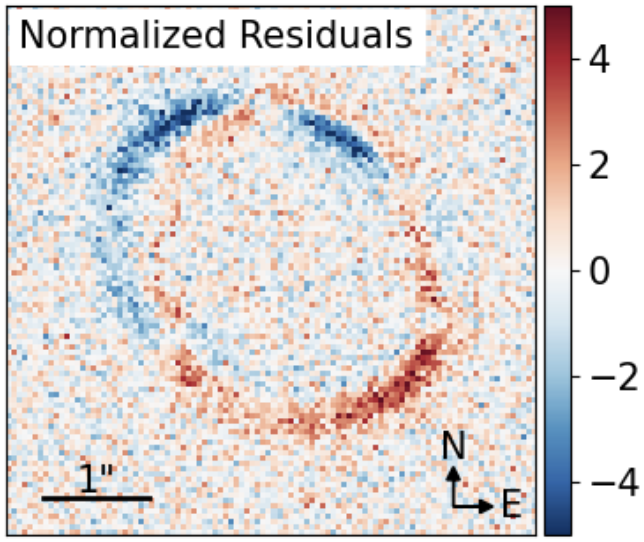} \\ $\chi^2_\nu = 1.546$ \\ $H_0 = 62.78$ \\ $\Delta H_0 = -10.3\%$ } & \makecell{ \includegraphics[height=0.9in]{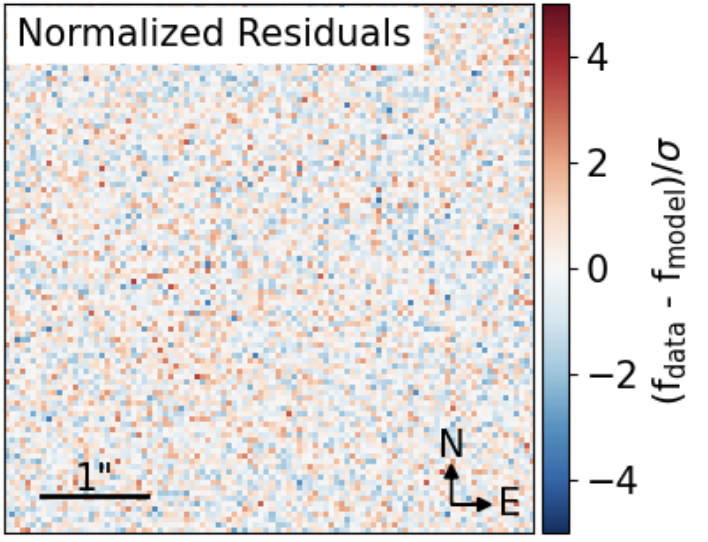} \\ $\chi^2_\nu=1.024$ \\ $H_0 = 73.12$ \\ $\Delta H_0 = +4.45\%$} \\
\hline
\end{tabular}
\label{tab: effect of source galaxy brightness}

\end{table*}

\section{Discussion}
\label{sec:discussion}

%\sdnotetwo{this is where comparisons to a) Van der vyere would be relevant. and also previous work with N-body sims,e.g. Ding+, Enzi+. Some of the substructure works e.g. Stacey, Nightingale}

\vspace{0.1mm}

Our results indicate that angular structures are not a rare correction to smooth elliptical lenses, but a common feature of massive early-type galaxies in the TNG population. The improvement in fit achieved by the EPL + multipole model relative to the EPL-only model shows that a purely elliptical description is often incomplete, even before one attempts to model the lensed source itself. In that sense, the multipole terms are not merely mathematical refinements, but capture departures from elliptical symmetry present in galaxy mass distributions.

\subsection{Multipole Distribution from IllustrisTNG}

The multipole amplitudes inferred here should be interpreted in the broader context of galaxy formation and the $\Lambda$CDM paradigm. In a hierarchical universe, ETGs are assembled through mergers, tidal interactions, and accretion. These processes leave behind low-order angular asymmetries in the projected mass distribution, such as lopsidedness ($m=4$) and triangularity ($m=3$) and boxy/discyness ($m=1$), which we find to be prevalent in the IllustrisTNG samples. By using self-consistent hydrodynamic simulations, we demonstrate that these higher-order modes are emergent properties of galaxy evolution. The fact that we find non-negligible $a_1$, $a_3$, and $a_4$ components suggests that lens models built only from smooth ellipticity plus external shear may absorb this internal structure only imperfectly. This is especially relevant because external shear can mimic some large-scale angular distortions \citep{etherington_strong_2024} without necessarily representing the true internal asymmetry of the deflector. 

Furthermore, the relevance of these distributions extends to dark matter substructure studies \citep[e.g.,][]{nightingale_scanning_2024, stacey_complex_2024}. Both smooth multipoles and localised subhaloes can perturb image positions, fluxes, and residuals. If smooth azimuthal perturbations in the main deflector are not explicitly accounted for, the resulting image residuals may be misinterpreted as evidence for CDM substructure, potentially contributing to false-positive detections \citep{oriordan_angular_2024}. In that sense, the multipole distributions derived here can serve as an astrophysical prior on lens asymmetry before residual anomalies are interpreted as evidence for substructure. This is not a substitute for direct subhalo inference, but it is a necessary step toward reducing false positives.

We should mention that the three orthogonal projections of the same TNG galaxy are not fully independent draws. They are best interpreted as distinct line-of-sight realisations of the same underlying population rather than separate galaxies. This means the effective statistical independence of the sample is lower than the raw number of projections suggests, and the quoted distributions should be interpreted accordingly. Even so, the qualitative conclusion remains unchanged: angular complexity is common enough to matter for lens modelling.

\subsection{Time-delay cosmographic inference of the Hubble constant}

\subsubsection{Bias on $H_0$ from multipole omission}

Our time-delay cosmography experiments show that omitting multipoles can bias individual $H_0$ at the several-percent level for realistic perturbation amplitudes. This is comparable to the $\lesssim 2\%$ precision target required for a $5\sigma$ test of the Hubble tension, so the effect is not negligible. 

A useful comparison can be made with \citet{vyvere_tdcosmo_2022}, who examined the impact of fourth-order multipoles. While we agree that $a_4$ introduces a directional bias, it is noteworthy that the sign of our $H_0$ bias relative to the $a_4$ amplitude appears to be flipped compared to their findings. We find that $H_0$ is generally negatively biased for positive $a_4$ (discy) configurations. This discrepancy likely stems from the transition from the circular potential-based multipoles used by their paper following the convention of \citet{xu_how_2014} to the elliptical isodensity multipoles of \citet{paugnat_elliptical_2025} used in this work. The factor of $(1-m^2)^{-1}$ in the Poisson equation implies that a positive mass perturbation in convergence corresponds to a negative perturbation in the lensing potential. Moreover, \citet{paugnat_elliptical_2025} demonstrated that circular multipoles on an elliptical deflector can force unphysical parametric compensations in the Einstein radius and shear to fit image positions. By using the elliptical convention, our results provide a more physically consistent representation of how $H_0$ is biased by astrophysical boxy/discyness.

\subsubsection{Interpretation of Population-Level Biases}

The multipole-weighted $H_0$ distributions presented in Section~\ref{sec: results of impact of multipole mismatch} may provide insights into how unmodelled angular complexity affects the precision of time-delay cosmography on a population-wide scale. Under the assumption that the distributions of $a_1$, $a_3$, and $a_4$ in the real Universe are accurately represented by the observational sample of \cite{hao_isophotal_2006} or our TNG-derived sample, the net bias on $H_0$ appears to be at the sub-percent level. 
This suggests a partial statistical cancellation effect where the negative biases are offset by positive biases within a large enough sample. However, we emphasize that the true underlying distribution of mass multipoles in the Universe remains a significant unknown. Our analysis relies on the \citet{hao_isophotal_2006} SDSS catalogue, which traces stellar light isophotes, and the IllustrisTNG suite, which may still be subject to sub-grid physics uncertainties and resolution-dependent effects. If the true deflector population in strong lensing surveys exhibits a significant skew -- for instance, if lensing selection effects favour more discy galaxies due to gas-rich minor mergers at $z \sim 0.5$ or if the dark matter halo orientation systematically enhances lopsidedness -- the population-averaged bias could realistically exceed the 1\% level. As Figures \ref{fig: weighted H0 PDF 1} and \ref{fig: weighted H0 PDF m=4} demonstrate using possible distributions of $a_m$ that have positively or negatively-skewed medians, even a moderate shift in the frequency of specific multipole amplitudes can shift the median of the combined $H_0$ PDF by several percent, notably for $a_4$ which produces a more directional bias.

The response to multipoles is heavily modulated by the image configuration as well. For $m=3$ mode in particular, the fold and cusp configurations exhibit opposite responses, and therefore the population median and spread in $H_0$ are likely to be sensitive to the fraction of folds, cusp and cross strong lenses present in the cosmographic sample. For instance, a sample consisting of a large fraction of cusp lenses with mostly $a_3<0$ can result in an underestimation of $H_0$. Interestingly, such sensitivity to the location and spread of the $H_0$ distribution is not observed for $a_1$, where the biases are somewhat consistent with increasing amplitude strength till it exceeds 4\%. This variation could be due to compensation by other parameters, in particular a shift in the deflector centre. We do not think that the $m=1$ multipole will contribute to a significant systematic bias in $H_0$ at population-level for a large cosmographic sample.

% \subsubsection{Decoupling Measurement Noise and Fermat Potential Systematic Errors}
% A key question raised in interpreting the dispersion of the $H_0$ distributions is whether the observed width is dominated by the input observational time-delay uncertainties ($\sigma_{\Delta t} \sim 0.5\text{--}2$~days) or by an irreducible systematic scatter originating from the unmodelled mass distribution. Because our simulated time-delay noise is drawn from a symmetric Gaussian distribution, it broadens individual posteriors but introduces no net directional shift in the ensemble median. Consequently, the systematic offsets in median $H_0$ shown in Figures \ref{fig: Model mismatch Impact on H0} $-$ \ref{fig: am_weighted_H0_distributions_by_configuration} are driven by the failure of the standard EPL model to accurately recover the Fermat potential differences:
% \begin{equation}
%     \frac{\Delta H_{0}}{H_{0, \textrm{ fiducial}}} \approx \frac{\Delta \phi_{\rm inferred} - \Delta \phi_{\rm true}}{\Delta \phi_{\rm true}}.
% \end{equation}

\subsubsection{Absorption of complexity and undetected azimuthal structures}

We find that external shear acts as a highly effective mathematical "sink" for unmodelled internal multipoles 
%We demonstrate that even in a system with zero true environmental tidal field, a standard EPL+shear model will recover a non-zero shear to compensate for $m=1, 3,$ and $4$ mass perturbations. 
and confirms the suspicions raised by \citet{etherington_strong_2024} that shear can act as a "fudge factor" that mimics internal angular mass complexity. This has two major implications for cosmography. First, it suggests that the shear values reported in many high-precision lens models may be partially contaminated by internal galaxy structure, complicating efforts to cross-correlate lensing shear with independent environmental weak-lensing measurements. Second, because the shear magnitude is intimately tied to the radial mass distribution and the resulting Fermat potential, this absorption mechanism may be a major driver of the $H_0$ biases and hidden systematics identified in Table~\ref{tab: effect of source galaxy brightness}.

The scope of a "hidden bias" is particularly concerning as demonstrated in Section \ref{sec: effect of lensed arc observation}. 
The scatter of $\pm5-10$\% seen in Table \ref{tab: effect of source galaxy brightness} is concerning for the aspiration of resolving the Hubble tension. If a significant fraction of the lens population contains azimuthal complexity, and if our observations lack the depth to detect this complexity, the total error budget will be dominated by this unaccounted-for systematic. 
%This reinforces the necessity of adopting the population-based priors suggested by \cite{oh_joint_2025}. By including an astrophysical expectation of multipole strengths, we could marginalize over these hidden biases even when the data S/N is insufficient to detect them directly. 
As \cite{2025MNRAS.539..704Lange} demonstrated with JWST imaging, using higher S/N and multi-wavelength imaging data is often the only way to overcome these biases, as the increased azimuthal information forces the residuals to manifest if the mass model is under-parametrised.

%The strongest biases appear when the source arcs are weak or poorly resolved, because the multipole structure becomes harder to identify from the imaging data alone. This is particularly important for transients such as lensed supernovae, where the host galaxy may be faint or only partially visible. The result is that a formally acceptable fit can still conceal a non-trivial systematic bias in H0.

\subsection{Outlook for future work}

%\sdnotetwo{HST PSFs, other forms of angular structure?}

While this work provides a systematic characterisation of multipole perturbations in simulated massive elliptical galaxies, several natural extensions are required to bridge the gap between these results and high-precision cosmography.

The first is to replace the idealized Gaussian PSF and simplified noise model with more realistic observing conditions, especially for \textit{HST}, \textit{JWST}, and future \textit{Euclid}-like data. As demonstrated in Section \ref{sec: time-delay and H0 inference}, multipole signatures are extracted from subtle structures in the image residuals. In real observations the PSF is complex, spatially varying, and often undersampled. An imperfect PSF reconstruction can %either mimic higher-order azimuthal structures or
smear out genuine multipole signals, exacerbating the biases described in this work. Future studies should employ realistic, data-derived PSF kernels to quantify the interplay between deconvolution artifacts and mass complexity.

The frequency and magnitude of $a_1$, $a_3$, and $a_4$ multipoles may depend on the specific assembly history of the lens. It would be valuable to investigate how these distributions correlate with redshift, stellar mass, and local environment richness. Establishing an astrophysical prior based on galaxy demographics would allow for more rigorous marginalisation over mass complexity in automated pipelines, such as those being developed for LSST and \textit{Euclid}.
Furthermore, the degeneracy between the radial profile and angular structure could be significantly reduced by the inclusion of independent observables. Jointly modelling lensing with stellar kinematics or multi-filter imaging (exploiting the wavelength-independence of the mass model) remains the most promising path toward breaking the degeneracies identified here and achieving a total $H_0$ error budget below 2\% \citep{ding_time_2021}.

\section{Conclusion}
\label{sec:conclusion}

%The quad configuration is generated by positioning the source slightly off centre within the inner lensing caustic (Figure \ref{fig: Mock quad images}).
In this work, we have investigated the prevalence of higher-order angular structure in realistic massive elliptical galaxies and quantified its impact on the inference of the Hubble constant via strong gravitational time-delay cosmography. By utilising the IllustrisTNG hydrodynamic simulation suite, we moved beyond the idealised elliptical mass profiles typically used in the literature to explore how self-consistent astrophysical asymmetries affect cosmological constraints. Our primary conclusions are as follows:

\begin{enumerate}
    \item Higher-order multipoles are an emergent feature of massive early-type galaxies in a $\Lambda$CDM framework. We find that $m=1$ (lopsidedness), $m=3$ (triangularity), and $m=4$ (boxy/diskyness) components are present at non-negligible levels across the TNG population. In 54\% of our galaxies from the IllustrisTNG 100 and 85\% of samples from the TNG50 simulation suites, the addition of these multipole terms to a standard Elliptical Power Law (EPL) model is significantly preferred by the Bayesian information criterion.

    \item Ignoring these angular structures leads to systematic errors in the inferred $H_0$. Consistent with recent frameworks, we find that discy ($a_4 > 0$) mass distributions lead to a systematic underestimation of $H_0$ when modelled with pure EPL profiles and vice versa for boxy ($a_4 < 0$) distributions. While the population-averaged bias may remain small ($< 1\%$), the bias for individual systems can reach several percent. The odd-order multipoles ($m=1, 3$) introduce a significant stochastic scatter in the $H_0$ inference as well. These modes, which arise from hierarchical assembly and physical companions, represent a significant source of angular noise that broadens the effective uncertainty of individual lens measurements.
    
    \item Our tests under low-S/N conditions (faint host galaxy arcs) demonstrate that multipole structures can remain undetected in image residuals ($\chi^2_\nu \approx 1$) while still inducing $H_0$ biases of several percent. This hidden bias confirms that a statistically satisfactory fit to the imaging data is not a guarantee of cosmological accuracy, as the standard EPL+external shear model possesses parametric flexibility to absorb complex mass features at the cost of the correct Fermat potential difference.

    \item We explicitly quantify how low-order multipoles are absorbed by the external shear component in standard models. We find that the recovery of a non-zero effective shear in the absence of a true tidal field is a pervasive feature when internal complexity is ignored. This degeneracy is particularly strong for the $a_3$ and $a_4$ modes, indicating that a significant portion of the shear inferred in time-delay cosmography may be an astrophysical manifestation of internal mass asymmetries rather than purely environmental in origin.
\end{enumerate}

The characterisation of smooth angular complexity is also essential for the robust detection of dark matter substructure. If azimuthal perturbations are not explicitly accounted for in the main deflector, they can be misinterpreted as evidence for low-mass subhaloes. Our TNG-derived distributions may provide an astrophysical prior to distinguish between smooth baryonic/dark matter multipoles and the localised perturbations predicted by $\Lambda$CDM.

As time-delay cosmography enters the era of survey-scale samples from \textit{Euclid}, LSST, and \textit{Roman}, our results highlight a critical need for more flexible mass models. Achieving 2\% accuracy in 
$H_0$ will require moving beyond the standard EPL assumption and incorporating astrophysically informed priors on angular complexity. Future work incorporating multi-wavelength data and stellar kinematics will be vital to break the degeneracies identified here and provide a robust resolution to the expansion rate of the Universe.

%%%%%%%%%%%%%%%%%%%%%%%%%%%%%%%%%%%%%%%%%%%%%%%%%%
%\section*{Data Availability}

\section{Acknowledgements}
The authors are grateful to Sean McGee for providing valuable comments that helped improved this manuscript.
A.A.Z. acknowledges support from the Institute of Astronomy, Cambridge, through its 2025 Summer Research Bursary. S.D. acknowledges support from UK Research and Innovation (UKRI) under the UK government’s Horizon Europe funding Guarantee EP/Z000475/1. E.E.H. is supported by a Gates Cambridge Scholarship (\#OPP1144).

\section{Data Availability}

The code (Jupyter notebooks and python scripts) used for the analysis in this paper will be made available after peer-review or on reasonable request.

% is available at: \hyperlink{https://github.com/zaif651/TNG_Multipoles_Lensing.git}{zaif651/TNG$\_$Multipoles$\_$Lensing.git}



\bibliographystyle{mnras}
\bibliography{Lensing_fin} % if your bibtex file is called example.bib

%\begin{thebibliography}{99}
%\end{thebibliography}

%%%%%%%%%%%%%%%%%%%%%%%%%%%%%%%%%%%%%%%%%%%%%%%%%%

%%%%%%%%%%%%%%%%% APPENDICES %%%%%%%%%%%%%%%%%%%%%

\appendix

\section{Synthetic null tests of EPL--multipole model selection}
\label{app:epl_null_tests}

We performed controlled synthetic-map experiments to assess whether the
preference for the EPL+Multipole model could arise spuriously from particle
sampling, particle-to-grid mapping, adaptive smoothing, or the model-selection
procedure. We generated synthetic convergence maps from smooth elliptical
power-law (EPL) profiles, varying representative Einstein radii, logarithmic
slopes, axis ratios, orientations, and small centroid offsets. Particles
representing stellar, dark-matter, and gas populations were sampled from the
smooth projected profiles using population-dependent particle counts and
lognormal particle masses. Subpixel particle positions were then assigned,
and the particles were mapped onto a two-dimensional grid using the adaptive
triangular-shaped-cloud (ATSC) prescription with the chosen nearest-neighbour
parameter, $k_{\rm nn}$. The resulting convergence and uncertainty maps were
processed using the same annular mask and fitting procedure as the TNG maps.
By construction, these synthetic maps contain no intrinsic multipole
perturbations.

For each realization, we fitted both an EPL-only model and an EPL model including
the $m=1$, $m=3$, and $m=4$ angular perturbations. Both models were evaluated
using the same data pixels and diagonal Gaussian likelihood,
\begin{equation}
 \ln L(\boldsymbol{\theta})
 =
 -\frac{1}{2}\sum_{i=1}^{N_{\rm pix}}
 \left[
 \frac{\kappa_i-\kappa_i^{\rm model}(\boldsymbol{\theta})}
 {\sigma_i}
 \right]^2 ,
\end{equation}
where $\kappa_i$ and $\sigma_i$ denote the mapped convergence and its assigned uncertainty in pixel $i$. We evaluated the Bayesian information criterion (BIC) using the same convention as in the TNG analysis given in Equation \ref{eqn: BIC definition},
% \begin{equation}
%  {\rm BIC}=k\ln N_{\rm pix}-2\ln L_{\max},
% \end{equation}
and defined
\begin{equation}
 \Delta{\rm BIC}
 =
 {\rm BIC}_{\rm EPL}
 -
 {\rm BIC}_{\rm EPL+Multipole}.
\end{equation}
Thus, $\Delta{\rm BIC}>0$ generally favours the EPL+Multipole prescription. The EPL and EPL+Multipole models have six and twelve free parameters, respectively; the additional six parameters must therefore yield a sufficient improvement in the maximum likelihood to offset the corresponding BIC penalty. Fits were
initialised using differential evolution and refined using MCMC sampling, with the maximum likelihood used for the BIC comparison.

\subsection{Synthetic Ensembles and Validation Result}

\begin{figure}
    \centering
    \includegraphics[width=1.\linewidth]{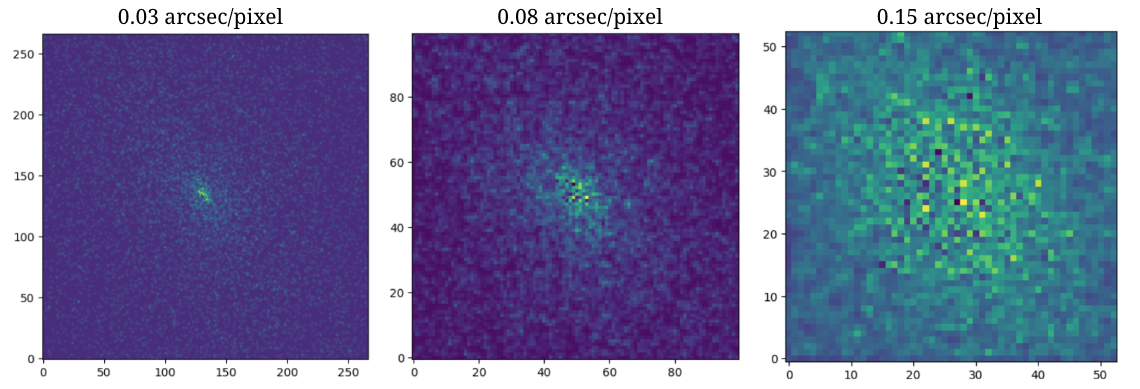}
    \caption{Synthetic Elliptic Power Law galaxy mass profiles sampled with TNG-like particles at 3 different pixel resolutions: $0.03^{\prime\prime}$/pixel, $0.08^{\prime\prime}$/pixel, $0.15^{\prime\prime}$/pixel, }
    \label{fig:placeholder}
\end{figure}

We generated ensembles of 40 true EPL realizations, varying the underlying EPL parameters, particle counts, and particle-sampling seeds between
realizations. We repeated the experiment for several map configurations, using pixel scales of $0.03$, $0.08$, and $0.15\,{\rm arcsec}$ and adaptive smoothing settings corresponding to nearest neighbour parameter $k_{\rm nn}=8$, $16$, and $32$. The results for each tested configuration are summarised in
Table~\ref{tab:epl-null-results}.

\begin{table}
\centering
\caption{Results of the synthetic EPL null tests. ``EPL $\rightarrow$ M''
denotes the number of true EPL realisations for which EPL+Multipole was
selected, and the false-multipole rate is this count divided by the number
of completed realizations.}
\label{tab:epl-null-results}
\begin{tabular}{lrrr}
\hline
Configuration & $N_{\rm EPL}$ & EPL $\rightarrow$ M &
False-multipole rate \\
\hline
$0.03''$, $k_{\rm nn}=8$  & 40 & 1  & $0.025$ \\
$0.03''$, $k_{\rm nn}=16$ & 40 & 2  & $0.05$ \\
$0.03''$, $k_{\rm nn}=32$ & 40 & 0  & $0.00$ \\
$0.08''$, $k_{\rm nn}=8$  & 40 & 2 & $0.05$ \\
$0.08''$, $k_{\rm nn}=16$ & 40 & 1  & $0.025$ \\
$0.08''$, $k_{\rm nn}=32$ & 40 & 3  & $0.075$ \\
$0.15''$, $k_{\rm nn}=8$  & 40 & 0  & $0.00$ \\
$0.15''$, $k_{\rm nn}=16$ & 40 & 0  & $0.00$ \\
$0.15''$, $k_{\rm nn}=32$ & 40 & 0  & $0.00$ \\
\hline
\end{tabular}
\end{table}

Across the 8 complete configurations, the more complex EPL+Multipole model was selected in 9 of the 120 true EPL realisations via the BIC ($7.5$\% false positive preference). The false-multipole
rate within each complete configuration ranged from zero to $0.075$ at most. Even in cases where a false positive preference was made for EPL+Multipole model, the $\Delta$BIC was low ($\lesssim$ 100) that it does not lend to a substantially strong evidence for the more complex profile.

These results indicate that the synthetic EPL maps do not routinely produce a preference for EPL+Multipole under the tested configurations. In particular, no false multipole selections occurred in four of the 8 complete ensembles, while the other two yielded low rates of detections.
The tests therefore provide an empirical check that particle sampling and ATSC map construction do not invariably generate a spurious preference for
the more flexible model. However, the observed rates are conditional on the synthetic profiles, particle-sampling prescriptions, uncertainty estimates, and fitting configurations considered here. They should not be interpreted
as universal false-positive probabilities for the TNG population.

\subsection{Sensitivity and Limitation}

The null tests address whether the adopted map-making and fitting pipeline can spuriously favour EPL+Multipole when the underlying projected profile is a smooth EPL. They do not establish that the standard BIC is exactly calibrated for maps with spatially correlated pixels. In particular, the diagonal likelihood treats the pixel uncertainties as independent, whereas particle-to-grid assignment and adaptive smoothing can induce correlations between neighbouring pixels. Consequently, the BIC values should be interpreted as a consistent model-comparison diagnostic within the adopted
pipeline, rather than as formal evidence calibrated under an independent-pixel assumption. An estimate of the effective number of independent spatial elements from residual autocorrelation could provide
additional diagnostic information, but should not be substituted mechanically for $n$ in the BIC penalty. We therefore use the synthetic tests as an empirical robustness check, while retaining the
caveat that the standard BIC comparison does not explicitly model the full spatial covariance of the mapped convergence field.

\section{Sensitivity of multipole inferences to map resolution}
\label{app:resolution_tests}

To assess the sensitivity of our inferred angular structure to the
resolution of the projected convergence maps, we repeated the lens-model
comparison at pixel scales of $0.03''$, $0.08''$, and $0.15''$. At each
resolution, we compared the elliptical power-law (EPL) model with the
EPL+Multipole model, which includes the $m=1$, $m=3$, and $m=4$ convergence
multipole terms. The same set of 1791 successful fits for TNG50 and TNG100 was available at
each resolution. We quantified the preference for the more complex model both at the level of individual paired fits and at the galaxy level,
where the latter accounts for the multiple projections associated with
individual galaxies. We also examined the normalised $a_m \frac{\sqrt{q}}{\theta_E} \cos(m\phi_m)$ among fits for which the EPL+Multipole model was
preferred, in order to assess whether the inferred multipole structure exhibits a systematic directional preference.

\begin{table}
    \centering
    \caption{Model preference as a function of convergence-map pixel
    scale. The preferred-pair fraction measures the fraction of individual
    paired comparisons favouring EPL+Multipole, whereas the preferred-galaxy
    fraction measures the corresponding fraction at the galaxy level.
    The number of preferred fits is reported for the individual paired
    comparisons.}
    \label{tab:resolution_preference}
    \begin{tabular}{lccc}
        \multicolumn{4}{c}{TNG100} \\
        \hline
        Pixel scale
        & Preferred fits
        & \makecell{Preferred-pair \\ fraction}
        & \makecell{Preferred-galaxy \\ fraction} \\
        \hline
        $0.03''$ & 866/1542 & $56.2\%$ & $56.0\%$ \\
        $0.08''$ & 780/1542 & $55.4\%$ & $55.9\%$ \\
        $0.15''$ & 918/1542 & $59.5\%$ & $60.1\%$ \\
        \hline
    \end{tabular}

    \begin{tabular}{lccc}
        \multicolumn{4}{c}{TNG50} \\
        \hline
        Pixel scale
        & Preferred fits
        & \makecell{Preferred-pair \\ fraction}
        & \makecell{Preferred-galaxy \\ fraction} \\
        \hline
        $0.03''$ & 231/249 & $85.9\%$ & $92.8\%$ \\
        $0.08''$ & 240/249 & $91.2\%$ & $96.4\%$ \\
        $0.15''$ & 246/249 & $96.0\%$ & $98.8\%$ \\
        \hline
    \end{tabular}
\end{table}

The results are summarised in Tables B1. 
Overall, the tests show that both the relative preference for including multipoles and the distribution of their signed coefficients depend to some extent on the map resolution, but is not significantly discrepant between different map resolutions. The model-preference fractions do not vary monotonically with pixel scale, indicating that the results cannot be characterised simply as becoming more or less robust with coarser sampling. The distribution of the $a_1, a_3$ and $a_4$ strengths
We find that approximately 50\% of TNG100 and 85\% of TNG50 galaxy projections do consistently prefer the more power-law with multipole model across the three resolutions. 
% These results should therefore be regarded as a resolution
% sensitivity test rather than, on their own, evidence for or against
% intrinsic directional structure in the lens population.

% \begin{figure}
%     \centering
%     \includegraphics[height = 3.6in, angle=90]{Figures/Model_Convergence_Fit_Results/Multipole-EPL-correlation.png}
%     \caption{}
% \end{figure}

% \newpage

% \begin{figure}
%     \centering
%     \includegraphics[height = 3.7in, angle=90]{Figures/Model_Convergence_Fit_Results/Multipole-Delta_EPL-correlation.png}
%     \caption{}
% \end{figure}

%%%%%%%%%%%%%%%%%%%%%%%%%%%%%%%%%%%%%%%%%%%%%%%%%%

% Don't change these lines
\bsp	% typesetting comment
\label{lastpage}
\end{document}